\documentclass[prl, twocolumn, superscriptaddress]{revtex4-1}
\usepackage{amsmath}
\usepackage[english]{babel}
\usepackage[pdftex]{graphicx}
\usepackage{color}
\usepackage{ulem}
\usepackage{comment}

\usepackage[sort&compress]{natbib}
\usepackage{hyperref}

\graphicspath{{images/}{./}}

\usepackage{amsmath}
\usepackage{amssymb}
\usepackage{mathtools}
\usepackage{braket}
\usepackage{booktabs}

\renewcommand{\vec}[1]{\ensuremath{\boldsymbol{#1}}}

\newcommand{\comA}[1]{\textcolor{black}{#1}}

\newcommand{\comD}[1]{\textcolor{black}{#1}}

\newcommand{\abs}[1]{\left| #1 \right|}

\renewcommand{\deg}{^{\circ}}
\renewcommand{\ap}{\alpha}
\renewcommand{\th}{\theta}

\newcommand{\sgn}{\,\mbox{\rm sgn}}

\begin{document}

\title{Topical Review: The rise of Klein tunneling in low-dimensional materials and superlattices}
\author{Yonatan Betancur-Ocampo}
\email{ybetancur@fisica.unam.mx}
\affiliation{Instituto de F\'isica, Universidad Nacional Aut\'onoma de M\'exico, Ciudad de México, Mexico}

\author{Guillermo Monsivais}
\email{monsi@fisica.unam.mx}
\affiliation{Instituto de F\'isica, Universidad Nacional Aut\'onoma de M\'exico, Ciudad de México, Mexico}

\author{Vít Jakubský}
\email{ jakubsky@ujf.cas.cz}
\affiliation{Nuclear Physics Institute, Czech Academy of Science, 250 68 \v Re\v z, Czech Republic}

\begin{abstract}
 We review recent advances in Klein and anti-Klein tunneling in one- and two-dimensional materials. Using a general tight-binding framework applied to multiple periodic systems, we establish the criteria for the emergence of Klein tunneling based on the conservation of an effective reduced pseudo-spin. The inclusion of higher-order terms in the wave vector leads to nontrivial matching conditions for wave scattering at interfaces. We further examine the emergence of multiple types of Klein tunneling in two-dimensional materials beyond graphene, including phosphorene and borophene, as well as in one-dimensional systems such as Su–Schrieffer–Heeger lattices. Finally, we discuss how these tunneling phenomena can be tested in both synthesized and artificial lattices, including elastic metamaterials, optical, photonic, phononic, and superconducting platforms, demonstrating the universality of Klein tunneling across different wave natures and length scales.
\end{abstract} 

\maketitle

\section{Introduction: Overview on the origin of Klein tunneling}

Since 1929, Oskar Klein predicted one of the most outstanding tunneling phenomena in the context of high-energy physics \cite{Klein1929}. This phenomenon, called today Klein tunneling, consists of a non-resonant perfect transmission of massless particles through a gradient electrostatic potential, where the kinetic energy can be less than the potential height \cite{Beenakker2008,Allain2011,Katsnelson2006}. Such an effect does not necessarily involve crossing classical forbidden regions via evanescent waves, like a Landauer-Zener tunneling in condensed matter \cite{He2020b,Landauer1994}. Interestingly, Klein tunneling has not been tested in high-energy physics experiments so far. However, the first experimental realization of Klein tunneling was performed in condensed matter with the discovery of graphene \cite{Katsnelson2006, Young2009}. In contrast to the Klein tunneling originally predicted in high-energy physics, this perfect transmission emerges in graphene within a low-energy regimen, when electrons pass from one band (conduction or valence band) to another under normal incidence, conserving an associated pseudo-spin lattice \cite{Katsnelson2006,Beenakker2008,Allain2011}. It is worth noting that Klein tunneling is not exclusive to spin-1/2 charged particles that normally impinge on an external electrostatic potential, as it is possible to realize analogues of Klein tunneling for phonons or photons in artificial lattices \cite{Jiang2020,Zhang2022c,Zhu2023,Wu2024}. 

Klein tunneling can also occur for massive particles with enlarged pseudo-spin beyond spin 1/2 and independent of the incidence angle, effect called as super-Klein tunneling \cite{Bercioux2009,Shen2010,Dora2011,Lan2011,Urban2011,Xu2016,Bercioux2017,BetancurOcampo2017,ContrerasAstorga2020,Cunha2020,Chen2019,Wang2020,Hao2021,Liu2022,Kim2019,Wang2022,Liu2023,Duan2023,Kim2019a,BetancurOcampo2018,Jakubsky2023,Zeng2022,Korol2018,CrastodeLima2020,Nandy2019,Kim2020,Filusch2020,Zhu2023,Xu2021,Kim2020,Sirota2021,Majari2021a,Chen2024,Wu2024}. This omnidirectional perfect transmission is performed in two-dimensional lattices with unit cells containing three atoms, among them: $\alpha$-$\tau_3$ lattices \cite{Zhou2021a,Oriekhov2021,Wang2021a,Ding2024a,Mandhour2026,Mandhour2020,Fadil2025,Liu2025,Iurov2022,Majari2021a,Liu2023,Iurov2021a,Hao2022}, Lieb lattices \cite{Wan2017,Graf2021,Lara2025,Lima2020,Feng2020a,Hwang2021,Banerjee2021,Han2022,Flannigan2021,Huhtinen2020,Zelaya2022,Jakubsky2023,Liu2020,Liu2021a,Hanafi2022,MejiaCortes2020,Cao2015,Li2022b,Ali2020,He2020b,Xu2014,Whittaker2021,Lima2022,Dauphin2016,Bouzerar2021,Kusdiantara2022,RomanCortes2021,Yamazaki2020,Feng2021,Goldman2011,Abramovici2021,Pires2021,Scafirimuto2021,Xia2018}, and honeycomb superlattices \cite{Cunha2021,Dragoman2021,Lan2011,Guzman2023,Bezerra2020,Wang2010,AgrawalGarg2013,BrionesTorres2014,Guzman2018}. In such systems, electrons behave as particles with pseudo-spin-one, where a flat band lies between two dispersive bands \cite{Bercioux2017,Zhang2025,Pang2025}. The study of materials with flat bands has increased recently due to its outstanding effects on electronic properties \cite{Wang2021,Zhang2022,Wang2021a,Wang2020a,Li2022,Zhang2022a,Zhang2022b,Wen2023,BetancurOcampo2017a,NavarroLabastida2023}. Super-Klein tunneling has been observed recently in acoustic experiments with triangular lattices of Willis resonant scatterers \cite{Zhu2023} and phononic Lieb lattice \cite{Wu2024}. 

Several variants of Klein tunneling have been identified. One notable example is anomalous Klein tunneling, which corresponds to a non-resonant perfect transmission with a pronounced angular dependence. This phenomenon arises in two-dimensional materials such as borophene 8-pmmn, uniaxially strained graphene \cite{Xie2019,BetancurOcampo2018,Huang2023,Xu2023}, twisted bilayer graphene \cite{Bahlaoui2026}, and photonic graphene, the latter of which has been realized experimentally \cite{Zhang2022c}. In Kekulé graphene superlattices, a different type of Klein tunneling is predicted, where perfect transmission is accompanied by a valley flip \cite{Garcia2022}. The opposite phenomena, known as anti-Klein tunneling, corresponds to a perfect backscattering under normal incidence, which has been predicted in bilayer graphene and checkerboard lattices \cite{Majari2023, Septembre2023, Cunha2020}.  The omnidirectional version named anti-super-Klein tunneling has been studied in phosphorene and few-layer black phosphorus  \cite{BetancurOcampo2019, BetancurOcampo2020, MolinaValdovinos2022, Hua2024}. Beyond the Klein tunneling, more phenomena with flat bands have been predicted, such as Andreev reflection \cite{Beenakker2008,Feng2020,Zeng2022,Septembre2023}, chiral magnons \cite{Sui2025}, topological charge pump \cite{Wang2021}, super skew scattering \cite{Wang2021a}, Seebeck and Nernst effects \cite{Duan2023}, enhanced magneto-optical effect \cite{Chen2019}, Anderson localization \cite{Kim2019}, and electron-beam collimation \cite{Wang2022,Xu2016}. The wide variety of predicted and observed Klein tunneling phenomena is derived from the unusual electronic and transport properties of novel low-dimensional materials.

 Another related topic is electron optics, which arises from the wave-particle duality with outstanding technological applications, such as the electron microscope. The rise of two-dimensional materials has recently given electron optics a new avenue. The experimental observation of negative refraction in graphene paves the way for the design of Veselago lenses \cite{Lee2015,Chen2016}, since electrons are refracted according to a generalized Snell’s law, where gating serves as the refractive index to tune the electrostatic potential and Fermi level \cite{ParedesRocha2021,BetancurOcampo2019,AgrawalGarg2013,Chen2016,Bai2018,Sajjad2011,Allain2011,Cheianov2007,Lassaline2025,Ildarabadi2025,Butanov2025,Chen2023a}. In this way, electron optics enables the simulation of refraction phenomena, such as light propagation through metamaterials \cite{Li2021a,Ali2020,Veselago1968,Pendry2000}. Controlling electron flow in two-dimensional materials can lead to the proposal of a series of nanodevices, among them superlenses \cite{BetancurOcampo2018a,Hills2017,Heinisch2013,BetancurOcampo2018,BetancurOcampo2017,Cheianov2007,PadillaOrtiz2025,Dragoman2024,Banerjee2024}, valley beam splitters and filters \cite{GarciaPomar2008,Garcia2022,Settnes2016,Zhai2018,ContrerasAstorga2020a,Grujic2014,Chauwin2022,Zhai2011,Hao2021,Wang2018,Prabhakar2019,Tan2020,Filusch2021,Islam2017,Parui2025,Niu2022,Wang2024g,Schaibley2016,Vanderstraeten2024}, collimators \cite{Park2008,AgrawalGarg2013,Xu2013,Yang2021a,Xu2023,Wang2022}, fiber-optic guidings \cite{Williams2011,ContrerasAstorga2020a,Wang2020b,BetancurOcampo2020,Sheffer2022,Jiang2017,Mladenovic2022}, interferometers \cite{Khan2014,Bercioux2020}, reflectors \cite{Graef2019,Zhou2021a,Zhou2021,Lyu2024,Zeng2022,Chen2024a,Septembre2023,Beenakker2008,BetancurOcampo2019,Xu2014,Xu2023,Wang2020c,Feng2020b}, and use the valley degree of freedom as conveyor of quantum information \cite{Atteia2021,GarciaPomar2008,Garcia2022,Zhai2011,Schaibley2016}. Klein and anti-Klein tunneling may help to improve the efficiency of these electron optics tools.

Recently, one-dimensional lattices have been widely investigated in artificial systems using effective models based on tight-binding approach  \cite{CaceresAravena2022,Coutant2021,Dietz2018,Torrent2013,MartinezArgueello2022,Stegmann2017,RamirezRamirez2020,Casteels2016,Majari2021,Belopolski2017,Torrent2013,Freeney2022,RamirezRamirez2020,BetancurOcampo2024,Dietz2018,MartinezArgueello2022,Liu2022,Suesstrunk2015,Bellec2013,Drost2017,Zhang2024,Huda2020,Barbier2008,Azcona2021,Jakubsky2024,Orso2021,Liu2013,Tilleke2020,Mukherjee2015,Hao2023,Jones2025,Zhang2022e,Yang2016,Zhang2022b}, where the main study is focused on topological phase transitions \cite{Hasan2010,Asboth2016,Shen2017}. Nevertheless, Klein tunneling has been explored scarcely in 1D chains due to the difficulty of creating an abrupt $pn$ junction in polyacetylene and other molecules to test perfect tunneling. Artificial lattices enable us to explore atypical Klein tunneling in various branches of physics, including optics, acoustics, and elastic waves. 

 \comA{In contrast to existing reviews on Klein tunneling \cite{Allain2011,Beenakker2008}, we show that several types of Klein and anti-Klein tunneling beyond graphene arise not only in two-dimensional materials but also in artificial systems. These phenomena are governed by the conservation of an effective pseudo-spin-1/2, which provides a unifying explanation across multiple platforms. We address Klein tunneling and its variants using both the tight-binding approach and supersymmetric quantum mechanics. The derivation of the effective pseudo-spin requires imposing non-standard boundary conditions in the wave functions, stemming from the inclusion of all higher-order terms in the Bloch Hamiltonian expansion, which leads naturally to the generalization of previous approaches. Assuming a ballistic regime, where the coherence length and mean free path exceed the device size, we analyze the transmission probability of quasiparticles (electrons, phonons, or photons) across junctions and stratified potential media. To illustrate the present framework, we demonstrate that Klein tunneling emerges in one-dimensional Su-Schrieffer-Heeger lattices described by the full Bloch Hamiltonian, in both trivial and topological phases. The concept of a reduced pseudo-spin 1/2 further provides a unified framework to explain all known Klein tunneling phenomena reported in the literature, including anomalous KT, super KT, anti-KT, anti-super-KT, and valley-cooperative KT. Finally, the rapid development of artificial crystals and engineered materials opens promising avenues for the experimental realization of currently unobserved Klein tunneling regimes.} 
 
 In this Review, we analyze the various versions of Klein tunneling from a general perspective. For that, the first section is dedicated to the application of the tight-binding approach to a crystal of $n$ atoms in the unit cell, where a general Bloch Hamiltonian is derived. This framework is employed in the analysis of Klein tunneling of a rectangular electric barrier. The continuity of the wave functions of the Bloch Hamiltonian is obtained from general considerations in the second section, because the Bloch Hamiltonian involves higher-order terms in the wave vector $\vec{k}$. In the third and fourth sections, we establish the analogs of Fresnel coefficients in condensed matter and metamaterials by applying the continuity wave function for step potentials, barriers, and stratified media under the reformulation of the transfer matrix method.  

In the fifth section, we derive the laws of electron optics in anisotropic media from the conservation of energy and linear momentum within the framework of a Dirac-type Hamiltonian. The sixth section is devoted to the study of super-Klein tunneling in pseudo-spin one systems and in the presence of a periodic chain of scatterers, using the formalism of supersymmetric quantum mechanics.
From the previous theoretical frameworks, the seventh section analyzes the emergence of Klein, anti-Klein, anomalous, and valley-cooperative Klein tunneling. Several manifestations of these effects are examined in two-dimensional materials such as graphene, borophene, phosphorene, as well as in artificial crystals including phononic Lieb and dice lattices.

\section{Tight binding approach of low-dimensional lattices}
The tight-binding approach is based on the consideration of a set of atoms in the molecule, nanostructure, polymeric chain, or $d$-dimensional crystal for the latter electron in the atomic orbital \cite{Mallick2021,Graf2021,RamirezRamirez2020,Mizoguchi2021,Ogata2021,Katsura2021,Bellec2013,Mallick2020,BetancurOcampo2024,Goerbig2011,Pereira2009a,Papaconstantopoulos1997,TaghizadehSisakht2015,Stegmann2017}. This electron interacts with the total energy potential due to the other electrons and nuclear ions. Each atom in the site is located at the minimum of this potential, like a quantum well. Therefore, the electron is confined in the potential well with the possibility to hop to another atom site across the tunnel effect. This scenario is not exclusive to electrons in molecules or crystals. The undulatory behavior of the confined electron in the potential well can also be emulated by other types of waves, such as elastic, electromagnetic, or sound waves, which are resonating within a cavity. If the cavities possess sidewalls with the possibility of leakage, the waves escape across evanescent modes to arrive at the neighborhood of the other cavity. In this way, artificial molecules, nanostructures, and $d$-dimensional crystals can be emulated \cite{RamirezRamirez2020,Stegmann2017,Zhang2022c,Polini2013,Palmer2020,CaceresAravena2022,Dietz2018,Hanafi2021,Nakatsugawa2024,Reisner2021,Ali2020,Mukherjee2015,Majari2021,Kim2020a,Tang2022,MartinezArgueello2022a,BetancurOcampo2024a,Freeney2022}.

For the particular situation of describing the electron propagation in a crystal, the wave function is 

\begin{equation}\label{BWF}
    \psi_j(\vec{r},\vec{k}) = \sum_{\vec{R}_\ell} \textrm{e}^{i\vec{k}\cdot(\vec{R}_\ell +\vec{\delta}_j)} \phi_j(\vec{r} -\vec{\delta}_j -\vec{R}_\ell),
\end{equation}

\noindent which satisfies the Bloch theorem, where $\vec{k}$ is the wave vector, the index $j = 1,2,\ldots,n$ indicates the position inside the unit cell, and $\vec{R}_\ell$ are lattice vectors, the index $\ell$ labels the unit cell positions. The vectors $\vec{\delta}_j$ correspond to the atomic positions within the unit cell. The function $\phi_j$ is the outer orbital of the $j$-th atom. In the following, we consider only a finite piece of a crystal of $N$ unit cells and suppose that the expression in Eq. \eqref{BWF} remains valid, where $\ell$ goes from 1 to $N$.

To get the matrix representation of the TB Hamiltonian, we split the Hamiltonian in two parts

\begin{equation}
    \hat{H} = \hat{H}_\textrm{at} + \hat{V}_\textrm{cr},
\end{equation}

\noindent where $\hat{H}_\textrm{at}$ is the atomic Hamiltonian acting on the site $j$:

\begin{equation}
    \hat{H}_\textrm{at}\phi_j(\vec{r}) = \epsilon_j\phi_j(\vec{r})
\end{equation}

\noindent and $\hat{V}_\textrm{cr}$ is the electrostatic potential of the electron with the whole crystal. Therefore, the matrix elements are

\begin{equation}
    H_{jm} = \int\limits_V \psi^*_j(\vec{r})(\hat{H}_\textrm{at} + \hat{V}_\textrm{cr})\psi_m(\vec{r}) dV.
\end{equation}

\noindent Substituting the Bloch wave function of Eq. \eqref{BWF} in the $H_{jm}$ elements, we have

\begin{equation}\label{Hjm}
    H_{jm}(\vec{k})=\sum_{\vec{R}'_\ell,\vec{R}_\ell} \textrm{e}^{i\vec{k}\cdot(\vec{R}_\ell -\vec{R}'_\ell + \vec{\delta}_m-\vec{\delta}_j)}\left[\epsilon_j W^{\ell\ell'}_{jm} + T^{\ell\ell'}_{jm}\right],
\end{equation}

\noindent where we define

\begin{equation}
    W^{\ell\ell'}_{jm}(\vec{R}_\ell) \equiv \int\limits_V\phi^*_j(\vec{r}-\vec{\delta}_j-\vec{R}'_\ell)\phi_m(\vec{r}-\vec{\delta}_m-\vec{R}_\ell)dV,
\end{equation}

\noindent which is the integral that quantifies the degree of overlapping of the orbital at the site $j$ for the unit cell $\ell$ with respect to another located at the site $m$ in the unit cell $\ell'$. While the element

\begin{equation}
    T^{\ell\ell'}_{jm} = \int\limits_V\phi^*_j(\vec{r}-\vec{\delta}_j-\vec{R}'_\ell)V_\textrm{cr}(\vec{r})\phi_m(\vec{r}-\vec{\delta}_m-\vec{R}_\ell)dV
\end{equation}

\noindent is the hopping integral and must be interpreted as a probability amplitude of the electron that hops from the atom $j$ in the unit cell $\ell$ to the atomic site $m$ in the unit cell $\ell'$. 

The quantity $\Delta\vec{R}=\vec{R}_\ell - \vec{R}'_\ell$ is also a lattice vector, and the sum in Eq. \eqref{Hjm} has $N$ repeated terms due to the $N$ unit cells in the crystal. Therefore, the TB Hamiltonian can be written as

\begin{equation}\label{TBH}
    H_{jm}(\vec{k}) = N(\epsilon_jW_{jm}(\vec{k}) + T_{jm}(\vec{k})) = N h_{jm}(\vec{k}).
\end{equation}

\noindent The overlap matrix elements are

\begin{equation}
    W_{jm}(\vec{k}) = \sum_{\Delta\vec{R}}w_{jm}^{(\Delta\vec{R})}\textrm{e}^{i\vec{k}\cdot(\Delta\vec{R}+\vec{d}_{mj})},
\end{equation}

\noindent with $\vec{d}_{mj}=\vec{\delta}_m-\vec{\delta}_j$ and

\begin{equation}
    w_{jm}^{(\Delta\vec{R})} = \int\limits_V \phi^*_j(\vec{r})\phi_m(\vec{r}-\Delta\vec{R}-\vec{d}_{mj})dV.
\end{equation}

\noindent The hopping matrix elements are 

\begin{equation}
    T_{jm}(\vec{k}) = \sum_{\Delta\vec{R}}t_{jm}^{(\Delta\vec{R})}\textrm{e}^{i\vec{k}\cdot(\Delta\vec{R}+\vec{d}_{mj})},
\end{equation}

\noindent where the hopping parameter between the atoms with positions $\vec{\delta}_j$ and $\vec{\delta}_m+\Delta \vec{R}$ is

\begin{equation}
    t_{jm}^{(\Delta\vec{R})} = \int\limits_V \phi^*_j(\vec{r})V(\vec{r})\phi_m(\vec{r}-\Delta\vec{R}-\vec{d}_{mj})dV.
\end{equation}

An approximation in the tight-binding approach is to neglect the overlap among atoms, where $W_{jm} \approx \delta_{jm}$, herein $\delta_{jm}$ is the Kronecker delta. Therefore, the effective TB Hamiltonian in Eq. \eqref{TBH} is

\begin{equation}\label{TBH2}
    h^{(TB)}_{jm}(\vec{k}) \approx \epsilon_j\delta_{jm} + T_{jm}(\vec{k}).
\end{equation}
It is also customary to retain the nearest neighbor interaction only; that is, the hopping amplitude $t_{jm}^{(\Delta R)}$ is set to zero for all non-nearest-neighbor sites.

\begin{figure}
\centering
\begin{tabular}{c}
(a) \qquad \qquad \qquad \qquad \qquad \qquad \qquad \qquad \qquad \qquad\\
    \includegraphics[width=0.8\linewidth]{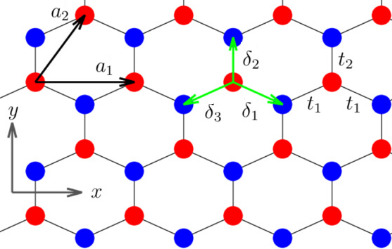}\\
(b) \qquad \qquad \qquad \qquad \qquad \qquad \qquad \qquad \qquad \qquad\\
    \includegraphics[width=0.8\linewidth]{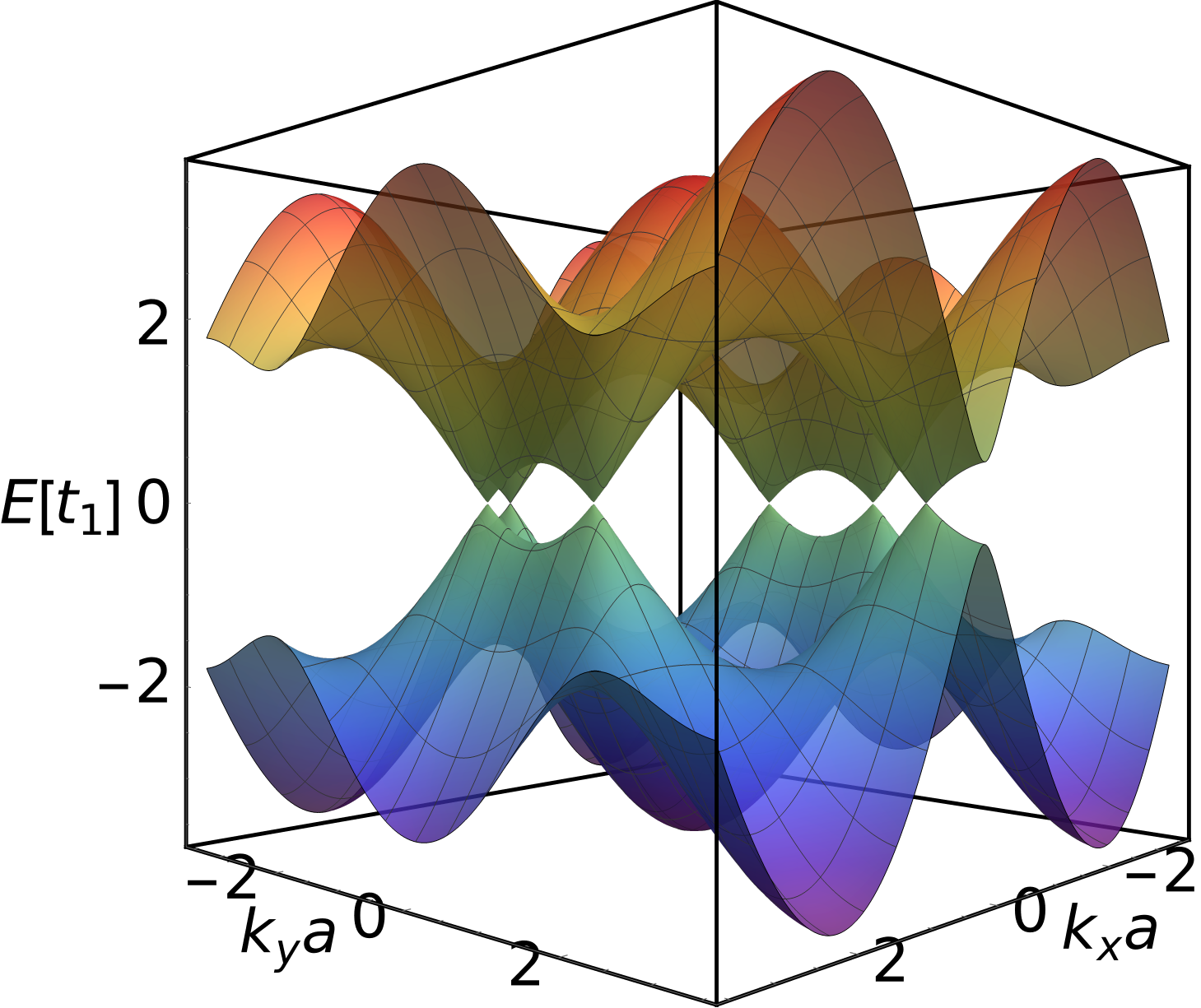}
\end{tabular}
    \caption{(a) Anisotropic hexagonal lattice that represents the strained graphene and simplified phosphorene structure, which is formed by two triangular sublattices. The arrows correspond to lattice vectors and nearest neighbors of atoms within the unit cell. The hopping parameters are $t_1$ and $t_2$ for quinoid deformation. (b) Electronic band structure with the values of $\epsilon_1 = \epsilon_2 = 0$, $t_1 = 1$, and $t_2 = 1.6$. The variation of hopping parameters can induce a gap opening for $t_2 \geq 2t_1$.}
    \label{fig:strgraph}
\end{figure}

 While two-dimensional materials, such as strained graphene and phosphorene, as shown in Fig. \ref{fig:strgraph}(a), the Bloch Hamiltonian is obtained taking into account two atoms in the unit cell to nearest neighbors \cite{BetancurOcampo2021,Naumis2017,BetancurOcampo2018,DiazBautista2020,DiazBautista2020a,Pereira2009,Pereira2009a,BetancurOcampo2019,BetancurOcampo2020,TaghizadehSisakht2015,Ezawa2014,Midtvedt2017,Rudenko2014}

\begin{equation}\label{BH2D}
    H_\textrm{2D}(\vec{k}) = \left(\begin{array}{cc}
      \epsilon_1   &  \sum^3_{j=1}t_j\textrm{e}^{-i\vec{k}\cdot\vec{\delta}_j}  \\
      \sum^3_{j=1}t_j\textrm{e}^{i\vec{k}\cdot\vec{\delta}_j}  & \epsilon_2 \\
      
    \end{array}\right),
\end{equation}

\noindent where we have introduced the following change of notation $\vec{d}_{1j} \rightarrow \vec{\delta}_j$ and $t_{1j}\rightarrow t_j$ with $j = 1, 2$ and $3$ for the nearest neighbor positions and hopping parameters, respectively. These relative positions without deformation are given by $\vec{\delta}_1 = (\sqrt{3}/2,-1/2)a$, $\vec{\delta}_2 = (0,1)a$, and $\vec{\delta}_3 = (-\sqrt{3}/2,-1/2)a$, with $a = 0.142$ nm the bond length of graphene \cite{CastroNeto2009}.  The diagonalization of the Bloch Hamiltonian \eqref{BH2D} for the anisotropic hexagonal lattice  gives us the electronic band structure, as seen in Fig. \ref{fig:strgraph}(b).

\begin{figure}
    \centering
    \begin{tabular}{c}
    (a) \qquad \qquad \qquad \qquad \qquad \qquad \qquad \qquad \qquad \qquad\\
    \includegraphics[width=0.75\linewidth]{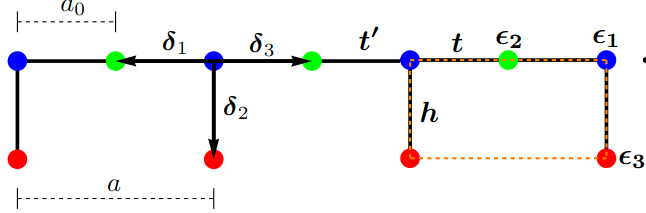}\\
    (b) \qquad \qquad \qquad \qquad \qquad \qquad \qquad \qquad \qquad \qquad\\
    \includegraphics[width=0.85\linewidth]{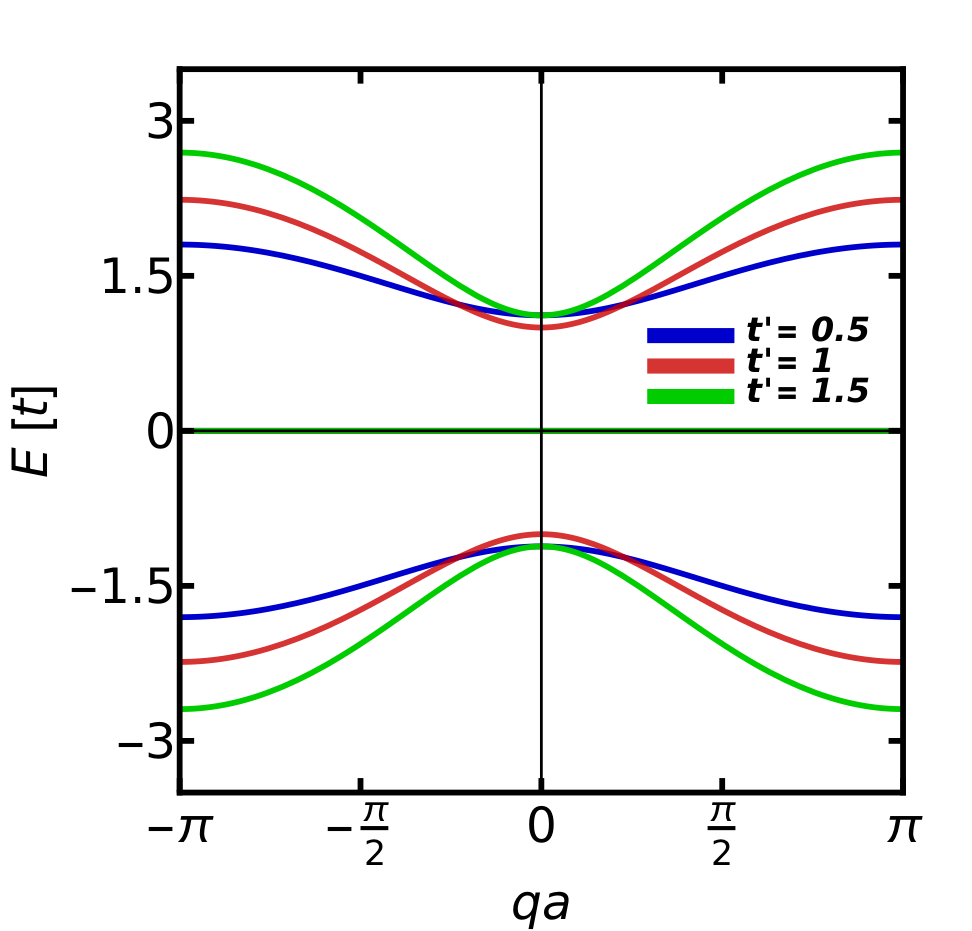}
    \end{tabular}
    \caption{(a) Bearded Su-Schrieffer-Heeger lattice and (b) its electronic band structure with the set of onsite energies $\epsilon_1 = \epsilon_2 = \epsilon_3 = 0$ and hopping parameters $t = h = 1$, and $t' = 0.5, 1, 1.5$.}
    \label{fig:BSSH}
\end{figure}

On the other hand, we can consider a chain with three atoms in the unit cell, see Fig. \ref{fig:BSSH}(a), which is known as the bearded Su-Schrieffer-Heeger (SSH) lattice \cite{BetancurOcampo2024,Jakubsky2024,Jones2025}

\begin{equation}
    h(\vec{k}) \approx \left(\begin{array}{ccc}
      \epsilon_1   &  T_{12}(\vec{k}) & T_{13}(\vec{k})\\
      T_{12}^*(\vec{k})  & \epsilon_2 & T_{23}(\vec{k})\\
      T_{13}^*(\vec{k}) & T_{23}^*(\vec{k}) & \epsilon_3
    \end{array}\right),
\end{equation}

\noindent where the functions $T_{ij}(\vec{k})$ are expressed in terms of hopping parameters to nearest neighbors

\begin{eqnarray}
    T_{12}(\vec{k})& \approx & t_{12}\textrm{e}^{i\vec{k}\cdot\vec{\vec{d}}_{21}} + t'_{12}\textrm{e}^{i\vec{k}\cdot\vec{\vec{d}}'_{21}},\nonumber\\
    T_{13}(\vec{k})& \approx & t_{13}\textrm{e}^{i\vec{k}\cdot\vec{\vec{d}}_{13}}\nonumber\\
    T_{23}(\vec{k})& \approx & t_{23}\textrm{e}^{i\vec{k}\cdot\vec{\vec{d}}_{23}}.
\end{eqnarray}

\noindent The expression $T_{12}(\vec{k})$ corresponds to the coupling between the atoms at sites 1 and 2. The coupling $t_{12}$ reflects the hopping to the site 1 from the site 2 within the unit cell, whereas $t_{12}'$ corresponds to the hopping from the site 2 to the adjacent cell. If the lattice is anisotropic, the parameters $t_{12}$ and $t'_{12}$ are different. The functions $T_{13}(\vec{k})$ and $T_{23}(\vec{k})$ corresponds to couplings within the unit cell,  where $t_{13}$ and $t_{23}$ are the nearest-neighbor hoppings. In the bearded SSH lattice in Fig. \ref{fig:BSSH}(a), we set $\vec{k}=(k_x,0)$ and the hopping parameters $t_{12} = t$, $t'_{12} = t'$, $t_{13} = h$, and $t_{23} = 0$. Fixing the relative position vectors $\vec{d}_{21} \rightarrow \vec{\delta}_3 =a_0(1,0)$, $\vec{d}'_{21} \rightarrow \vec{\delta}_1 = a_0(-1,0)$, and $\vec{d}_{13} \rightarrow \vec{\delta}_2 = a_0(0,1)$, we obtain the simplified Hamiltonian

\begin{equation}\label{BH1D}
    H_\textrm{1D}(k) = \left(\begin{array}{ccc}
      \epsilon_1   &  t \textrm{e}^{ika/2}+ t'\textrm{e}^{-ika/2} & h\\
      t\textrm{e}^{-ika/2} + t'\textrm{e}^{ika/2}  & \epsilon_2 & 0\\
      h & 0 & \epsilon_3
    \end{array}\right).
\end{equation}

\noindent This Hamiltonian has been used to study Klein tunneling in bearded SSH lattices \cite{BetancurOcampo2020}. The electronic band structure, obtained from the eigenenergies of the Hamiltonian in Eq. \eqref{BH1D}, depends on the choice of hopping parameters, as shown in Fig. \ref{fig:BSSH}(b). 

Tuning the hopping parameters, for instance, with the application of strain on the material \cite{Pereira2009,Naumis2017,BetancurOcampo2018}, is possible to change either smoothly or drastically the electronic and transport properties, and in some cases, the induction of topological phase transitions \cite{Dey2019,Thatcher2022,Shen2022}. In artificial systems, such as elastic metamaterials, the modulation of hopping parameters can be performed straightforwardly, where it is possible a great control in the phase transition \cite{BetancurOcampo2024a,ManjarrezMontanez2025}.

\section{General matching conditions of wave functions from a Bloch Hamiltonian}
For interfaces in a material with abrupt changes in the electrostatic potential, the matching conditions are obtained by integrating the Schr\"odinger equation into a narrow range $(-\epsilon,\epsilon)$ around the abrupt jump in the electrostatic potential $V(x)$ at $x =0$. Choosing as an example the Bloch Hamiltonian in Eq. \eqref{BH1D} of the bearded SSH lattice, we get
\begin{widetext}
\begin{eqnarray}\label{ISE}
\lim_{\epsilon \rightarrow 0}\int\limits^\epsilon_{-\epsilon} [H(q) + V(x)]\vec{\Psi}(x) dx & = &  E\lim_{\epsilon \rightarrow 0}\int\limits^\epsilon_{-\epsilon}\vec{\Psi}(x) dx\\
    \lim_{\epsilon \rightarrow 0}\int\limits^\epsilon_{-\epsilon}\left(\begin{array}{ccc}
        \epsilon_1+V(x) & \hat{g}^*(q) & h \\
        \hat{g}(q) & \epsilon_2 +V(x)& 0 \\
        h & 0 & \epsilon_3+V(x)
    \end{array}\right)\left(\begin{array}{c}
   \psi_1(x)\\
   \psi_2(x)\\
   \psi_3(x)\end{array}\right)dx & = & E \lim_{\epsilon \rightarrow 0}\int\limits^\epsilon_{-\epsilon}\left(\begin{array}{c}
    \psi_1(x)\\
    \psi_2(x)\\
    \psi_3(x)\end{array}\right) dx.
\end{eqnarray}
\end{widetext}

\noindent The operator $\hat{g}(q)$ is defined by

\begin{eqnarray}\label{gf}
   \hat{g}(q) & = & i\left(-t\textrm{e}^{-ia\hat{q}/2} + t'\textrm{e}^{ia\hat{q}/2}\right) \nonumber\\
    & = & i\sum^\infty_{n = 0}\frac{a^n}{2^nn!}[t' - (-1)^n t]\frac{d^n}{dx^n}
\end{eqnarray}

\noindent and

\begin{equation}
    \hat{g}^*(q)=-i\sum^\infty_{n = 0}\frac{a^n}{2^nn!}[-t + (-1)^n t']\frac{d^n}{dx^n}
\end{equation}

\noindent where the linear momentum is $\hat{q} = -i\frac{d}{dx}$. We performed the change $k = q + \pi/a_0$ for mirror symmetry bands respect to vertical axis, as shown in Fig. \ref{fig:BSSH}(b).

With the limit $\epsilon \rightarrow 0$ and considering the absence of singularities in $\psi_j(x)$ and $V(x)$, where $j = 1,2$ and 3, several integrals are equal to zero, but the equations with $\hat{g}(q)$ and $\hat{g}^*(q)$

\begin{eqnarray}\label{int}
    \lim_{\epsilon \rightarrow 0}\int\limits^\epsilon_{-\epsilon}\hat{g}(q)\psi_1(x)dx & = & 0 \nonumber\\
    \lim_{\epsilon \rightarrow 0}\int\limits^\epsilon_{-\epsilon}\hat{g}^*(q)\psi_2(x)dx & = & 0 
\end{eqnarray}

\noindent need and adequate treatment. Substituting the operator functions $\hat{g}(q)$ and $\hat{g}^*(q)$ in the integrals in Eq. \eqref{int}, the matching conditions are given by

\begin{eqnarray}\label{TCC}
    \left.\hat{Q}(a)\psi_1(x)\right|_{x=0^-} & = &\left.\hat{Q}(a)\psi_1(x)\right|_{x=0^+} \nonumber\\
     \left.\hat{Q}(-a)\psi_2(x)\right|_{x=0^-} & = &\left.\hat{Q}(-a)\psi_2(x)\right|_{x=0^+},
\end{eqnarray}

\noindent where the operator $\hat{Q}(a)$ is defined as

\begin{equation}\label{Qs}
    \hat{Q}(a) =  \sum^\infty_{n = 0}\frac{a^n}{2^n(n+1)!}[t' + (-1)^n t]\frac{d^n}{dx^n}.
\end{equation}

\noindent The matching conditions, such as the Dirac Hamiltonian case, can be obtained by considering only
the first term ($n = 0$) in the series \eqref{Qs}.

The matching conditions \eqref{TCC} shall be imposed on the wave functions 

\begin{equation}\label{PsiI}
    \vec{\Psi}_\textrm{I}(x) = \vec{u}(q_\textrm{in})\textrm{e}^{iq_\textrm{in}x} + {A_\textrm{r}}\vec{u}(q_\textrm{r})\textrm{e}^{iq_\textrm{r}x}, 
\end{equation}
\noindent in region I, and

\begin{equation}\label{PsiII}
    \vec{\Psi}_\textrm{II}(x) = A_\textrm{t}\vec{u}(q_\textrm{t})\textrm{e}^{iq_\textrm{t}x} 
\end{equation}

\noindent as the transmitted wave function in region II. The calculations can be facilitated by these useful properties of $\hat{Q}(a)$

\begin{eqnarray}
    \hat{Q}(a)\textrm{e}^{i\lambda x} & = & \sum^\infty_{n = 0}\frac{(i\lambda a)^n}{2^n(n+1)!}[t' + (-1)^n t]\textrm{e}^{i\lambda x} \nonumber\\
    & = & -2\left[\frac{g(\lambda) + i(t-t')}{\lambda a}\right]\textrm{e}^{i\lambda x}
\end{eqnarray}

\noindent and

\begin{equation}
    \hat{Q}(-a)\textrm{e}^{i\lambda x} = -2\left[\frac{g^*(\lambda) - i(t-t')}{\lambda a}\right]\textrm{e}^{i\lambda x},
\end{equation}

\noindent where $\lambda = q_\textrm{in/r/t}$. By using these relations for the wave functions \eqref{PsiI} and \eqref{PsiII}, we have

    \begin{eqnarray}
    \left.[u_1(q_\textrm{in})\hat{Q}(a)\textrm{e}^{i q_\textrm{in}x} + A_r u_1(q_\textrm{r})\hat{Q}(a)\textrm{e}^{i q_\textrm{r}x}]\right|_{x=0^-} & = & \nonumber\\
    \left.[A_t u_1(q_\textrm{t})\hat{Q}(a)\textrm{e}^{i q_\textrm{t}x}]\right|_{x=0^+} &&\nonumber\\
    \left.[u_2(q_\textrm{in})\hat{Q}(-a)\textrm{e}^{i q_\textrm{in}x} + A_r u_2(q_\textrm{r})\hat{Q}(-a)\textrm{e}^{i q_\textrm{r}x}]\right|_{x=0^-} & = & \nonumber\\
    \left.[A_t u_2(q_\textrm{t})\hat{Q}(-a)\textrm{e}^{i q_\textrm{t}x}]\right|_{x=0^+}, && 
\end{eqnarray}

\noindent which can be arranged in a matrix form as
\begin{equation}\label{AryAt}
    \vec{w}(q_\textrm{in}) + A_\textrm{r}\vec{w}(q_\textrm{r}) = A_\textrm{t}\vec{w}(q_\textrm{t}).
\end{equation}

\noindent The states $\vec{w}(q_{\textrm{in}/\textrm{r}/\textrm{t}})$ are defined by

\begin{align}\label{wtild}
   \vec{w}(q_{\textrm{in}/\textrm{r}/\textrm{t}}) & =\nonumber\\
   & (f(q_\textrm{in/r/t})u^{(1)}(q_\textrm{in/r/t}),f^*(q_\textrm{in/r/t})u^{(2)}(q_\textrm{in/r/t})),
\end{align}

\noindent where the function $f(q_\textrm{in/r/t})$ is given by

\begin{equation}\label{laf}
    f(q_\textrm{in/r/t}) = \frac{g(q_\textrm{in/r/t}) + i(t -t')}{q_\textrm{in/r/t}a}.
\end{equation}

In general, for a single interface Eq. \eqref{wtild} remains valid. Repeating the same procedure from Eq. \eqref{ISE} to Eq. \eqref{laf}  with a different Bloch Hamiltonian, the result can have a different form for the function $f(\vec{k})$. For two and three-dimensional lattices, the integration of Schr\"odinger equation in \eqref{ISE} is performed for the non-conserved  component of the wave vector $k_\perp$, which is perpendicular to the interface. In the next sections, these general matching conditions for the wave functions will be applied to multiple systems that allow us to get unusual transport phenomena related to Klein tunneling effect.

\section{Fresnel-like coefficients: Reflection and transmission of Heaviside and barrier potentials}
\begin{figure}
    \centering
    \begin{tabular}{c}
    (a) \qquad \qquad \qquad \qquad \qquad \qquad \qquad \qquad \qquad \qquad \qquad \qquad\\
    \includegraphics[width=0.9\linewidth]{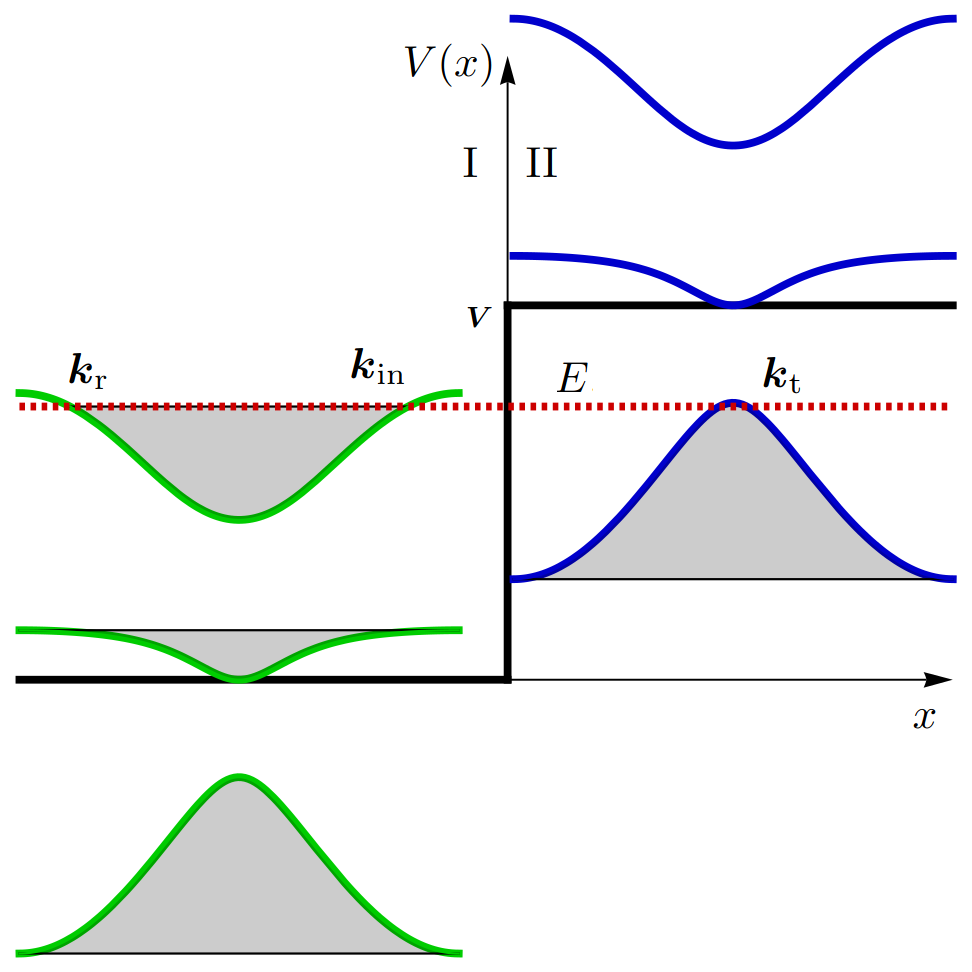}\\
    (b) \qquad \qquad \qquad \qquad \qquad \qquad \qquad \qquad \qquad \qquad \qquad \qquad\\
    \includegraphics[width=0.9\linewidth]{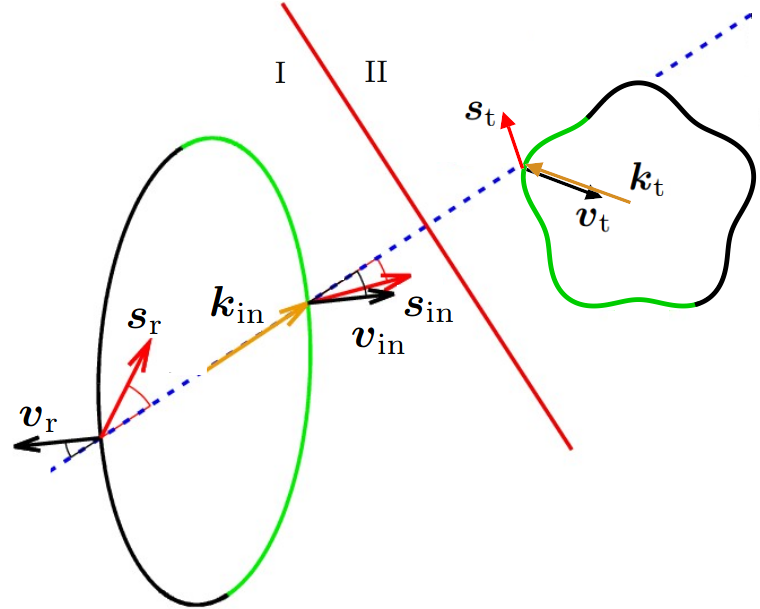}
    \end{tabular}
    \caption{(a) Electronic band structure of a $pn$ junction for an electrostatic step potential. The curves correspond the energy bands in regions I and II. The dashed red line indicates the Fermi level, where gray regions are the occupied states. (b) Kinematical construction represents the conservation of energy (energy contours), linear momentum (dashed line), and current density (green semi-arcs). The arrows indicate the direction of linear momentum $\vec{k}_\textrm{in/r/t}$, pseudo-spin $\vec{s}_\textrm{in/r/t}$, and group velocity $\vec{v}_\textrm{in/r/t}$ for the scattering of electrons at the interface (red line).}
    \label{fig:EOD}
\end{figure}

In two- and three-dimensional ballistic systems, electron wave propagation can be described within an electron-optics framework, where electronic states behave as optical-like beams. Electrostatic gates in such junctions define effective refractive indices for electrons \cite{Allain2011,AgrawalGarg2013,Chen2016,Bai2018,BetancurOcampo2019,ParedesRocha2021}, as illustrated by the typical electronic band structure shown in Fig. \ref{fig:EOD}(a). The ballistic regime achieved in experimental settings can lead to negative refraction of electrons \cite{Chen2016,Lee2015}. When a linear interface separates two regions with different potentials, electrons tunnel by changing both group velocity and pseudo-spin and follow electron optical laws for the reflection and refraction phenomena, as well as the scattering probability \cite{Allain2011,BetancurOcampo2018,ParedesRocha2021}. The conservation of energy $E$, the component of wave vector $k_y$, and probability current density $j_x$ helps to establish these electron optic laws, which are schematically represented using a kinematical construction, see Fig. \ref{fig:EOD}(b). 

In a junction, the electron scattering in the interface can be modeled by a step potential, as shown in Fig. \ref{fig:EOD}(a). The regions I ($x<0$) and II ($x\geq 0$) have the electrostatic potential 0 and $V$, respectively. The wave functions for both regions are written in terms of the eigenvectors of a general Bloch Hamiltonian as in Eqs. \eqref{PsiI} and \eqref{PsiII}, where the quantities $A_\textrm{r}$ and $A_\textrm{t}$ are the wave amplitudes, while $\vec{k}_\textrm{r}$ and $\vec{k}_\textrm{t}$ are the wave vectors of the reflected and transmitted wave, respectively. The wave vectors $\vec{k}_\textrm{in}$, $\vec{k}_\textrm{r}$, and $\vec{k}_\textrm{t}$ are evaluated in the eigenvector $\vec{u}(\vec{k}_{\textrm{in}/\textrm{r}/\textrm{t}})$ following the conservation of energy and linear momentum, as shown in Fig. \ref{fig:EOD}(b) of the kinematical construction. Propagation modes occur if electrons with Fermi energy $E$ from one of the bands cross to another band when there are allowed states. The direction of group velocity is given by the gradient of the energy band. With positive (negative) curvature, the group velocity points outwards (inwards) of the energy contour. 

Returning to Eq. \eqref{AryAt} and solving the $2 \times 2$ linear equation system, the reflection coefficient is

\begin{eqnarray}\label{Refl}
    R(\vec{k}_\textrm{in}) &=& |A_\textrm{r}|^2 = \frac{|\vec{w}(\vec{k}_\textrm{in})\times\vec{w}(\vec{k}_\textrm{t})|^2}{|\vec{w}(\vec{k}_\textrm{r})\times\vec{w}(\vec{k}_\textrm{t})|^2}\nonumber\\
    & = &\frac{|\vec{w}(\vec{k}_\textrm{in})|^2\sin^2\gamma(\vec{k}_\textrm{in},\vec{k}_\textrm{t})}{|\vec{w}(\vec{k}_\textrm{r})|^2\sin^2\beta(\vec{k}_\textrm{r},\vec{k}_\textrm{t})},
\end{eqnarray}

\noindent where $\times$ indicates vectorial product of the states $\vec{w}(\vec{k}_\textrm{in})$ and $\vec{w}(\vec{k}_\textrm{t})$. The relative pseudo-spin angle between the incident and transmitted states is defined by

\begin{equation}\label{dphi}
    \gamma(\vec{k}_\textrm{in},\vec{k}_\textrm{t}) \equiv \arccos\left(\frac{|\vec{w}^*(\vec{k}_\textrm{in})\cdot\vec{w}(\vec{k}_\textrm{t})|}{|\vec{w}(\vec{k}_\textrm{in})||\vec{w}(\vec{k}_\textrm{t})|}\right).
\end{equation}

\noindent For the reflected and transmitted states, the relative pseudo-spin angle is

\begin{equation}
    \beta(\vec{k}_\textrm{r},\vec{k}_\textrm{t}) \equiv \arccos\left(\frac{|\vec{w}^*(\vec{k}_\textrm{r})\cdot\vec{w}(\vec{k}_\textrm{t})|}{|\vec{w}(\vec{k}_\textrm{in})||\vec{w}(\vec{k}_\textrm{t})|}\right).
\end{equation}

\noindent While that the transmission coefficient is given by
\begin{equation}\label{Transm}
    T(\vec{k}_\textrm{in}) = 1 - R(\vec{k}_\textrm{in}) 
\end{equation}

\noindent due to the current density conservation $J_x$. The reflection and transmission coefficients  \eqref{Refl} and \eqref{Transm}, respectively, can be related in terms of the hopping parameters and on-site energies of the tight-binding approach. To obtain exact or numerical wave vectors $\vec{k}_\textrm{in},  \vec{k}_\textrm{r}$, and $\vec{k}_\textrm{t}$ involved in the scattering process, we use the complete energy bands of the Bloch Hamiltonian. 

\comA{In electronic systems, such as synthesized materials among them graphene, silicene, or borophene, the experimental observation of Klein tunneling requires special conditions. The most important is a ballistic regime, which is achieved when both the coherence length and mean free path exceed the characteristic size of the clean sample \cite{Cayssol2009,Young2009,Varlet2014}. \comD{In graphene, the electron mean free path can reach values greater than 20 $\mu$m \cite{Domaretskiy2025}.} The electron de Broglie wavelength must be larger than the lattice bond length to justify the continuum approximation.} More generally, when the wavelength satisfies $\lambda = 2\pi/k > d$, where $d$ is the smoothness characteristic length over which the potential varies, the barriers can be considered as abrupt. \comD{Electron wavelengths within the linear dispersion of graphene range from 10 to 500 nm, corresponding to carrier densities on the order of 10$^{10}$-10$^{12}$ cm$^{-2}$ \cite{Huard2007}. Sharply defined $pn$ junctions at the atomic scale exhibit characteristic lengths of $d \approx 1$ nm \cite{Bai2018}.} 

\comA{The smoothness in the junction causes multiple reflections, manifested as an increase in resistance due to the classically forbidden zones \cite{ParedesRocha2021,Cheianov2007,Allain2011}. However, the spatial profile of a smooth junction does not affect the pseudo-spin conservation, which is the fundamental requirement for Klein tunneling \cite{Allain2011}. Atomically sharp $pn$ junctions can be engineered through the formation of monolayer vacancy islands on the Cu surface \cite{Bai2018}. In artificial crystals, the ballistic regime is likewise essential. A key advantage of these systems is the direct manipulation of the separation of artificial atoms and evanescent couplings by avoiding dislocations or impurities that otherwise reduce the mean free path. The introduction of disorder naturally affects the mean free path, making it shorter than the device size and thereby suppressing Klein tunneling \cite{Cayssol2009}. Several theoretical works have analyzed the disorder effects in both Klein and anti-Klein tunneling regimes \cite{Vigh2013,BetancurOcampo2020,Liu2020,Liu2023a,Guzman2023}. Notably, in phosphorene $pnp$ junctions exhibiting anti-super-Klein tunneling, disorder does not destroy the perfect waveguiding behavior \cite{BetancurOcampo2020}.}

\begin{figure}
    \centering
    \includegraphics[width=1\linewidth]{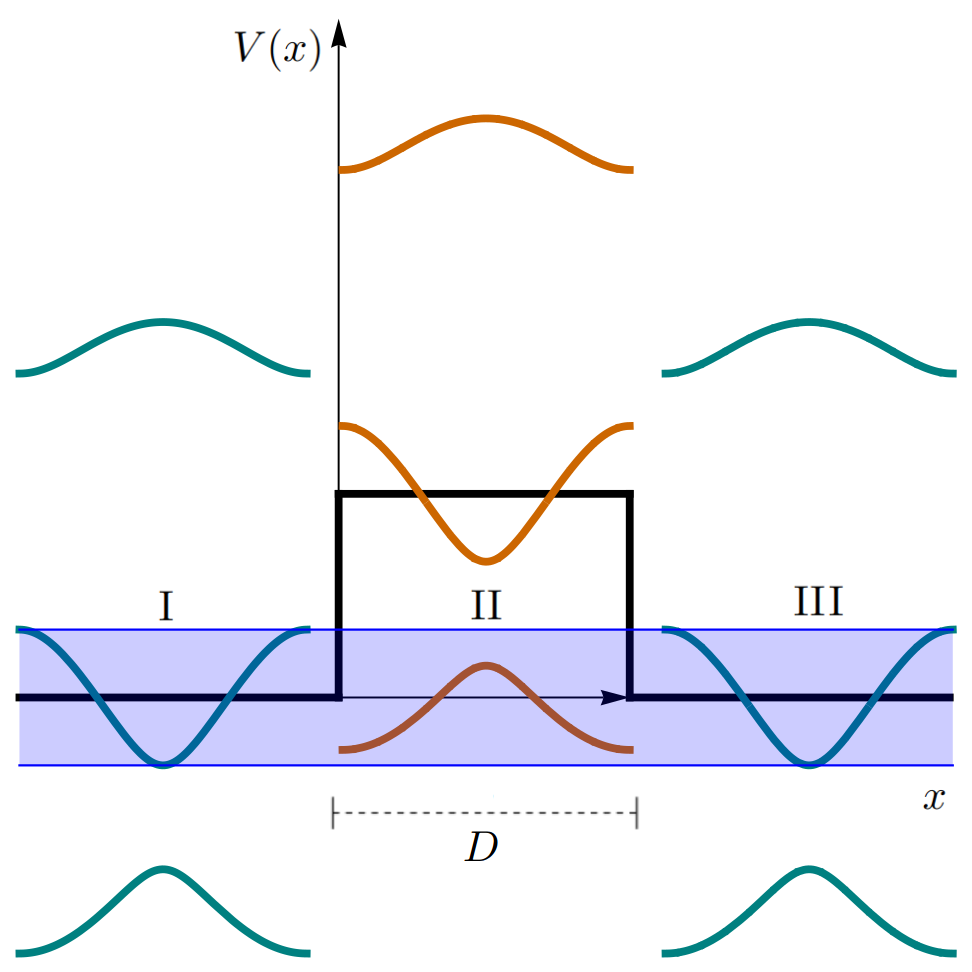}
    \caption{Electronic band structure of a bipolar $npn$ junction, which can be modeled through an electrostatic potential barrier. The curves correspond to energy bands; the blue area is the range of transmission from one of the conduction bands to the valence band.}
    \label{fig:bipolarjunc}
\end{figure}

Let us examine the dynamics of quasi-particles in the presence of interfaces induced by localized fluctuations either in the hopping parameter or in the external electric field. We consider two distinct scenarios. In the first one, the interface arises from spatial inhomogeneity in the hopping parameter $\omega$, which manifests itself as a rectangular barrier of the width $D$, as shown in Fig. \ref{fig:bipolarjunc}
\begin{equation}
t_1=\begin{cases}\tau_1,\quad x\in(-\infty,0]\cup[D,\infty)\\
\tilde{\tau}_1,\quad x\in(0,D).\end{cases}\label{t1rectangular}\end{equation}
In the second one, a rectangular junction $pnp$ with the following electric field will be considered. 
\begin{equation}
V_{el}=\begin{cases}0,\quad x\in(-\infty,0]\cup[D,\infty)\\
V,\quad x\in(0,D).\end{cases}\label{Vrectangular}\end{equation}

The eigenvectors of the Bloch Hamiltonian $H(\vec{k})$ can be written in $x$-representation in the following form for each of the three regions,

\begin{eqnarray}\label{ansatz}
\vec{\Psi}_\textrm{I}(\vec{r}) & = & \vec{u}(\vec{k}_\textrm{in})\textrm{e}^{i\vec{k}_\textrm{in}\cdot\vec{r}} + A_\textrm{r}\vec{u}(\vec{k}_\textrm{r})\textrm{e}^{i\vec{k}_\textrm{r}\cdot\vec{r}} \nonumber\\
\vec{\Psi}_\textrm{II}(\vec{r}) & = & A_a\vec{u}(\vec{k}_a)\textrm{e}^{i\vec{k}_a\cdot\vec{r}} + A_b\vec{u}(\vec{k}_b)\textrm{e}^{i\vec{k}_b\cdot\vec{r}} \nonumber\\
\vec{\Psi}_\textrm{III}(\vec{r}) & = & A_\textrm{t}\vec{u}(\vec{k}_\textrm{t})\textrm{e}^{i\vec{k}_\textrm{t}\cdot\vec{r}}, 
\end{eqnarray}
 The momenta $\vec{k}_\textrm{in}$ and $\vec{k}_\textrm{r}$ correspond to  incident and reflected waves, respectively, while $\vec{k}_a$ and $\vec{k}_b$ are the wave vectors within the barrier. The explicit values of $\vec{k}_a$ and $\vec{k}_b$ depend on the nature of the barrier, i.e. they either depend on $\tilde{\tau}_1$ or on $V$, see Eqs. (\ref{t1rectangular}) and (\ref{Vrectangular}).

With the two interfaces, in $x = 0$ and $x = D$, the reflection and transmission coefficients $R$ and $T$ are obtained by solving the following $4 \times 4$ linear equation systems, which are the result of applying the matching conditions \eqref{TCC}

\begin{eqnarray}\label{EqB}
    \vec{w}(\vec{k}_\textrm{in}) + A_\textrm{r}\, \vec{w}(\vec{k}_\textrm{r})  =  A_a\, \vec{w}(\vec{k}_a) & + & A_b\,\vec{w}(\vec{k}_b), \nonumber\\
A_a\,\textrm{e}^{ik_aD}\vec{w}(\vec{k}_a) + A_b\,\textrm{e}^{ik_bD}\vec{w}(k_b) & = & A_t\,\textrm{e}^{ik_tD}\vec{w}(\vec{k}_t).\nonumber\\
& & 
\end{eqnarray}
\noindent Using the Cramer formula, the amplitude $A_t$ is

\begin{equation}\label{At}
    A_t = \frac{\left|\begin{array}{cc}
       |\vec{w}(\vec{k}_\textrm{r}) \, \vec{w}(\vec{k}_b)|  & |\vec{w}(\vec{k}_\textrm{in}) \, \vec{w}(\vec{k}_b)| \\
          |\vec{w}(\vec{k}_a) \, \vec{w}(\vec{k}_\textrm{r})|  & |\vec{w}(\vec{k}_a) \, \vec{w}(\vec{k}_\textrm{in})|
    \end{array}\right|}{\left|\begin{array}{cc}
       |\vec{w}(\vec{k}_\textrm{r}) \, \vec{w}(\vec{k}_b)|  & |\vec{w}(\vec{k}_\textrm{t}) \, \vec{w}(\vec{k}_b)|\textrm{e}^{-ik_aD} \\
          |\vec{w}(\vec{k}_a) \, \vec{w}(\vec{k}_\textrm{r})|  & |\vec{w}(\vec{k}_a) \, \vec{w}(\vec{k}_\textrm{t})|\textrm{e}^{-ik_bD}
    \end{array}\right|},
\end{equation}

\noindent where  $|\vec{w}(\vec{k}_\textrm{r}) \, \vec{w}(\vec{k}_b)|$ is the determinant of the $2\times2$ matrix that has $\vec{w}(\vec{k}_\textrm{r})$ and $\vec{w}(\vec{k}_\textrm{b})$ as its columns. It is important to note that in Eq. \eqref{At}, we have two nested determinants.

It is possible to get the simplest expression of Eq. \eqref{At}, taking into account that $k_{x,r} = -k_{x,\textrm{in}}$ and $k_{x,a} = -k_{x,b}$ for propagation modes. Therefore, the spinors are $\vec{w}(\vec{k}_\textrm{r}) = \vec{w}(\vec{k}_\textrm{in})^*$ and $\vec{w}(\vec{k}_b) = \vec{w}(\vec{k}_\textrm{a})^*$. Defining $z_1 = |\vec{w}(\vec{k}_\textrm{in})\, \vec{w}(\vec{k}_a)|$ and $z_2 = |\vec{w}(\vec{k}_\textrm{in})\, \vec{w}(\vec{k}_a)^*|$, we have for the transmission

\begin{equation}\label{TCoeff}
    T = \frac{(|z_1|^2-|z_2|^2)^2}{(|z_1|^2-|z_2|^2)^2\cos^2k_aD + (|z_1|^2+|z_2|^2)^2\sin^2k_aD},
\end{equation}

\noindent which is the generalization of the transmission found for chiral tunneling in graphene \cite{Katsnelson2006}. The Fabry-Pérot resonances are identical to the conventional one of a constructive interference pattern of electromagnetic waves when $k_aD = n\pi$ with $n = 0,1,2,\ldots$. 

\section{General transfer matrix method applied to low-dimensional lattices}
\begin{figure}
    \centering
    \includegraphics[width=1\linewidth]{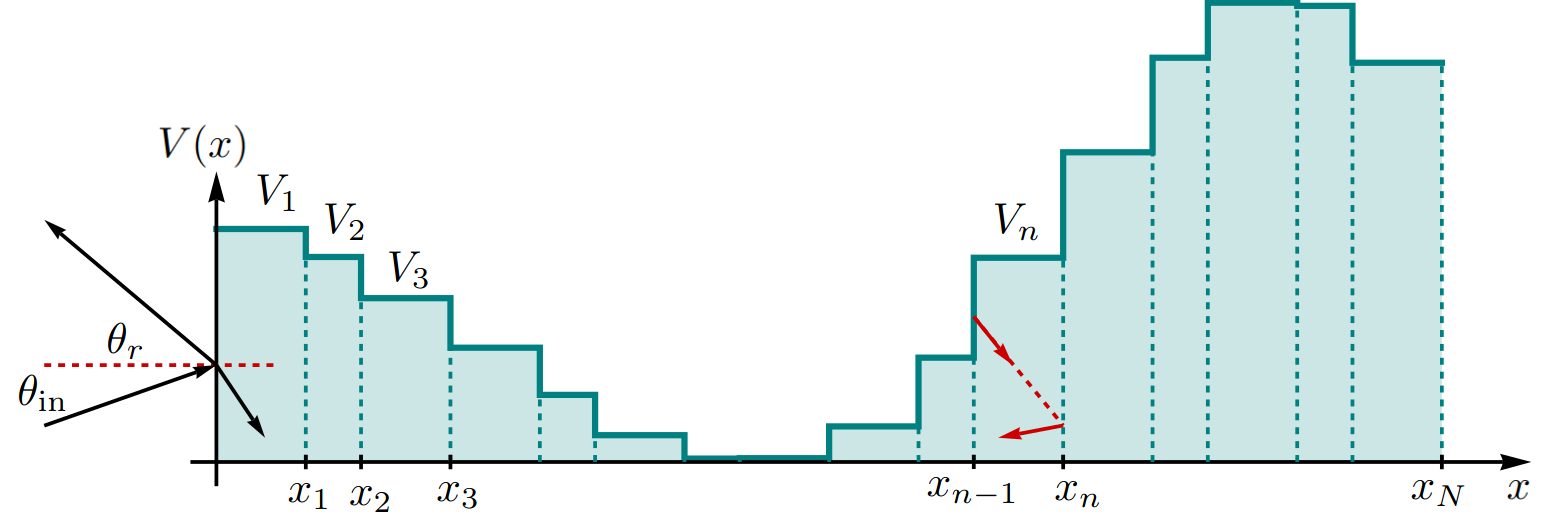}
    \caption{Arbitrary electrostatic potential modeled as a series of potential barriers, where the transfer matrix method can be used.}
    \label{fig:Arbpot}
\end{figure}

 To study the electron scattering of anisotropic materials under stratified electrostatic potentials, (see Fig. \ref{fig:Arbpot}), we model the system as a sequence of adjacent potential barriers with width $x_n - x_{n-1}$ and heights $V_n$, where $n$ denotes the barrier index \cite{Walker1994,BrionesTorres2014,CarreraEscobedo2014,Xu2015,Bezerra2020,Phan2021,DiazBautista2024,Carrillo2021,GarciaCervantes2015,GarciaCervantes2022,MolinaValdovinos2022,RodriguezVargas2012}. \comD{Experimentally, individual barrier widths $D_n = x_n-x_{n-1}$ typically range from 5 nm to 1 $\mu$m \cite{Cayssol2009,Young2009,Lee2015,Bai2018}, while barrier heights are on the order of 0.05-0.3 eV. Current graphene samples have characteristic sizes of 10-20 $\mu$m, although recent reports indicate the centimeter scale samples are also achievable \cite{Huhtasaari2025}. Therefore, devices comprising $N \geq 100$ barriers, with minimum widths of approximately 5 nm and a total system size of $L \approx 20 \mu$m, are experimentally feasible.}
 
 The general matrix transfer method starts by writing the wave function ansatz for the barrier $n$ as

\begin{equation}
    \vec{\Psi}^n(\vec{r}) = A_{n}\vec{u}(\vec{k}^a_n)\textrm{e}^{i\vec{k}^a_n\cdot\vec{r}}+ B_{n}\vec{u}(\vec{k}^b_n)\textrm{e}^{i\vec{k}^b_n\cdot\vec{r}},
\end{equation}

\noindent where $A_{n}$ and $B_{n}$ are the amplitudes of the wave inside the barrier. For the first region, $n = 0$, the incident wave has an amplitude $A_0 = 1$ and wave vector $\vec{k}^a_0=\vec{k}_\textrm{in}$, for the reflected part, the amplitude is $B_0 = A_r$ and wave vector $\vec{k}^b_0=\vec{k}_\textrm{r}$. With $n = N+1$ in the last region, $B_{N+1} = 0$ because there is no reflection, and the transmitted amplitude is $A_{N+1} = A_t$.

Applying the matching conditions of Eq. \eqref{TCC} at $x = x_{n}$ that separates the regions with wave functions $\vec{\Psi}^n(\vec{r})$ and $\vec{\Psi}^{n+1}(\vec{r})$, we have

\begin{equation}
    \left(\begin{array}{c}
          A_{n+1}\\
          B_{n+1}
    \end{array}\right) = M^{-1}_{n+1 n} M_{n n}\left(\begin{array}{c}
          A_{n}\\
          B_{n}
    \end{array}\right),
\end{equation}

\noindent where the matrices $M_{m l}$ are

\begin{equation}
    M_{m l} =\left[\begin{array}{cc}
      \vec{w}(\vec{k}^a_n)  & \vec{w}(\vec{k}^b_n)
    \end{array}\right]\left(\begin{array}{cc}
      \textrm{e}^{ik^a_{x,n}x_l}  & 0\\
      0 & \textrm{e}^{ik^b_{x,n}x_l}
    \end{array}\right),
\end{equation}

\noindent which taken into account the conservation of $k_y$ parallel to the interfaces at $x = x_n$ and the substitution $\vec{u}(\vec{k}_n)\rightarrow\vec{w}(\vec{k}_n)$ given by Eq. \eqref{wtild}. It is possible to get a $2\times2$ linear equation system involving only $A_\textrm{r}$ and $A_\textrm{t}$ for the first and last regions by defining the matrix of the stratified potential media

\begin{equation}
\Lambda \equiv \prod^N_{n=1}\mathcal{M}_{n}\Theta_n\mathcal{M}_n^{-1},
\end{equation}

\noindent where the multiplication is performed from left to right, increasing the index $n$. The matrix $\mathcal{M}_n$ is
\begin{equation}
 \mathcal{M}_n = \left[\begin{array}{cc}
      \vec{w}(\vec{k}^a_n)  & \vec{w}(\vec{k}^b_n)
    \end{array}\right]   
\end{equation}

and

\begin{equation}
    \Theta_n= \left(\begin{array}{cc}
      \textrm{e}^{-ik^a_{x,n}(x_n-x_{n-1})}  & 0\\
      0 & \textrm{e}^{-ik^b_{x,n}(x_n-x_{n-1})}
    \end{array}\right).
\end{equation}

\noindent The matrix $\Lambda$ contains all the information about the electron crossing a stratified electrostatic potential medium. The effect of $\Lambda$ is similar to replacing the stratified media by a single interface at $x = 0$, and therefore, the transmission is identical to $pn$ junctions, as discussed in the previous section. The equation of a $pn$ junction-like is

\begin{equation}
    \vec{w}(\vec{k}_\textrm{in})+A_r\vec{w}(\vec{k}_\textrm{r}) = A_t\textrm{e}^{ik_{x,t}x_{N+1}}\Lambda\vec{w}(\vec{k}_t),
\end{equation}

\noindent which is identical to Eq. \eqref{AryAt}. Hence, the reflection coefficient is similar to Eq. \eqref{Refl}

\begin{equation}\label{Refl_MTM}
    R(\vec{k}_\textrm{in})= |A_\textrm{r}|^2 = \frac{|\vec{w}(\vec{k}_\textrm{in})\times\Lambda\vec{w}(\vec{k}_\textrm{t})|^2}{|\vec{w}(\vec{k}_\textrm{r})\times\Lambda\vec{w}(\vec{k}_\textrm{t})|^2}.
\end{equation}

\noindent It is worth mentioning that this transfer method embodies known results for $pn$ junctions of graphene, transistors, and one-dimensional superlattices~\cite{Allain2011,Katsnelson2006,Park2008}. From Eq. \eqref{Refl_MTM}, it is clear that the perfect transmission is reached when the effective incident one-half pseudo-spin $\vec{w}(\vec{k}_\textrm{in})$ is parallel to the transmitted pseudo-spin $\Lambda\vec{w}(\vec{k}_\textrm{t})$, while total reflection appears if $|\vec{w}(\vec{k}_\textrm{in})\times\Lambda\vec{w}(\vec{k}_\textrm{t})| = |\vec{w}(\vec{k}_\textrm{r})\times\Lambda\vec{w}(\vec{k}_\textrm{t})|$. Therefore, the relative angle of the states $\vec{w}(\vec{k}_\textrm{in})$ and $\Lambda\vec{w}(\vec{k}_\textrm{t})$ can indicate the emergence of Klein and anti-Klein tunneling as

\begin{equation}\label{gamma}
\gamma(\vec{k}_\textrm{in},\vec{k}_\textrm{t}) = \arccos\left(\frac{|\vec{w}^*(\vec{k}_\textrm{in})\cdot\Lambda\vec{w}(\vec{k}_\textrm{t})|}{|\vec{w}(\vec{k}_\textrm{in})||\Lambda\vec{w}(\vec{k}_\textrm{t})|}\right),
\end{equation}

\noindent where $\gamma(\vec{k}_\textrm{in},\vec{k}_\textrm{t}) = 0$ and $\gamma(\vec{k}_\textrm{in},\vec{k}_\textrm{t}) = \pi/2$ correspond to Klein and anti-Klein tunneling, respectively.

\comA{The modified transfer matrix method presented here can also be employed to model wave transmission across smooth $pn$ junctions. Thus, the first region, $x <0$, has a constant potential $V_0$. The smooth region, extended over the interval $0 < x < d$, can be modeled as a stratified medium composed of successive potential barriers whose heights increase linearly. While the last region, $x > d$, has a constant potential $V$.}

\section{The emergence of Klein and anti-Klein tunneling}
The absence of backscattering on electric barriers in the normal direction was understood as
a consequence of a topological singularity identified with a
Dirac point, see \cite{Ando1998}, \cite{Ando2000}, or as a result
of the pseudo-spin conservation \cite{Katsnelson2006}. 
A key point in the absence of back-scattering is the conservation of pseudo-spin, that explains the perfect transmission of massless Dirac fermions in potential barriers \cite{Allain2011}. The Heisenberg picture indicates that the conservation of pseudo-spin along the direction $x$ is given by

\begin{equation}
    \frac{d \sigma_x}{dt} = \frac{i}{\hbar}[\hat{H},\sigma_x].
\end{equation}
\noindent
The Hamiltonian must depend straightforwardly on $\sigma_x$, and therefore, the pseudo-spin $s_x = \hbar\sigma_x/2$ is a constant of motion. This fact occurs under normal incidence of electrons in an electrostatic potential in graphene, where the Hamiltonian is written as $H = v_F\sigma_xp_x+V(x)$, being $v_F$ the Fermi velocity, $p_x$ the linear momentum, and $V(x)$ the electrostatic potential \cite{Katsnelson2006}. 

The absence of backscattering can also be explained by a unitary transformation that maps the subsystem with perfect transmission into the free-particle dynamics \cite{Jakubsky2011}. Indeed, let us consider the two-dimensional Dirac fermion in the presence of an electric barrier that has translational symmetry described effectively by a two-dimensional Dirac equation. The fermions bouncing on the barrier in the normal direction ($k_y=0$) are governed by the following operator

\begin{equation}\label{Hv}
H_{k_y=0}=e^{-i k_y y}(-i\sigma_x\partial_x-i\sigma_y\partial_y+V(x))e^{ik_y y}|_{k_y=0}.
\end{equation}

The operator can be mapped to the free-particle Hamiltonian via the unitary transformation  

\begin{equation}\label{u}
 U=e^{i\alpha\sigma_x}=\cos\alpha\, +
 i\sin\alpha\,\sigma_x,\quad U^{\dagger}=U^{-1},
\end{equation}
dependent on the interaction potential,
\begin{equation}
 \alpha(x)=\frac{1}{v_F}\int^x V(\tau)d\tau.
\end{equation}
Indeed, there holds
\begin{equation}
UH_{k_y=0}U^{-1}=-i\sigma_1\partial_x-i\sigma_2\partial_y.
\end{equation}
The relation justifies the absence of reflection in the system governed by $H_{k_y=0}$ as there are no back-scattered waves in the free-particle system. The unitary mapping can be also used for deformation of both the vector potential and effective mass \cite{Jakubsky2015}.

However, related phenomena, such as super-Klein tunneling of massive pseudo-spin-one particles, do not seem to follow from the pseudo-spin conservation. The spin-orbit Hamiltonian for super-Klein tunneling is $H = v_F\vec{S}\cdot\vec{p} + U$, where $U$ is the mass term and $\vec{S}$ is a spin-one operator \cite{BetancurOcampo2017}. If we apply again the Heisenberg picture, the omnidirectional perfect transmission involves both components in the linear momentum $\vec{p}$, then

\begin{equation}
     \frac{d \vec{S}}{dt} = \frac{i}{\hbar}[\hat{H},\vec{S}] \neq 0,
\end{equation}

\noindent which leads to the non-conservation for the pseudo-spin one operator. The usual explanation for super-Klein tunneling is due to the special matching condition of the wave function at the interface. In the present, we established a general condition that embodies the emergence of Klein tunneling in graphene, super-Klein tunneling in pseudo-spin one systems, and anti-Klein tunneling of bilayer graphene and phosphorene, through the definition of a shortened pseudo-spin 1/2 from the wavefunction continuity. The shortened spinor appears when the boundary conditions are applied at the interface for the eigenvectors of the Bloch Hamiltonian, as seen in Eq. \eqref{wtild}. The perfect transmission (reflection) occurs when the incident spinor $\vec{w}_i$ is parallel (perpendicular) to the transmitted state $\Lambda\vec{w}_t$, giving rise to the Klein (anti-Klein) tunneling, as seen in Eqs. \eqref{Refl} and \eqref{Refl_MTM}.

\section{Types of Klein tunneling beyond graphene: anomalous, anti-Klein, and super-Klein tunneling}

\begin{table*}[t]
\centering
\textcolor{black}{
\caption{Comparative classification of different variants of Klein tunneling (KT) in condensed matter and artificial systems.}
\label{tab:KT_comparison}
\begin{tabular}{|p{3.2cm} |p{3.2cm}| p{3.8cm} |p{4.2cm} |p{3.0cm}|}
\hline
\toprule
\textbf{Type of KT} 
& \textbf{Physical features} 
& \textbf{Effective Hamiltonian} 
& \textbf{Material / System} 
& \textbf{Experimental status} \\
\hline
\midrule
1D KT 
& Perfect transmission and pseudo-spin conservation
& 1D Dirac and modified Dirac Hamiltonian
& Bearded SSH, SSH lattices, and possibly in another one-dimensional chains \cite{BetancurOcampo2024}
& Unobserved \\
\hline
\midrule
Conventional KT
& Perfect transmission at normal incidence
& 2D Dirac Hamiltonian
& Graphene and 2D phononic lattices \cite{Katsnelson2006,Allain2011,Beenakker2008,Jiang2020}
& Observed \cite{Young2009,Jiang2020}\\
\hline
\midrule
Anomalous KT 
& Perfect transmission for incidence angles outside the normal direction
& 2D Dirac-Weyl Hamiltonian with and without tilted Dirac cone term 
& Strained graphene, borophene, and anisotropic hexagonal lattices \cite{BetancurOcampo2018,Zhou2019a}
& Observed in photonic lattices \cite{Zhang2022c}\\
\hline
\midrule
Super KT 
& Omnidirectional perfect transmission 
& Spin-one Dirac Hamiltonian, spin-1/2 Dirac Hamiltonian, and Klein-Gordon equation 
& Dice, Lieb, $\alpha$-$\tau_3$ lattices, graphene, and electrostatic gratings \cite{Urban2011,BetancurOcampo2017,Kim2019a,ContrerasAstorga2020,Wang2022}
& Observed in phononic Lieb and triangular lattices for massless quasi-particles \cite{Zhu2023,Wu2024}. 
Unobserved for massless spin 1/2 particles and massive pseudo-spin one particles \\
\hline
\midrule
Anti-KT 
& Backscattering at normal incidence by pseudo-spin momentum locking. 
& Hybrid Schrödinger-Dirac Hamiltonian 
& Bilayer graphene and checkerboard lattice \cite{Katsnelson2006,Hua2024}
& Observed in bilayer graphene\cite{Varlet2014,Du2018}\\
\hline
\midrule
Anti-super-KT 
& Omnidirectional backscattering by pseudo-spin momentum locking 
& Hybrid Schrödinger-Dirac Hamiltonian 
& Phosphorene and black phosphorus 
& Unobserved \\
\hline
\midrule
Valley-cooperative KT 
& Perfect transmission and valley-flip transport 
& Dirac Hamiltonian with intervalley coupling 
& Kekulé graphene superlattices and artificial crystals \cite{Garcia2022} 
& Unobserved \\
\hline
\bottomrule
\end{tabular}}
\end{table*}

The previous sections established the methodologies for studying wave transmission in synthesized and artificial crystals under position-dependent potentials. In the following subsections, we present several manifestations of Klein and anti-Klein tunneling in low-dimensional systems for massless and massive particles. 

 \comA{Table \ref{tab:KT_comparison} summarizes the different types of Klein and anti-Klein tunneling predicted for electrons in 1D and 2D materials. Owing to the close analogies between these phenomena and wave propagation in periodic media, similar effects can be realized across a wide range of physical platforms, provided that the pseudo-spin degree of freedom and band structure can be suitably engineered. In condensed matter systems, clean and sharp $pn$ junctions at low temperatures have been experimentally realized in graphene and bilayer graphene, enabling the successful observation of Klein, anti-Klein tunneling, negative refraction, and electron optics \cite{Young2009,Bai2017,Chen2016,Lee2015,Bai2018}.}
 
 \comA{In contrast, related effects such as super-Klein, anomalous, and anti-super-Klein tunneling remain unobserved experimentally, despite significant advances in the field. In particular, the realization of super-Klein tunneling in a dice or Lieb lattice is challenging in condensed matter, mainly due to the lack of a stable 2D material with the required lattice geometry. Beyond electronic systems, super-Klein tunneling has been tested in phononic Lieb lattices \cite{Zhu2023,Wu2024}. While anomalous Klein tunneling has been experimentally observed in photonic graphene \cite{Zhang2022c}.} 
 
 \comA{The principal challenges in electronic systems stem from the synthesis and characterization of novel 2D materials, as well as the requirements to achieve a ballistic transport regime. In contrast, the versatility and rapid prototyping capabilities of metamaterials provide a promising and direct route toward the experimental realization of atypical and hitherto unexplored Klein tunneling phenomena.}

\subsection{Klein tunneling in one-dimensional chains}

\begin{figure*}\label{transmssh}
    \centering
    \begin{tabular}{ccc}
    (a) \qquad \qquad \qquad \qquad \qquad \qquad \qquad \qquad & (b) \qquad \qquad \qquad \qquad \qquad \qquad \qquad \qquad  & (c) \qquad \qquad \qquad \qquad \qquad \qquad \qquad \qquad\\
    \includegraphics[width=0.25\linewidth]{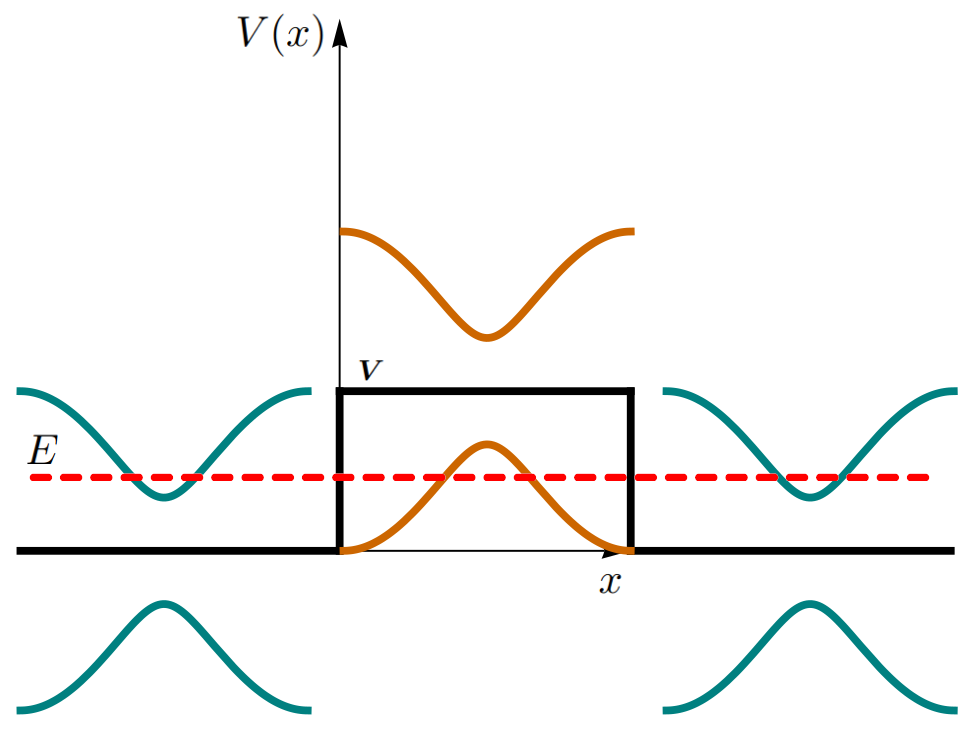} \qquad \qquad &
    \includegraphics[width=0.25\linewidth]{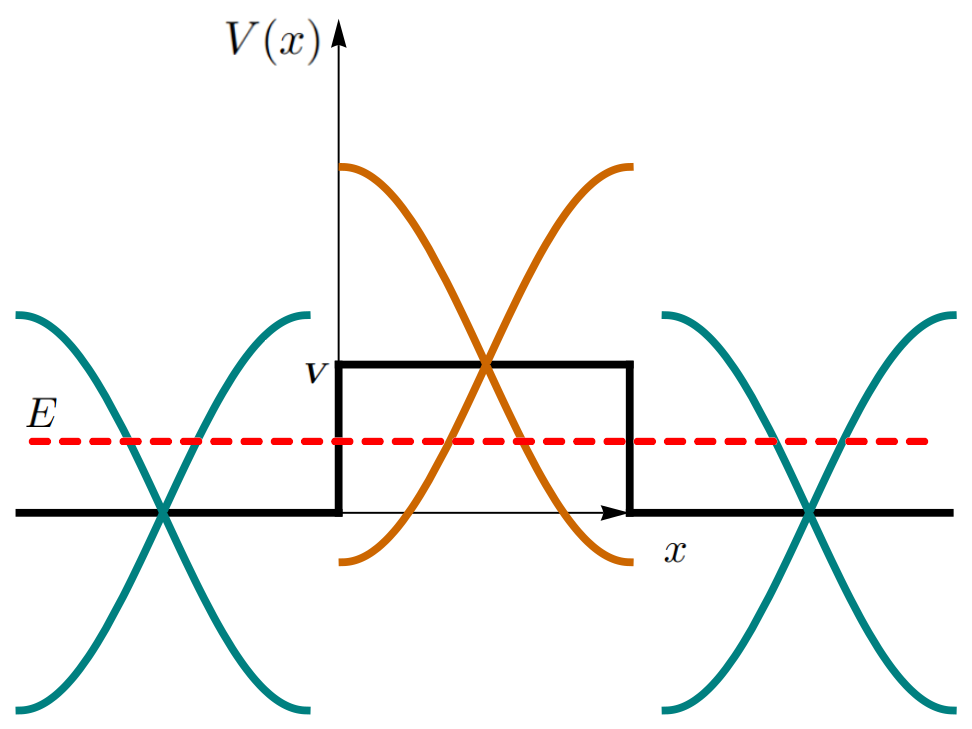} \qquad \qquad &
    \includegraphics[width=0.25\linewidth]{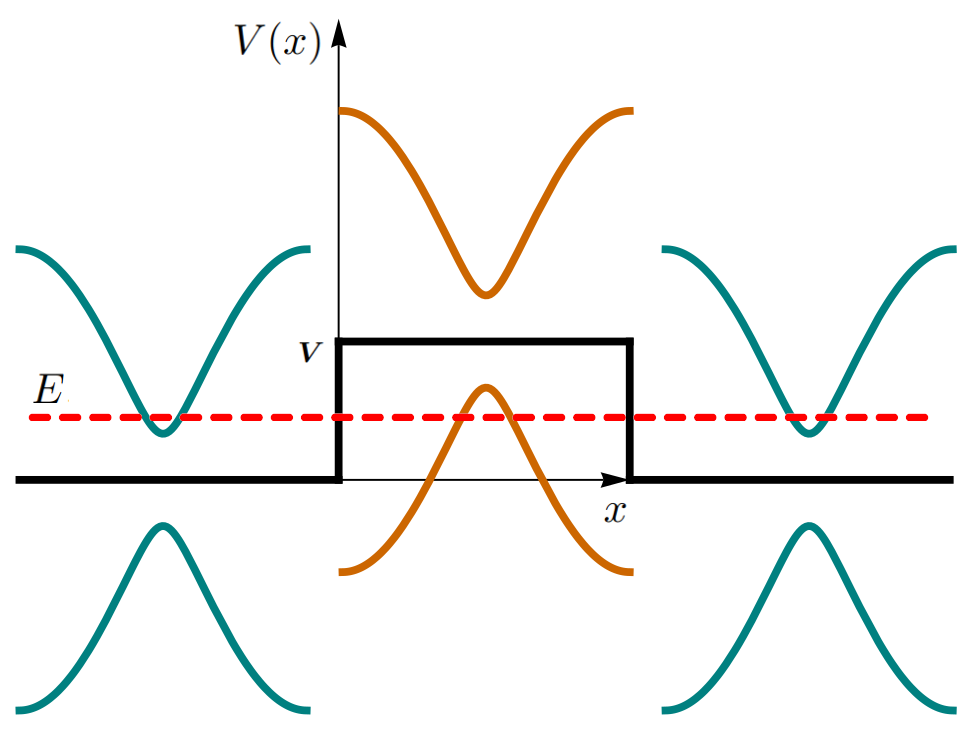}\qquad \qquad 
    \end{tabular}
    \begin{tabular}{ccc}
    (d) \qquad \qquad \qquad \qquad \qquad \qquad \qquad \qquad & (e) \qquad \qquad \qquad \qquad \qquad \qquad \qquad \qquad  & (f) \qquad \qquad \qquad \qquad \qquad \qquad \qquad \qquad\\
    \includegraphics[width=0.25\linewidth]{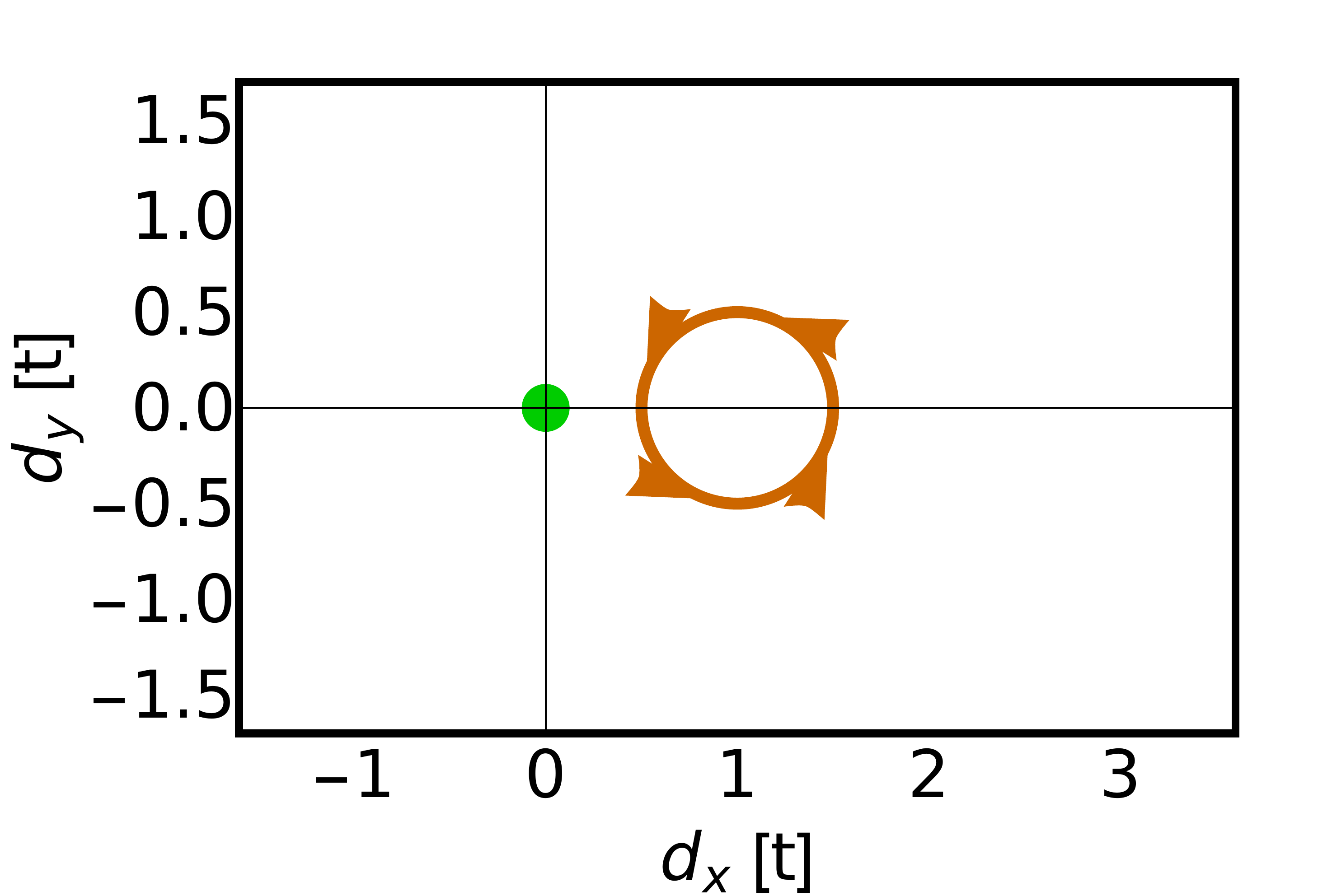} \qquad \qquad &
    \includegraphics[width=0.25\linewidth]{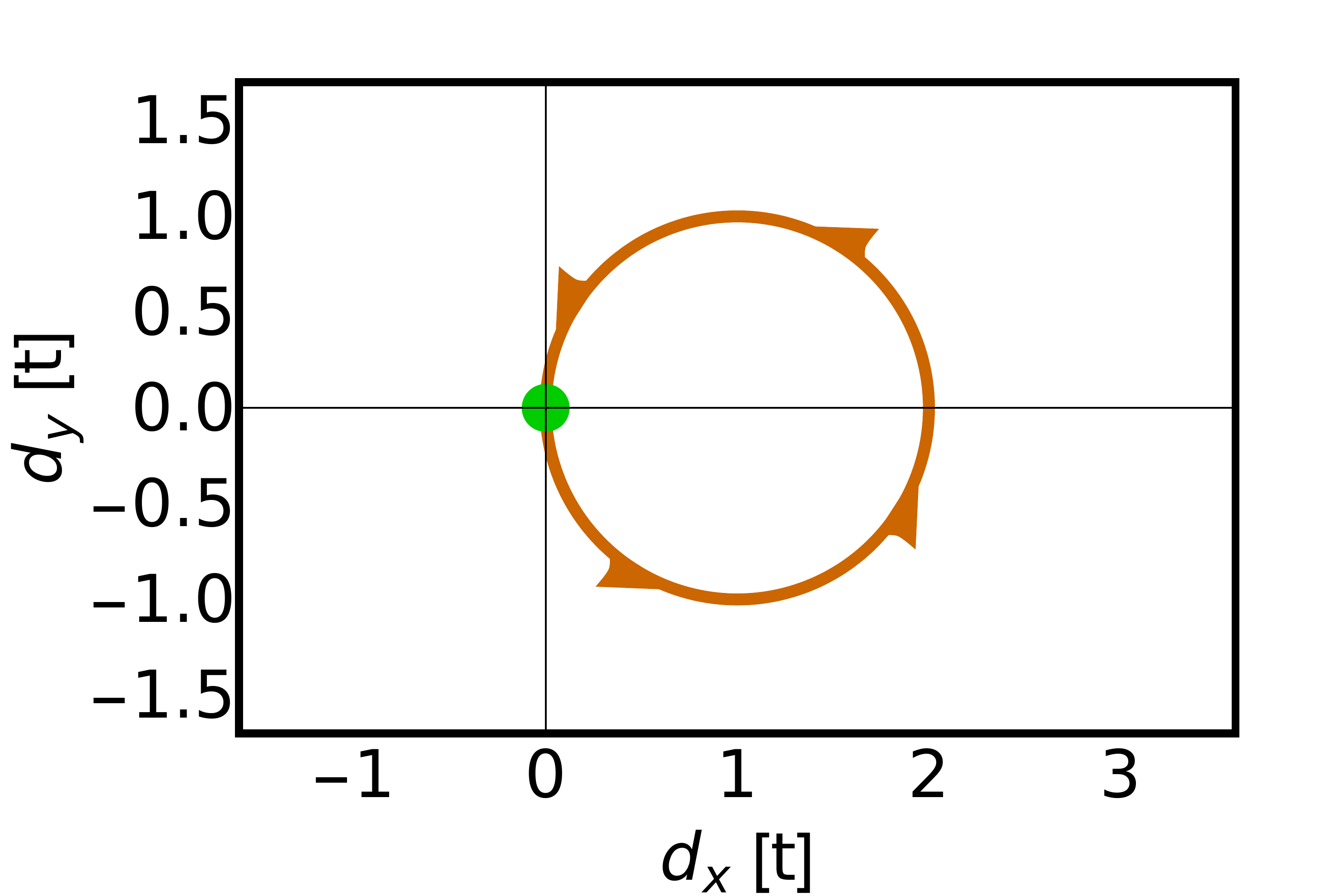} \qquad \qquad &
    \includegraphics[width=0.25\linewidth]{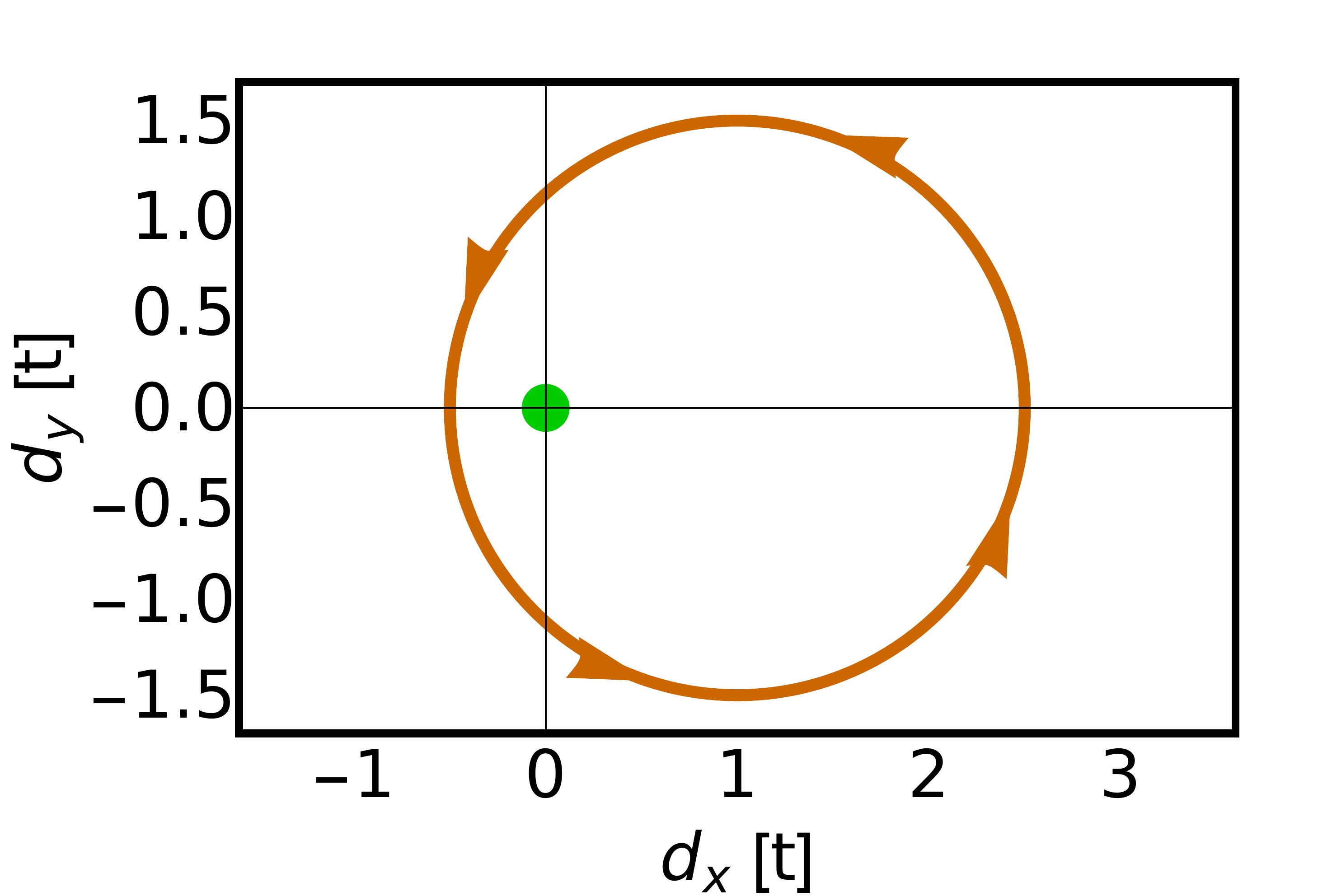}\qquad \qquad \\
    (g) \qquad \qquad \qquad \qquad \qquad \qquad \qquad \qquad & (h) \qquad \qquad \qquad \qquad \qquad \qquad \qquad \qquad  & (i) \qquad \qquad \qquad \qquad \qquad \qquad \qquad \qquad\\
    \end{tabular}
    \begin{tabular}{cccc}
    \includegraphics[width=0.3\linewidth]{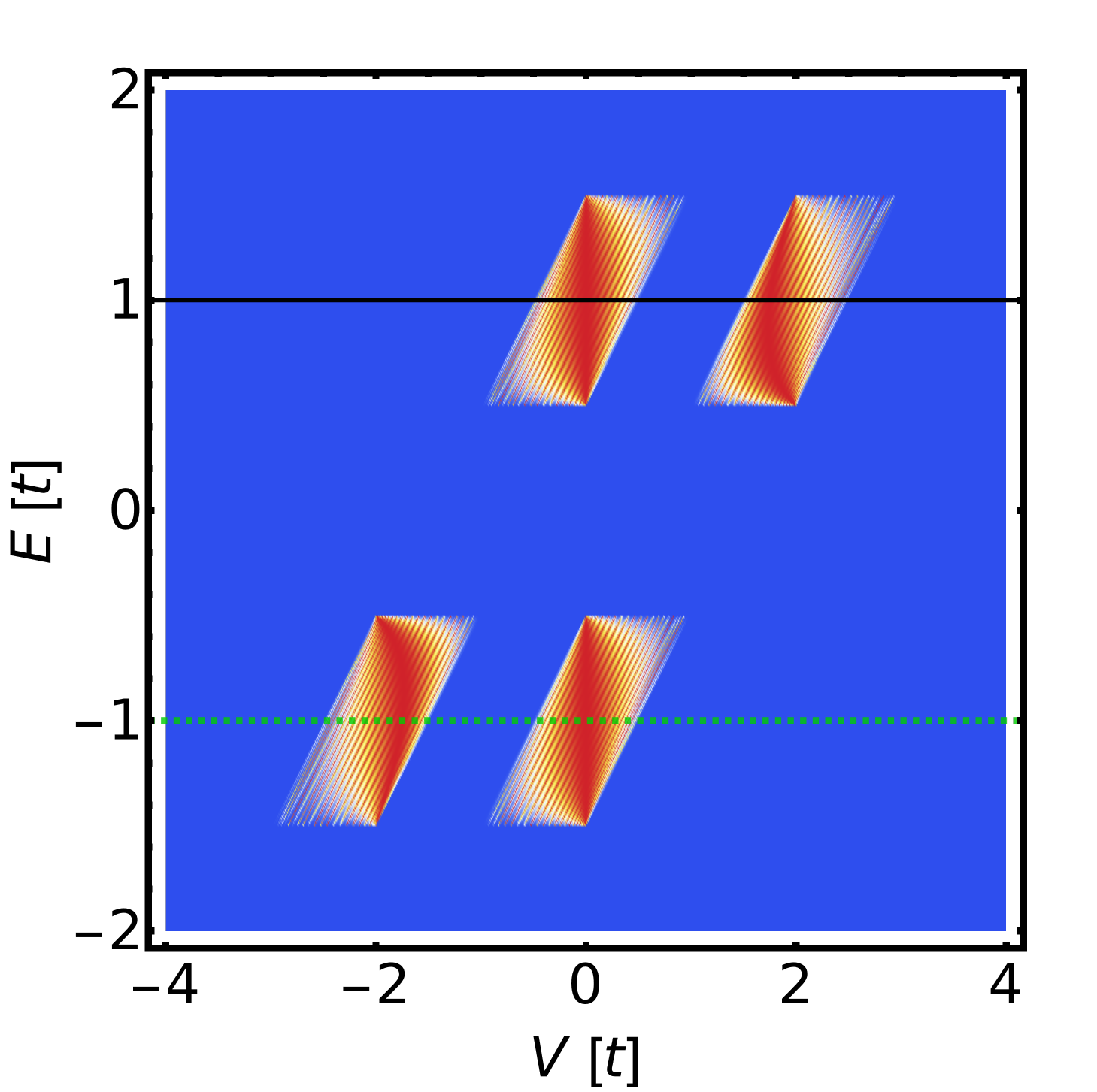}&
    \includegraphics[width=0.3\linewidth]{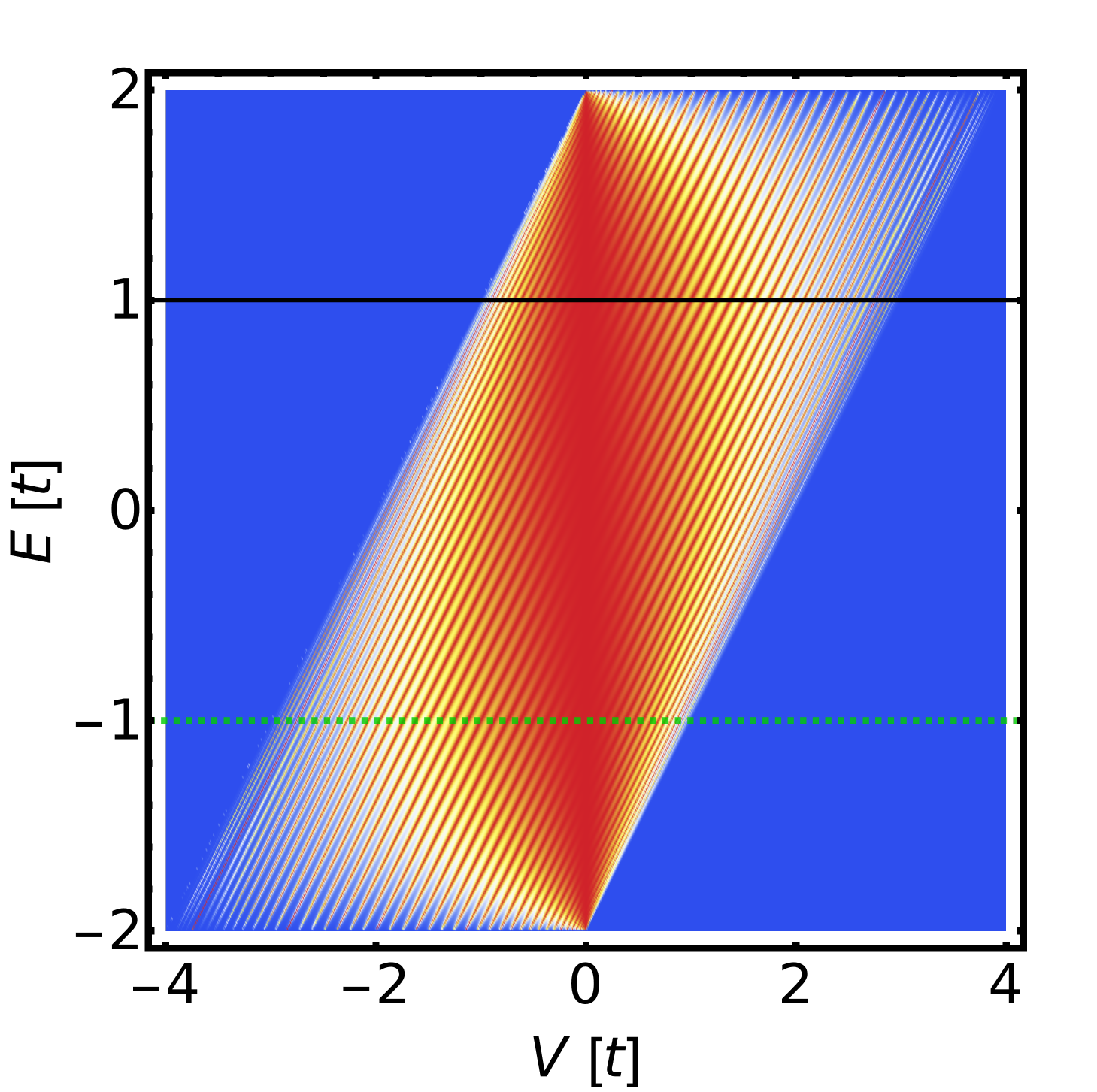}&
    \includegraphics[width=0.3\linewidth]{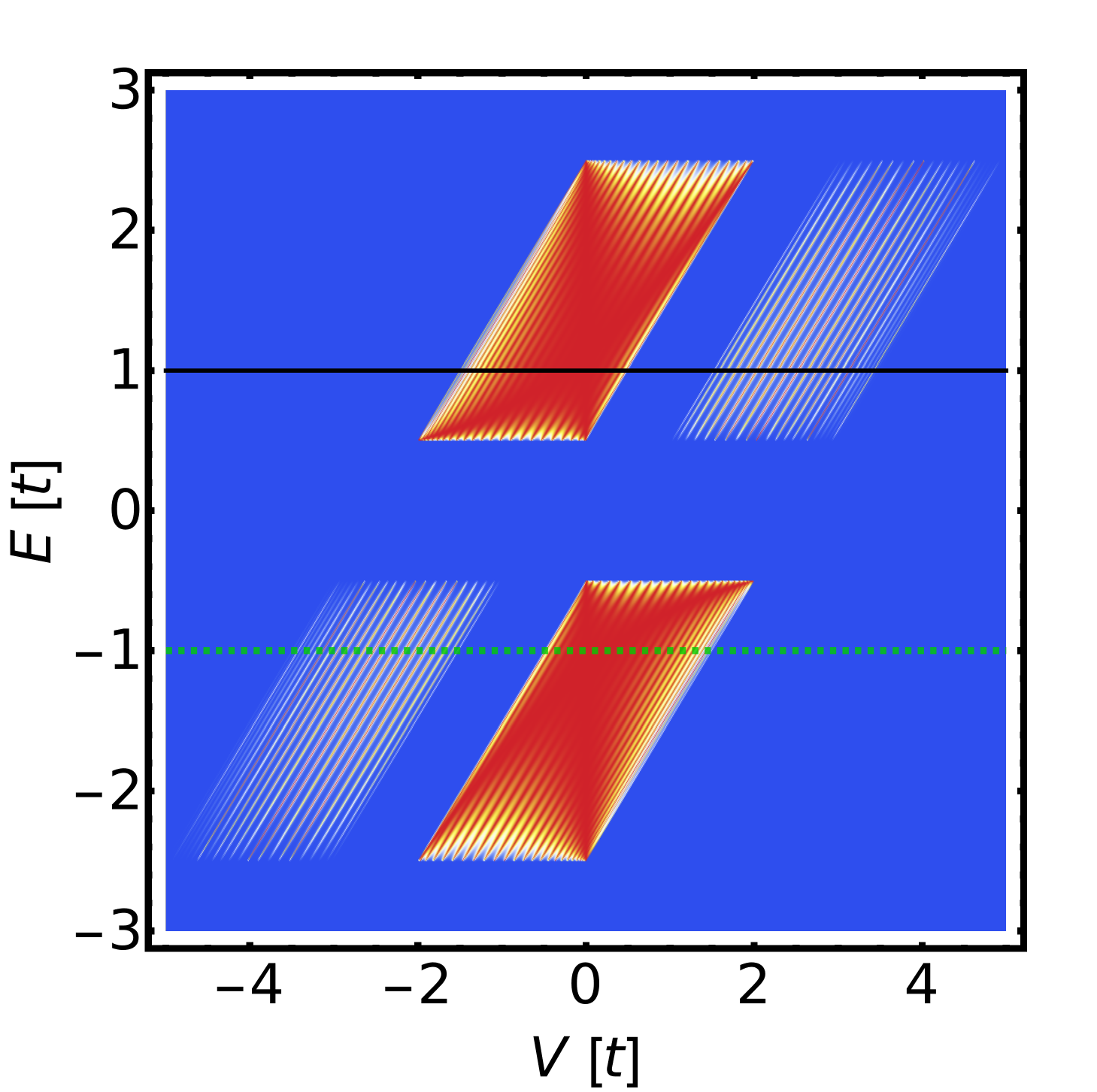}&
     \includegraphics[width=0.035\linewidth]{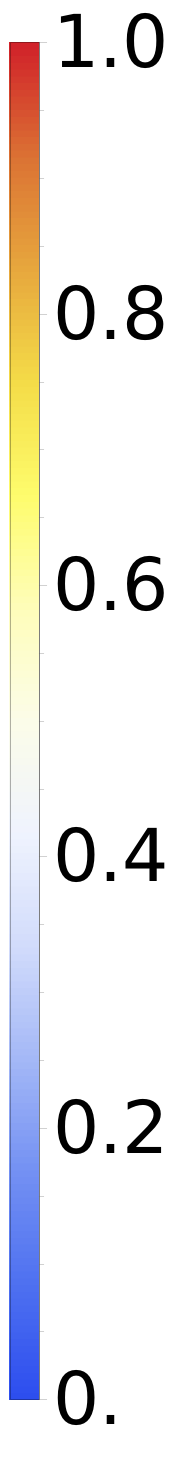}
    \end{tabular}
    \begin{tabular}{ccc}
    (j) \qquad \qquad \qquad \qquad \qquad \qquad \qquad \qquad & (k) \qquad \qquad \qquad \qquad \qquad \qquad \qquad \qquad  & (l) \qquad \qquad \qquad \qquad \qquad \qquad \qquad \qquad\\
    \includegraphics[width=0.27\linewidth]{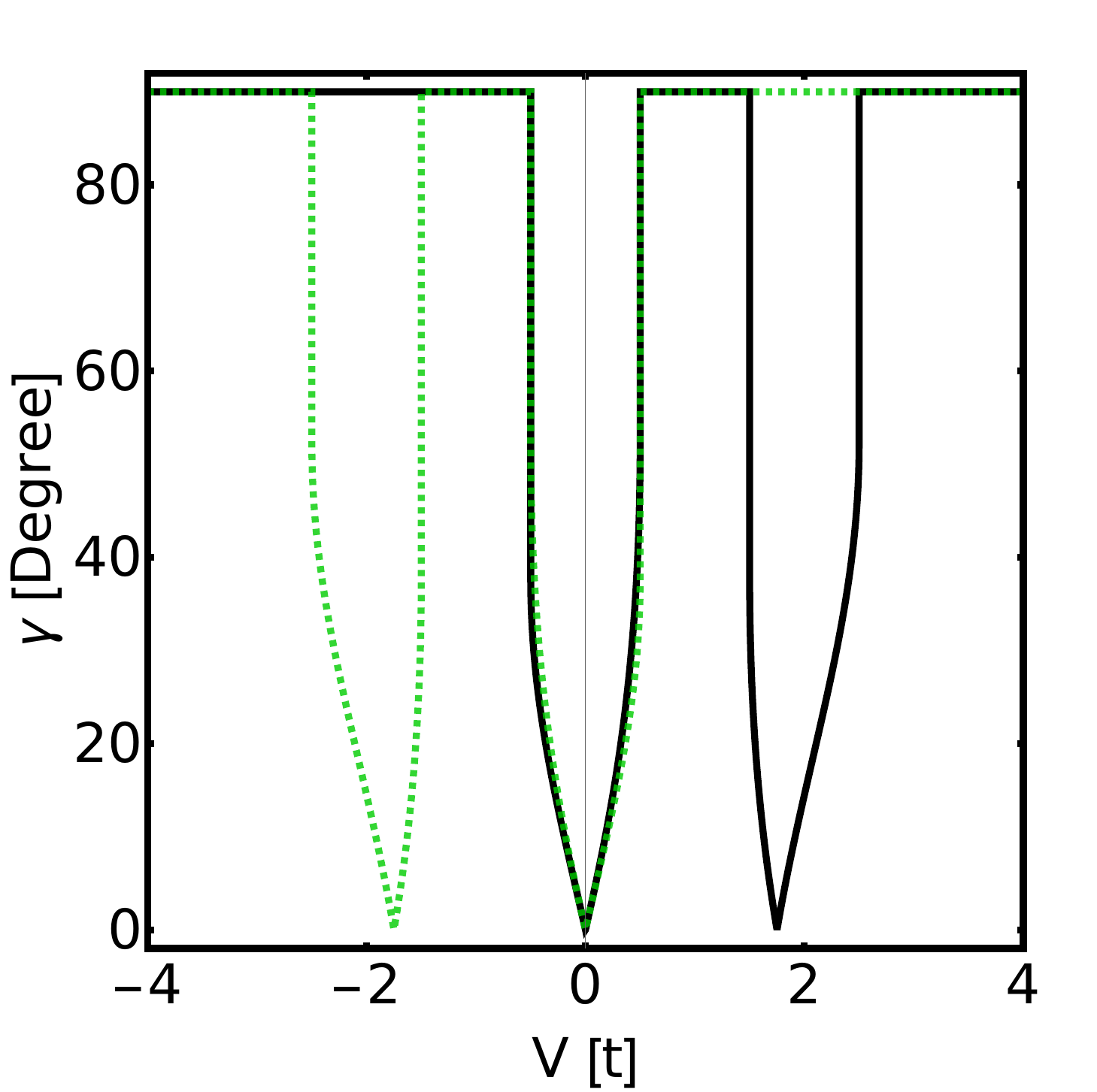} \qquad \qquad &
    \includegraphics[width=0.27\linewidth]{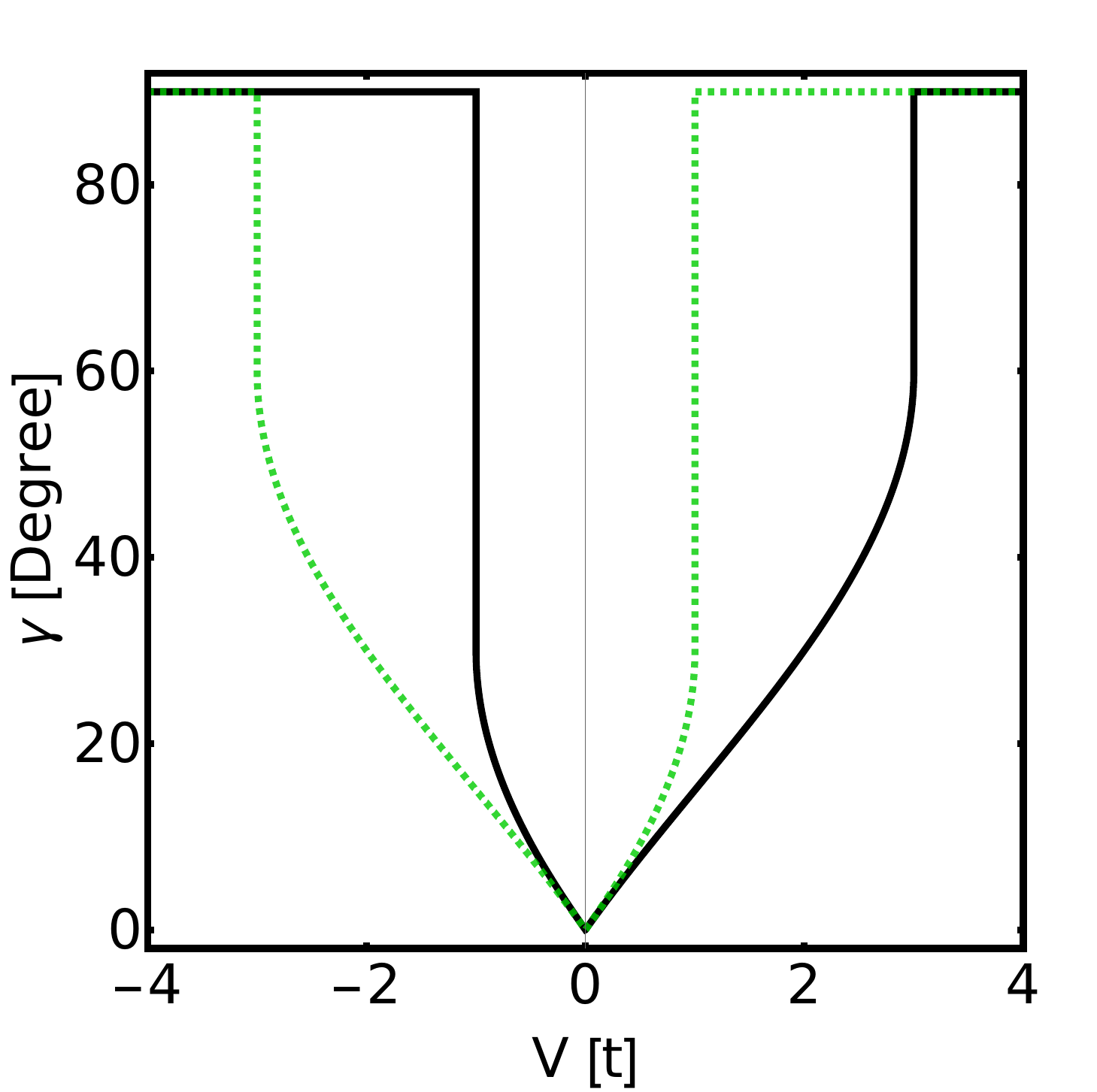} \qquad \qquad &
    \includegraphics[width=0.27\linewidth]{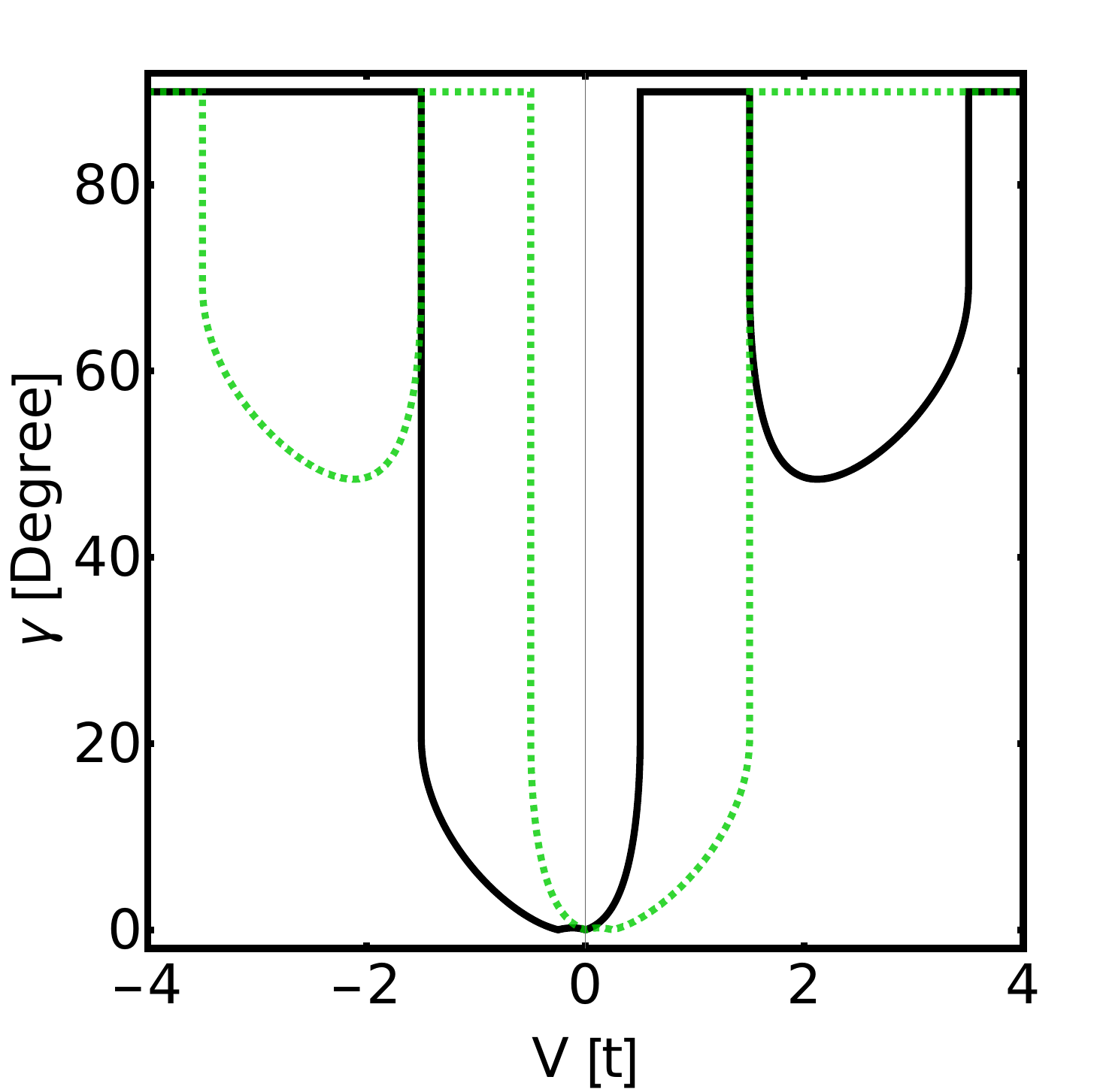}\qquad \qquad 
    \end{tabular}
    \caption{Electronic band structure of a bipolar junction based on Su-Schrieffer-Heeger chain. The curves indicate the energy bands for the trivial (a), critical (b), and topological phase (c). The dashed red line is the Fermi level $E$ and $V$ the barrier height, which can be negative for well potentials. Pseudo-spin loops in the Bloch plane for the trivial (d), critical (e), and topological phase (f). The green circle pointed out the origin of the Bloch plane. Electron transmission as a function of the Fermi level $E$ and barrier height $V$. (g) Trivial phase has the set of values $t = 1$ and $t' = 0.5$, (h) the parameters for the critical topological phase transition are $t = t' = 1$, and (i) topological phase with $t = 1$ and $t' = 1.5$. The barrier width is $D = 25$ in units of $a$. \comA{Panels (j)-(l) show the relative pseudo-spin orientation defined in Eq. \eqref{gamma} as a function of $V$ for the fixed energies $E = 1$ and $E = -1$. corresponding to the trivial (j), critical (k), and topological (l) phases.}}
    \label{fig:transmssh}
\end{figure*}

To study particle transmission in one-dimensional chains, such as Su-Schrieffer-Heeger (SSH) lattices in $pnp$ junctions,  we analyze the transmission as a function of the energy $E$ and electrostatic potential of the barrier $V$, as shown in Fig. \ref{fig:transmssh}(a)-(c). To identify possible signatures of Klein tunneling, we focus on areas of high transmission. It is well known that the SSH lattices undergo a topological phase transition upon tuning one of the hopping parameters: the system is in a trivial (topological) phase for $t < t'$ ($t > t'$) \cite{Su1979,Heeger1988,Heeger1988,Meier2016,Li2018,Lieu2018,Xie2019,Thatcher2022,Qian2020,Kim2020a,Hu2020,Coutant2021,Mukherjee2021,Li2022a,Manda2023,BetancurOcampo2024}, where the winding number is the topological invariant that characterizes these phases, see Fig. \ref{fig:transmssh}(d)-(f). This transition strongly influences both interband and intraband tunneling in the whole regimens of the bipolar junction from $nn'n$ to $pp'p$, as shown in Fig. \ref{fig:transmssh}. We compute the transmission in the trivial ($t < t'$), critical ($t = t'$), and topological ($t > t'$) phase using Eq. \eqref{TCoeff}, taking into account that the SSH chain is described by the Bloch Hamiltonian

\begin{equation}\label{Hssh}
    H_\textrm{SSH}(k) = \left(\begin{array}{cc}
       0  & t-t'\textrm{e}^{-ik} \\
       t-t'\textrm{e}^{ik}  & 0
    \end{array}\right),
\end{equation}

\noindent which warranties the simple relation of wave numbers $k_\textrm{in} = -k_\textrm{r}$. The length of the unit cell is $a_c = 1$. The matching conditions for the wave function with the Hamiltonian in Eq. \eqref{Hssh} gives rise to the effective pseudo-spin vectors in Eq. \eqref{TCC}, where the functions $f(k)$ are 

\begin{equation}
   f(k_{\textrm{in/r/t/}a/b}) = \frac{\textrm{e}^{i k_{\textrm{in/r/t/}a/b}}-1}{k_{\textrm{in/r/t/}a/b}}.
\end{equation}

\noindent The conservation energy $E$ allows us to relate $k_{\textrm{in/r/t/}a/b}$ in terms of $E$, the barrier height $V$, and hopping parameters

\begin{eqnarray}\label{ks}
    k_\textrm{in}(E) & = & -k_\textrm{r}(E) = k_\textrm{t}(E) \nonumber\\
    & = & \sgn(E)\arccos\left(\frac{t^2+t'^2-E^2}{2\,t\,t'}\right)\nonumber\\
    k_a(E,V) &=& -k_b(E,V) \nonumber\\
    & = & \sgn(E-V)\arccos\left(\frac{t^2+t'^2-(E-V)^2}{2\,t\,t'}\right).\nonumber\\
    &&
\end{eqnarray}

\noindent With these relations in Eq. \eqref{ks}, the transmission coefficient in Eq. \eqref{TCoeff} can be expressed as a function of $E$ and $V$.
The transmission features, as shown in Fig. \ref{fig:transmssh}(g)-(i), are explained by using the relative pseudo-spin direction, which is given by Eq. \eqref{gamma}. Although the SSH chain is one-dimensional, the orientation of the pseudo-spin is represented by the polar angles $\alpha$ and $\phi$ in the Bloch sphere. The electron scattering involves reduced components of the spinors as indicated in Eq. \eqref{EqB}, we can relate these spinors in the Bloch sphere for the incident, reflected, and transmitted states with the set of angles $(\alpha_\textrm{in}, \phi_\textrm{in})$, $(\alpha_\textrm{r}, \phi_\textrm{r})$, and $(\alpha_\textrm{t}, \phi_\textrm{t})$ as a function of $k_\textrm{in}$. Nevertheless, our interest is to get the relative angle $\gamma(k_\textrm{in},k_\textrm{t})$ between the incident and transmitted state in Eq. \eqref{gamma}, since that reflection and transmission depend on this angle. In the case of evanescent waves, total internal reflection is obtained. In such a situation, the definition of $\gamma(k_\textrm{in},k_\textrm{t})$ in Eq. \eqref{gamma} has a real value. However, to indicate this total reflection we can define the constant value of $\gamma = \pi/2$. For propagation states, the value of $\gamma = \pi/2$ in Eq. \eqref{dphi} is not excluded. If the pseudo-spin of incident and transmitted states is perpendicular, an anti-Klein tunneling appears \cite{Hua2024,Katsnelson2006}. One simple explanation of the perfect transmission that gives rise to Klein tunneling is the conservation of the pseudo-spin $\vec{w}$. The relative difference in the pseudo-spin angle in Eq. \eqref{gamma} indicates that perfect tunneling emerges when $\gamma = 0$, such as Klein tunneling in graphene \cite{Katsnelson2006}. 

In Fig. \ref{fig:transmssh}(g), Klein tunneling appears for the negative $nn'n$ and positive $pp'p$ regimen. The reddish region for the vertical line in $V = 0$ is the absence of scattering, while bluish regions emerge due to evanescent waves in the scattering process. The different regimes for interband ($npn$ and $pnp$) and intraband tunneling ($nn'n$ and $pp'p$) are separated by the energy band gap of the SSH chain in the trivial phase $2\Delta = 2|t-t'|$. Since the transmitted wave vector $k_a$ and $k_\textrm{t}$ have imaginary part, the relative pseudo-spin angle is $\gamma = 90\deg$ and the transmission is zero by total internal reflection. The interference patterns within the regimens are due to the Fabry-Pérot resonances $k_aD = n\pi$, where $D$ is the barrier width.

The hopping parameter $t'$  controls the topological phase transition in the SSH chain, and Klein tunneling for intraband tunneling disappears in the critical value $t = t'$, [see Fig. \ref{fig:transmssh}(h)]. Due to the closing of the band gap, the four regions for the transmission merge. In contrast, passing from the trivial to the topological phase, the Klein tunneling changes its location now in the interband tunneling regimens, as seen in Fig. \ref{fig:transmssh}(i). The four regimens $npn$, $nn'n$, $pnp$, and $pp'p$ are again separated in the topological phase by the opening in the gap. The most remarkable feature is that the topological phase transition interchanges the Klein tunneling from intraband to interband tunneling.

\comA{The reduced pseudo-spin $\gamma$ defined in Eq. \eqref{gamma} and shown in Fig. \ref{fig:transmssh}(j-l), indicates that perfect transmission occurs when the incident and transmitted pseudo-spin are parallel. In Fig. \ref{fig:transmssh}(g-i), the dashed green and black transmission curves, which correspond to the energies $E = 1$ and $E = -1$, respectively, exhibit maximum transmission at the minimum values of $\gamma$, consisting with pseudo-spin alignment. The bluish regions correspond to $\gamma = 90\deg$ by definition, as the total reflection originates from evanescent modes. In contrast, genuine anti-Klein tunneling appears when the pseudo-spins are perpendicular for propagating electronic states.} 

\subsection{Anomalous Klein tunneling in anisotropic two-dimensional materials}

\begin{figure}
    \centering
    \includegraphics[width=0.9\linewidth]{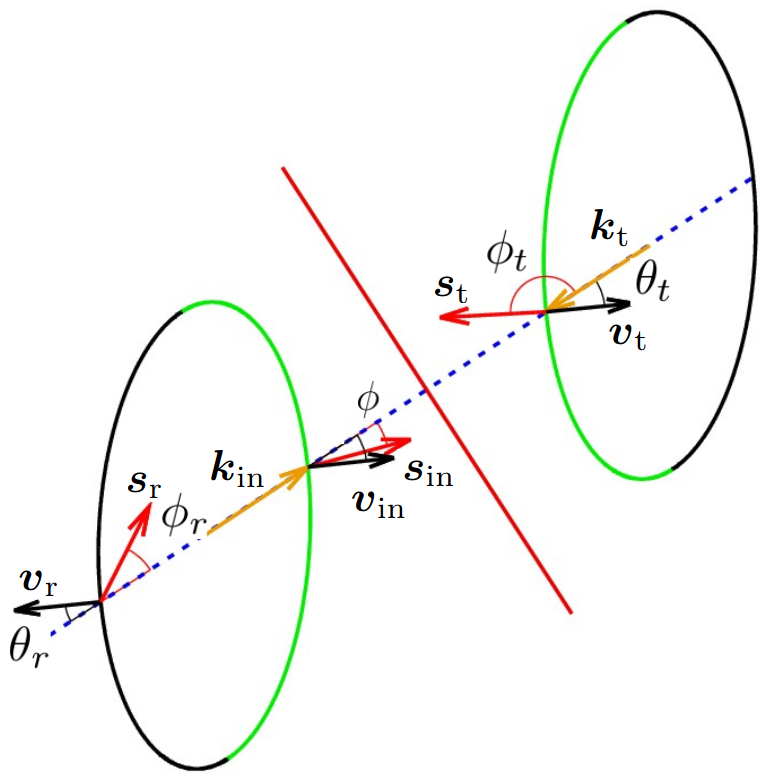}
    \caption{ Kinematical construction for the focusing condition, when $V = 2E$, which represents the conservation of energy (energy contours), linear momentum (dashed blue line), and current density (green semi-arcs). The arrows indicate the direction of linear momentum $\vec{k}_\textrm{in/r/t}$, pseudo-spin $\vec{s}_\textrm{in/r/t}$, and group velocity $\vec{v}_\textrm{in/r/t}$ for the scattering of electrons at the interface (red line).}
    \label{fig:KC}
\end{figure}

Anomalous Klein tunneling refers to perfect transmission occurring away from normal incidence, enabled by the conservation of pseudo-spin. This effect arises in anisotropic 2D materials, such as borophene and strained graphene \cite{Zhang2019,BetancurOcampo2021,BetancurOcampo2018,DiazBautista2024,Chen2023,Zhou2019a}. In this section, we analyze anomalous Klein tunneling in a general anisotropic hexagonal lattice subjected to an electrostatic and stratified potential. The study is addressed within a low-energy regimen using the matrix transfer formalism.

\begin{figure}
    \centering
    \includegraphics[width=0.75\linewidth]{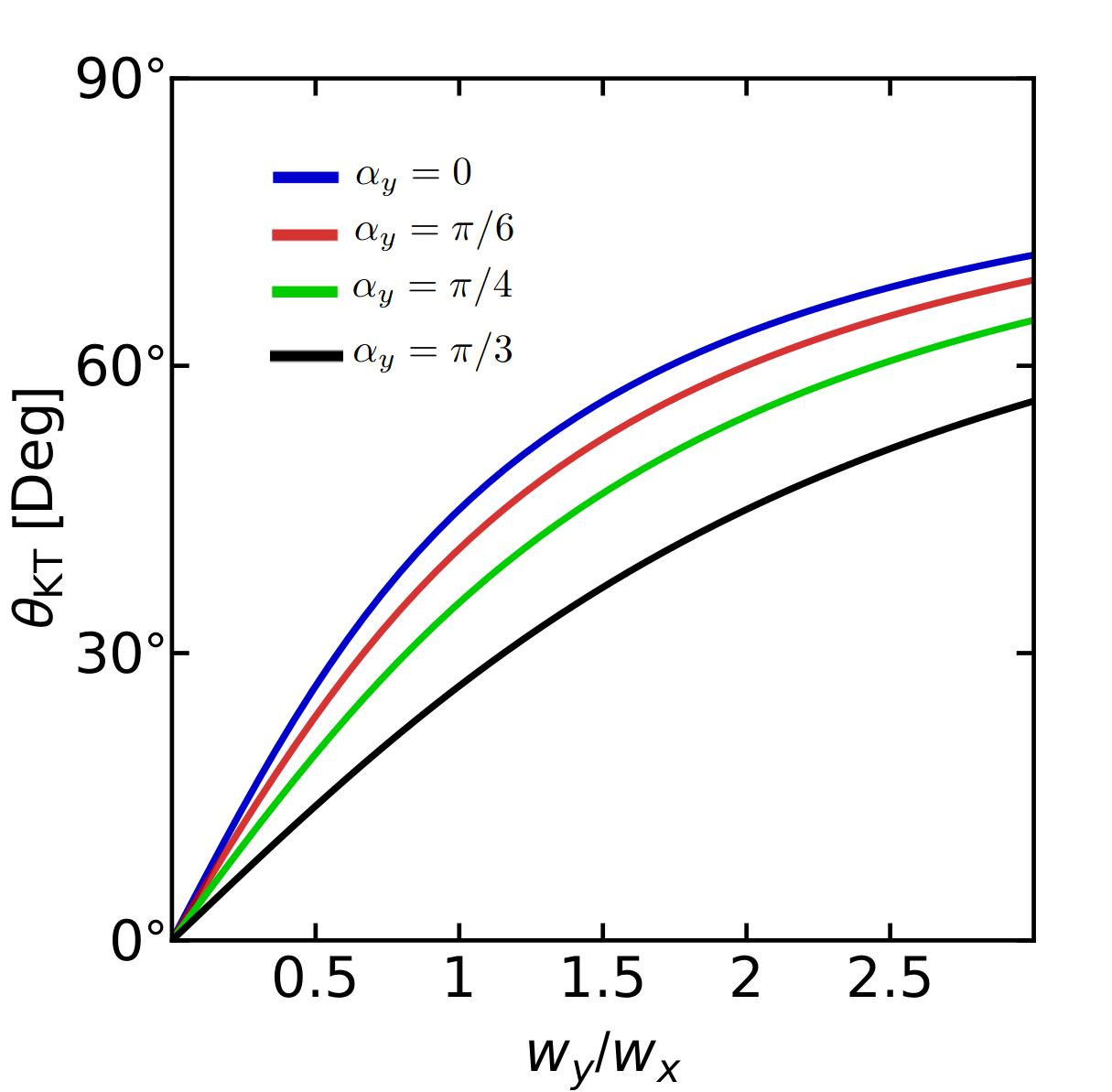}
    \caption{Anomalous Klein tunneling angle of Eq. \eqref{thKT_eq} as a function of $w_y/w_x$ ratio with the set of values for the velocity phases $\alpha_x = 0$ and $\alpha_y = 0, \pi/6, \pi/4$ and $\pi/3$.}
    \label{fig:AKT}
\end{figure}

\begin{figure*}
    \centering
    \begin{tabular}{ccc}
    (a) \qquad \qquad \qquad \qquad \qquad \qquad \qquad \qquad \qquad \qquad \qquad \qquad & (b) \qquad \qquad \qquad \qquad \qquad \qquad \qquad \qquad \qquad \qquad \qquad \qquad & \\
    \includegraphics[width=0.45\linewidth]{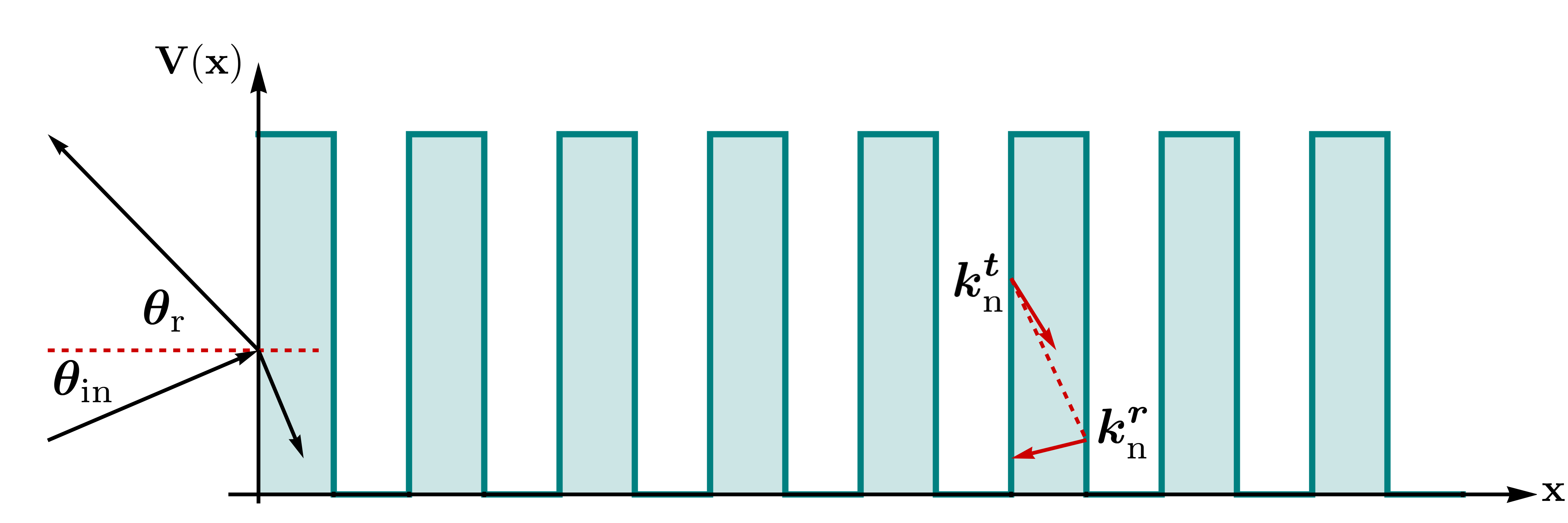}&
    \includegraphics[width=0.45\linewidth]{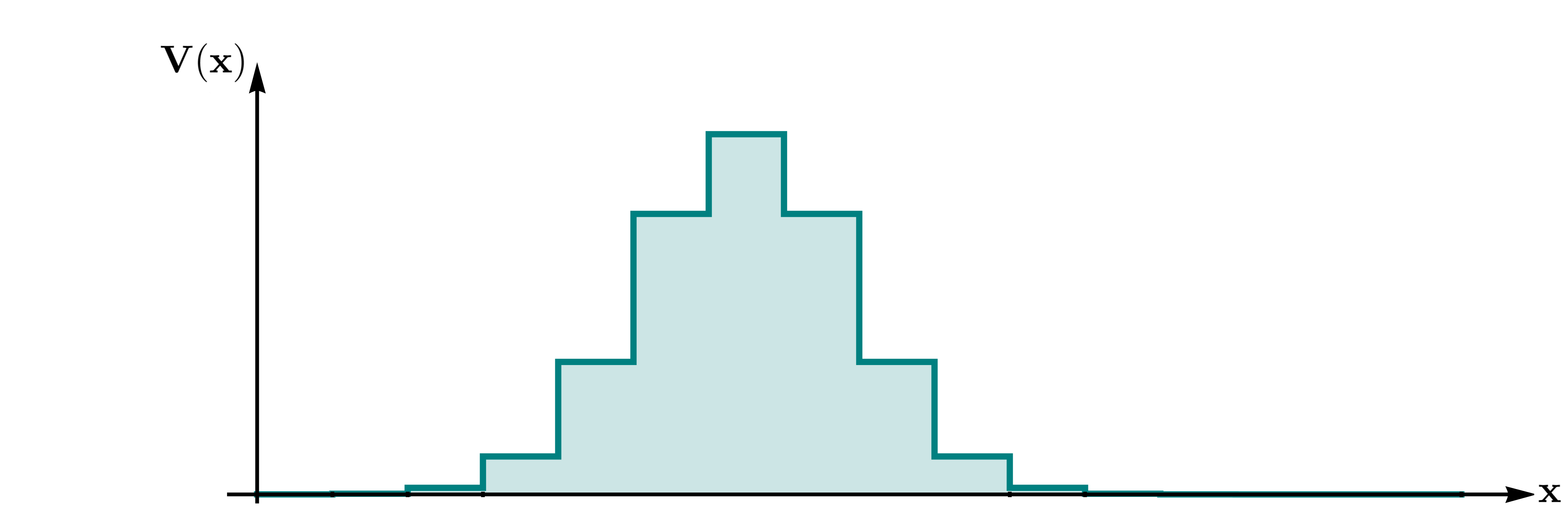}& \\
    (c) \qquad \qquad \qquad \qquad \qquad \qquad \qquad \qquad \qquad \qquad \qquad \qquad & (d) \qquad \qquad \qquad \qquad \qquad \qquad \qquad \qquad \qquad \qquad \qquad \qquad & \\
    \includegraphics[width=0.45\linewidth]{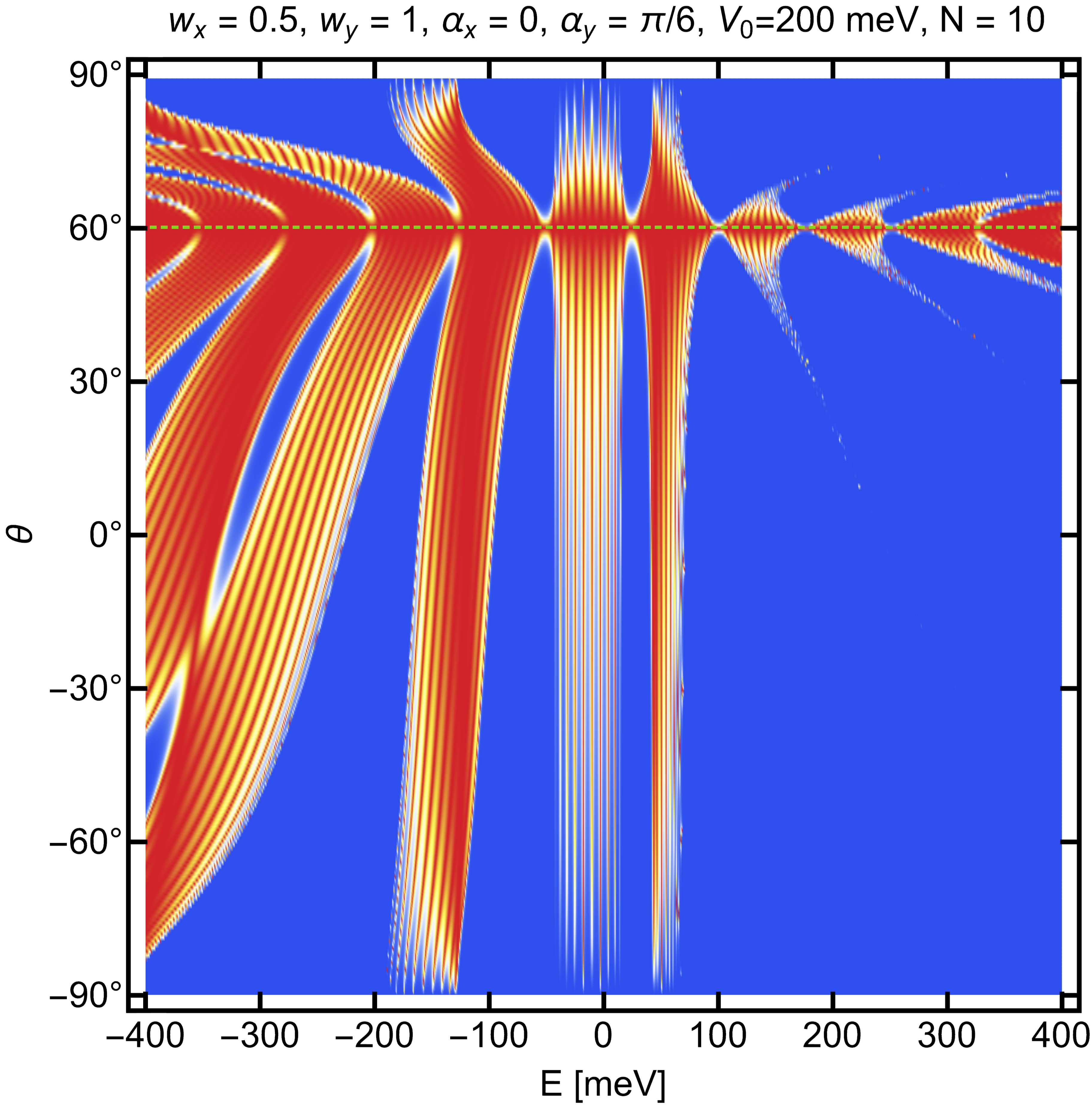}&
    \includegraphics[width=0.45\linewidth]{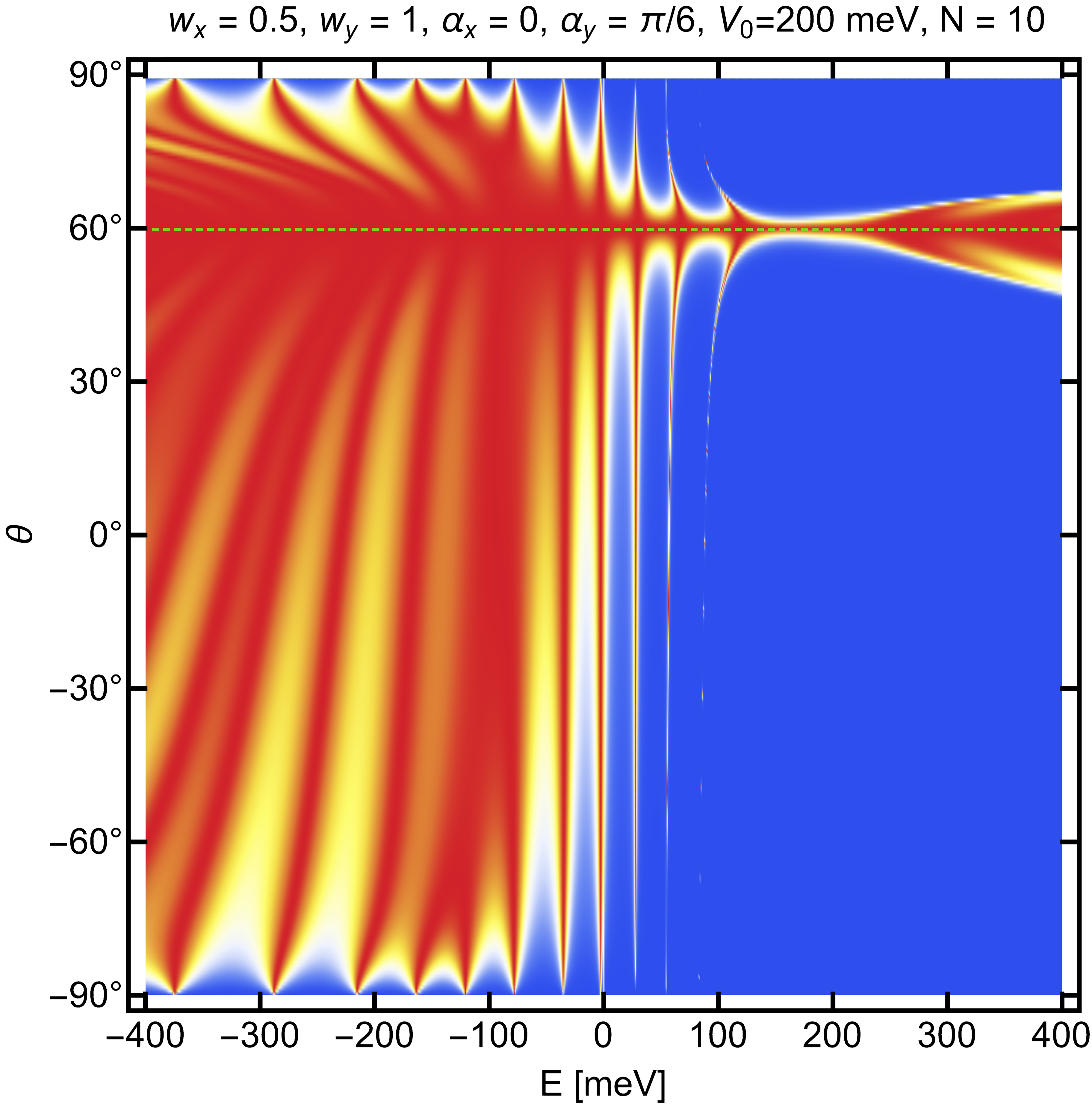} &
    \includegraphics[width=0.055\linewidth]{Barra.pdf}
    \end{tabular}
    \caption{Electron transmission for electrostatic potentials as a function of energy and incidence angle. (a) and (b) Periodic and Gaussian-like profile as a set of potential barriers. (c) and (d) Transmission coefficient corresponding to the stratified potential profile in (a) and (b), respectively. The anomalous Klein tunneling in $\theta = 60\deg$ (green dashed line) is independent of the shape of electrostatic potential, but the Fabry-Pérot resonances present drastic changes due to the electrostatic potential profile.}
    \label{fig:transm_str_media}
\end{figure*}

The Bloch Hamiltonian in Eq. \eqref{BH2D} is simplified by expanding up to first order in $\vec{k}$ around the Dirac point. We found the following Dirac-type Hamiltonian that describes spin-$1/2$ quasi-particles,
\begin{equation}\label{HW}
H=\begin{pmatrix}0&\mathbf{\omega}^*\cdot\mathbf{p}\\\mathbf{\omega}.\mathbf{p}&0\end{pmatrix}+V
\end{equation}
where $\mathbf{p}=(p_x,p_y)$ is the linear momentum, $\mathbf{\omega}=(w_xe^{i\alpha_x},w_ye^{-i\alpha_y})$ is the complex velocity, and $V$ is the constant potential. We note that the operator acquires the Dirac-type standard form for the values $\alpha_x=0$ and $\alpha_y=-\pi/2$. 

The dispersion relation for the simplified Hamiltonian in Eq. \eqref{HW} reads as follows

\begin{equation}\label{elliptC}
|E-V|=\sqrt{w_x^2p_x^2+w_y^2p_y^2+2w_xw_y p_xp_y\cos(\alpha_x+\alpha_y)},
\end{equation}

\noindent where the components $p_x$ and $p_y$ of linear momentum are parametrized as 
\begin{eqnarray}
 p_x & = &\pm s(w_y/w_x)\rho\sin(\alpha_x + \alpha_y \mp \chi)\nonumber\\
    p_y & = & \rho\sin\chi.
\end{eqnarray}
\noindent The minus in $p_x$ indicates the reflection for region I. The angle $\chi$ is a useful parameter that allows us to obtain simple expressions for reflection and Snell's law. The quantity $s = \pm 1$ is the band index. Starting from elliptical energy contours in Eq. \eqref{elliptC}, which are typical of anisotropic materials such as strained graphene, phosphorene, and borophene, the refraction index $\rho$ is shown to be the half-width of the ellipse (see Fig. \ref{fig:KC})

\begin{eqnarray}
\rho_j & = & |E - \delta_{2j}V|/[w_y\sin(\alpha_x + \alpha_y)],
\end{eqnarray}

\noindent where the Kronecker delta switches the potential value with $j = 1$ and 2 for regions I and II, respectively. The definition of pseudo-spin angle and conservation of $p_y$ allow us to derive the analogue of Snell's law for the interface created by the electric barrier. Indeed, the conservation of $p_y$ is given by $\rho_1\sin\chi_\textrm{in} = \rho_2\sin\chi_t$, where $\chi_\textrm{in}$ and $\chi_\textrm{t}$ are angles related to the incidence and refraction angles $\theta_\textrm{in}$ and $\theta_\textrm{t}$ from the group velocity, respectively. In general, for an anisotropic dispersion relation, reflected electron beams do not follow the typical reflection law $\theta_\textrm{in} = \theta_\textrm{r}$, due to the rotation of the elliptical Dirac cone. 

To obtain the appropriate direction of the electron beam, we calculate the group velocity $\mathbf{v}=(v_x,v_y)$, which is given by
\begin{eqnarray}
v_x&=&\partial_{p_x}E=\frac{w_x^2p_x+w_x w_y p_y \cos(\alpha_x+\alpha_y) }{E}, \nonumber\\ v_y&=&\partial_{p_y}E=\frac{w_y^2p_y+w_x w_y p_x \cos(\alpha_x+\alpha_y) }{E}. 
\end{eqnarray}
The direction of the group velocity is then specified by $\tan \theta=\frac{v_y}{v_x}$ and indicates the electron beam direction. It can be expressed in terms of $\chi$

\begin{equation}
\tan\theta = \pm\frac{w_y\cos[\chi \mp s(\alpha_x + \alpha_y)]}{w_x\cos\chi}.
\label{thchi}
\end{equation}

\noindent The sign minus corresponds to the reflection angle $\theta_r$. The substitutions $\theta \rightarrow \theta_t$, $\chi \rightarrow \chi_t$, and $s \rightarrow s'$ allow us to obtain the relation of the refraction angle $\theta_t$ and $\chi_t$ with the sign plus in Eq. \eqref{thchi}. Since $\chi = \chi_t = 0$ leads to the conservation of pseudo-spin, the angular shift of the anomalous KT is 

\begin{equation}
\theta_\textrm{KT} = \arctan\left[\frac{w_y}{w_x}\cos(\alpha_x + \alpha_y)\right].
\label{thKT_eq}
\end{equation}

\noindent  This result is independent of the Fermi level and potential profile. The angle for the KT can be tuned by the anisotropic parameters of the velocity $w_x$ and $w_y$. This special angle allows us to write the reflection and Snell's law in a straightforward way

\begin{equation}
\tan{\theta_\textrm{r}} = \tan\theta_\textrm{in} - 2\tan\theta_\textrm{KT}
\label{eoptlr}
\end{equation}
\begin{equation}
\tan\theta_\textrm{t} = ss'\tan\theta_\textrm{in} + (1 - ss')\tan\theta_\textrm{KT},
\label{eoptlrr}
\end{equation}

\noindent where the last expression corresponds to the special case of Snell's law when the focusing condition $\rho_1 = \rho_2$ is satisfied for elliptical energy contours with the same vertical half-width, as shown in Fig \ref{fig:KC}. These optical laws embody unusual effects such as negative reflection and refraction of electrons impinging on the interface \cite{BetancurOcampo2019,Cserti2007,Lee2015,BetancurOcampo2018}. Since the elliptical Dirac cone is rotated on both sides of the junction, the outgoing electron beam has a $v_y$ component different from zero.

In contrast, Klein tunneling in graphene occurs only for normal incidence~\cite{Katsnelson2006}. In anisotropic materials with rotated elliptical Dirac cones, however, the angle for Klein tunneling can be modulated through the complex velocities $\omega$. In Fig. \ref{fig:AKT}, the Klein tunneling angle depends on the velocity phases $\alpha_{x,y}$ and on the ratio $w_y/w_x$. Importantly, this perfect transmission is not resonant and independent of the electrostatic potential, as a direct consequence of the conservation of the pseudo-spin. This phenomenon, known as anomalous Klein tunneling, can be controlled via strain \cite{BetancurOcampo2018,BetancurOcampo2021,DiazBautista2024}.

The anomalous Klein tunneling persists from a single interface of a $pn$ junction or through stratified potential media. For instance, using a periodic electrostatic potential in Fig. \ref{fig:transm_str_media}(a), we show the transmission as a function of incidence angle $\theta_\textrm{in}$ and Fermi level energy $E$ in Fig. \ref{fig:transm_str_media}(c). The anomalous Klein tunneling appears in positive incidence, for the particular case treated here, $\alpha_x = 0$, $\alpha_y = \pi/6$, $w_x = 0.5$, and $w_y = 1$, giving rise to a deviation of $\theta_\textrm{KT} = 60\deg$ from Eq. \eqref{thKT_eq}. Nevertheless, for negative angles, the other set of resonances remains almost constant. This is due to the anisotropy of the dispersion relation. 

To show that the anomalous Klein tunneling is independent of the electrostatic potential, we change the periodic potential barrier by a set of subsequent barriers that follows a gaussian profile, see Fig.~\ref{fig:transm_str_media}(b) and (d). We can observe how the perfect transmission remains in the angle $\theta_\textrm{KT} = 60^\circ$, as previous periodic case. Moreover, the resonances are closer for positive incidence angles. Contrary to the case of negative angles, where resonances are separated from each other.  If we change the velocity phase $\alpha_y$ to negative values, the anomalous Klein tunneling appears at negative incidence angles. The transmission can be valley-independent because the values of $w_x$, $w_y$, $\alpha_x$, and $\alpha_y$ are considered the same for the whole material. The valley-dependence can appear in borophene and graphene when the system has inhomogeneity regions of hopping parameters \cite{Zhai2011}.

 When the electrostatic potential is periodic, as shown in Fig. \ref{fig:transm_str_media}(a) and (c), the transmission has alternating zones of minimum and maximum values as a function of the Fermi level. It takes higher values for larger incidence angles when $E < V$ due to the Fabry-Pérot interference. The resonances become more defined in the range of $E>V$, and their width becomes narrower. While $E<V$, a wide range of incidence angle values shows maximum transmission. Energy minigaps ($T=0$) emerge for electrons under oblique incidences due to destructive interferences. These minigaps follow a fractal structure if the number of barriers is increased. In contrast with Gaussian potential profiles, where energy minigaps are less, see Fig. \ref{fig:transm_str_media}(d). 

\subsection{Anti-Klein tunneling in two-dimensional materials}

\begin{figure}
    \centering
    \begin{tabular}{c}
         (a) \qquad \qquad \qquad \qquad \qquad \qquad \qquad \qquad \qquad \qquad \qquad \qquad\\
         \includegraphics[width=0.75\linewidth]{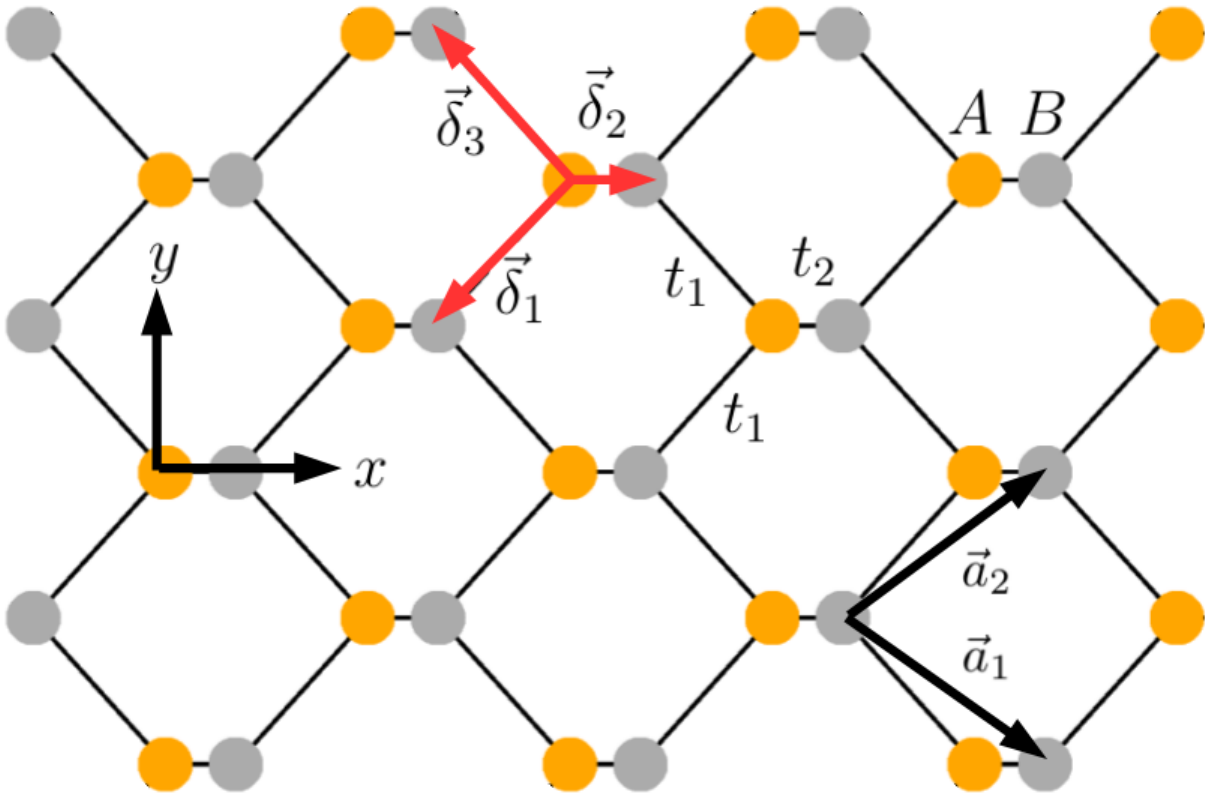} \\
         (b) \qquad \qquad \qquad \qquad \qquad \qquad \qquad \qquad \qquad \qquad \qquad \qquad\\
         \includegraphics[width=0.75\linewidth]{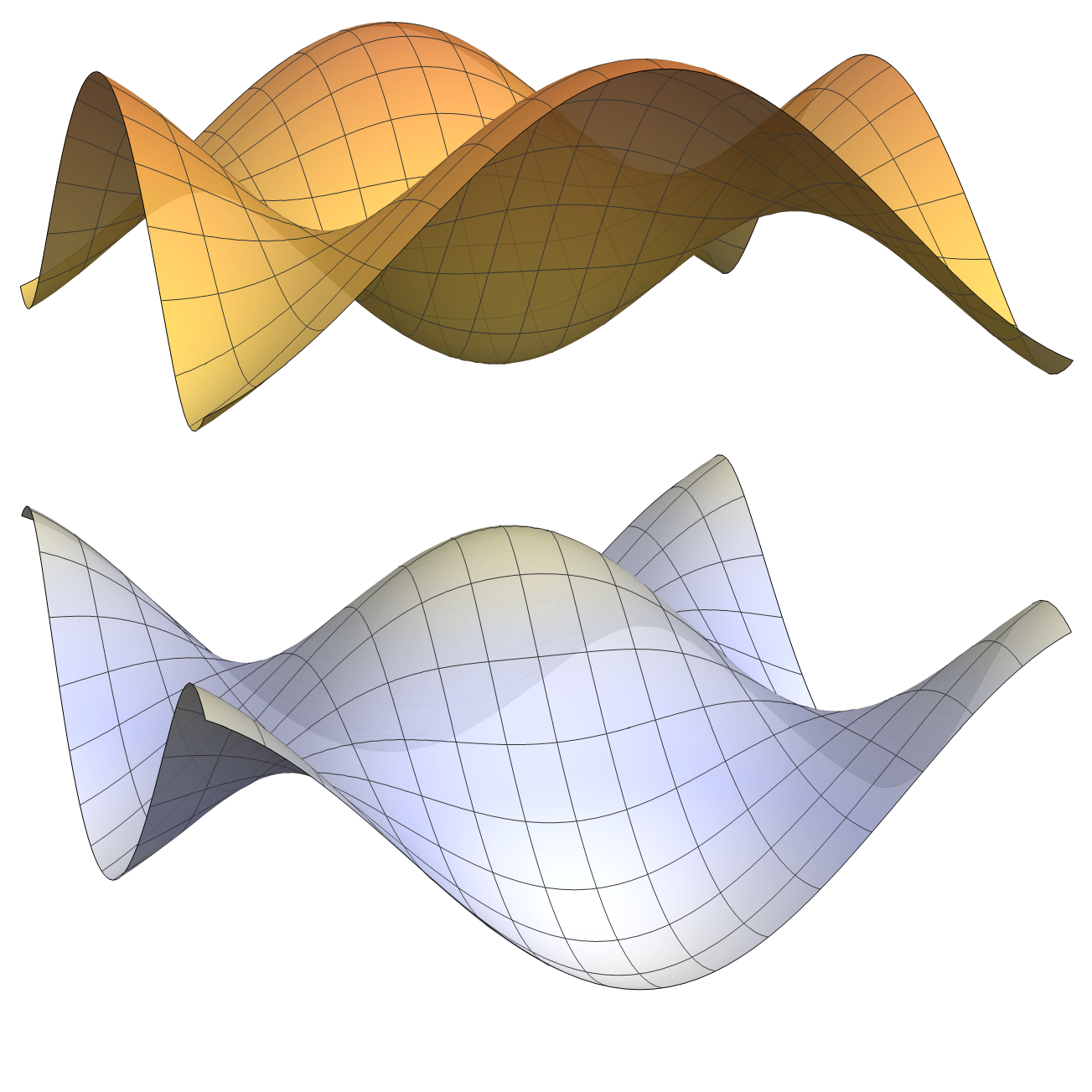}
    \end{tabular}
    \caption{(a) Top view of the lattice structure of phosphorene, where two non-equivalent sites are considered in the tight-binding approach. (b) The electronic band structure shows a band gap and a strong anisotropic dispersion relation.}
    \label{fig:phosphor}
\end{figure}

Anti-Klein tunneling is a total reflection by the pseudo-spin momentum locking. When compared to  the total internal reflection, which is due to evanescent waves, anti-Klein tunneling possesses allowed energy states. However, these states are not accessible because the effective pseudo-spin for incident waves is perpendicular to transmitted one states \cite{BetancurOcampo2019,Hua2024,LizarragaBrito2025,BetancurOcampo2020,Katsnelson2006,LamasMartinez2024,MolinaValdovinos2022,Du2018,Varlet2014,Dell’Anna2018}. One way to observe anti-Klein tunneling is under normal incidence of electrons in a $pn$ junction of bilayer graphene \cite{Katsnelson2006}. Herein, we studied the anti-Klein and anti-super-Klein tunneling of phosphorene under the special condition $V=2E$, where electrons go from
the conduction band in region I to the valence band in region II. The lattice and band structure of phosphorene are shown in Fig. \ref{fig:phosphor}. By expanding the tight-binding Hamiltonian in Eq. \eqref{BH2D} around the $\Gamma$ point in the center of the first Brillouin zone \cite{Midtvedt2017,BetancurOcampo2019,Rudenko2014}, we have

\begin{equation}\label{Hphosp}
    H_\textrm{ph}(\vec{k}) = \left(\Delta^2+\frac{k^2_x}{2m_x}+\frac{k^2_y}{2m_y}\right)\sigma_x + vk_x\sigma_y,
\end{equation}

\noindent which has an anisotropic dispersion relation, as shown in Fig. \ref{fig:phosphor}(b), where $\Delta$ is the semigap, $m_x$ and $m_y$ are the anisotropic masses, and $v$ is the velocity limit of pseudo-relativistic particles. These effective quantities are related to the hopping parameters $t_1$ and $t_2$ \cite{BetancurOcampo2019,ParedesRocha2021,LizarragaBrito2025,Ezawa2014,TaghizadehSisakht2015}. 

\begin{figure}
    \
    \begin{tabular}{c}
    (a) \qquad \qquad \qquad \qquad \qquad \qquad \qquad \qquad \qquad \qquad \qquad \qquad\\
    \includegraphics[width=0.75\linewidth]{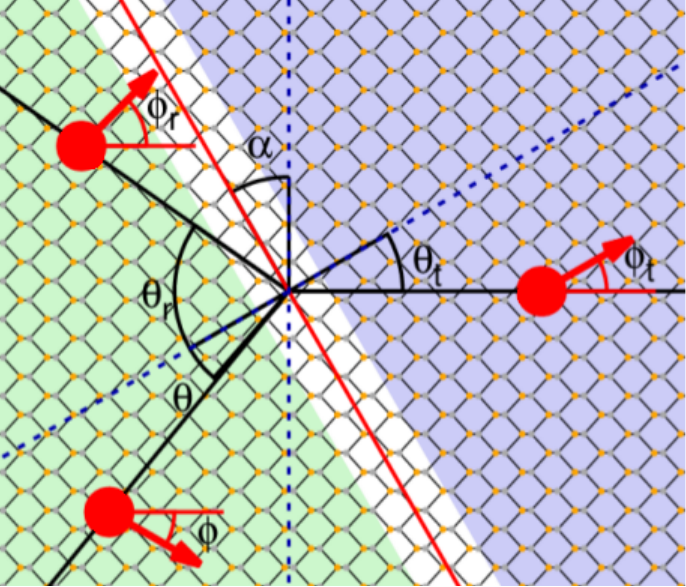}\\

    (b) \qquad \qquad \qquad \qquad \qquad \qquad \qquad \qquad \qquad \qquad \qquad \qquad\\
    \includegraphics[width=1\linewidth]{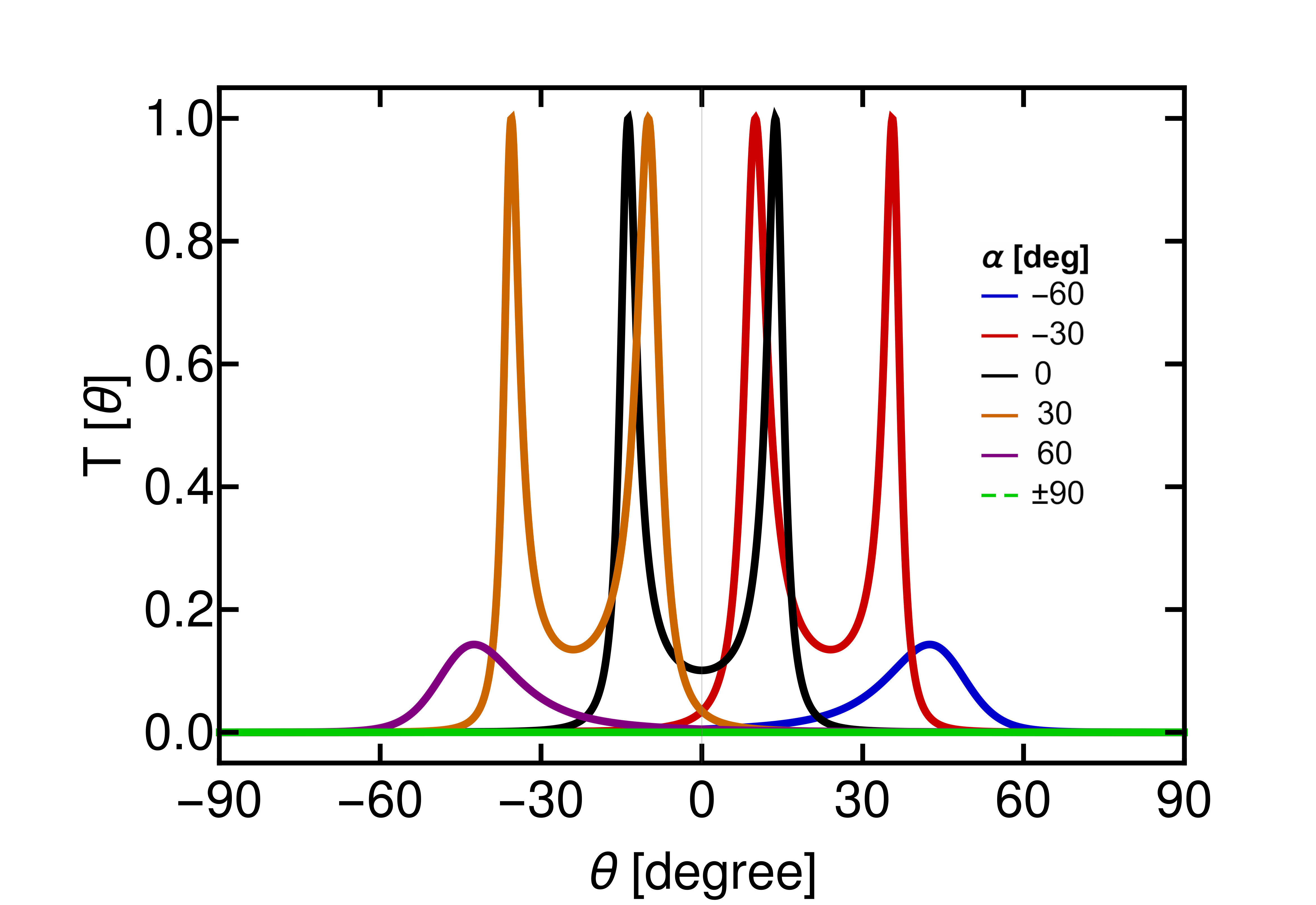}
    \end{tabular}
    \caption{(a) Electron scattering of a tilted phosphorene $pn$ junction with the schematic representation of the group velocity and pseudo-spin direction. (b) Transmission as a function of incidence angle of electrons impinging a potential barrier $V$ for different tilted angles $\alpha$.}
    \label{fig:Transm_barr_phosp}
\end{figure}

Due to the anisotropy in the dispersion relation, the reflection angle $\th_\textrm{r}$ as a
function of the incidence angle $\th_\textrm{in}$ does not satisfy the conventional reflection law, since the tilted $pn$ junction of phosphorene in Fig. \ref{fig:Transm_barr_phosp}(a) can have negative reflection \cite{BetancurOcampo2019}. The angle $\alpha$ controls the tilt of the junction. The typical reflection law $ \theta_\textrm{r} = \theta_\textrm{in} $ is found only for junctions parallel to the $x$ ($\ap= 0\deg$) and $y$ axes ($\ap= \pm 90\deg$), but in general $\th_\textrm{r}$ is a non-linear function of $\th_\textrm{in}$. Following an identical procedure for the reflection law in Eq. \eqref{eoptlr}, the reflection law is \cite{BetancurOcampo2019}

\begin{equation}
    \tan\theta_\textrm{r} = \tan\theta_\textrm{in} -2\tan\theta_\textrm{B},
\end{equation}

\noindent where the backscattered angle $\theta_\textrm{B}$, which corresponds to the case $\theta_\textrm{r} = -\theta_\textrm{in}$,  is defined by

\begin{equation}
    \tan\theta_\textrm{B} = \frac{(1 - \lambda)\tan\alpha}{1+ \lambda\tan^2\alpha},
\end{equation}

\noindent being $\lambda = (m_y/m_x)(1 + m_xv^2/\Delta)$, a term that depends on the effective parameters of the phosphorene Hamiltonian, as seen in Eq. \eqref{Hphosp}. This negative reflection appears for certain values of the tilt angle $\alpha$ in the incidence range $0 < \abs{\th_\textrm{in}} < \abs{\th_\textrm{M}}$, where $\th_\textrm{M}$ is the incidence angle for normal reflection. For $\abs{\th_i}>\abs{\th_M}$, anomalous reflection is observed, because $\th_\textrm{in}$ and $\th_\textrm{r}$ have the same sign but different values. Moreover, backscattering $\th_\textrm{r}= -\th_\textrm{in}$ emerges for certain incidence angles. The breaking of the reflection law in $pn$ junctions of phosphorene is independent of $V$, and therefore, it is valid for when $V \neq 2E$. 

Negative reflection can be viewed from the kinematical construction in Fig. \ref{fig:KC},  the conserved linear momentum parallel to the interface, which is represented by the dashed line, intersects the points for the conserved energy contours. The component of
group velocity parallel to the interface becomes different for incident, reflected, and transmitted waves. Only if $\ap=0\deg, \pm 90\deg$, these states have the conserved group
velocity by causing the conventional reflection $\theta_\textrm{in} = \theta_\textrm{r}$. It is worth noticing that atypical reflection is explained by the fact that the tilting of the junction breaks the mirror symmetry with respect to the $k_x$ and $k_y$ axes in momentum space.

We consider a bipolar junction based on phosphorene, where the transmission as a function of incidence angle $\theta_\textrm{in}$ is shown in Fig. \ref{fig:Transm_barr_phosp}(b) and it is calculated using Eq. \eqref{TCoeff}. By  changing the tilting of the junction $\ap$,  the non-resonant transmission is always less than one to difference of Klein tunneling, which is observed in graphene \cite{Katsnelson2006,Young2009}. Remarkably, zero transmission is observed for the angle $\ap=\pm 90\deg $ in the whole incidence angles. This omni-directional total reflection, called anti-super Klein tunneling, ocurrs when the effective states $\vec{w}(\vec{k}_\textrm{in})$ and $\vec{w}(\vec{k}_\textrm{t})$ are perpendicular, as indicated in Eq. \eqref{gamma}. In contrast to total internal reflection, it is not caused by the evanescent modes due to the band gap, since there exist states allowed energetically \cite{BetancurOcampo2019,ParedesRocha2021,LizarragaBrito2025}. The perfect transmission observed in Fig. \ref{fig:Transm_barr_phosp}(b) is due to Fabry-Pérot resonances with a tilted angle $\alpha \neq 90\deg$. The anisotropy of phosphorene and other materials, such as strained graphene \cite{BetancurOcampo2018} and black phosphorus \cite{LizarragaBrito2025}, is responsible of anisotropic transport \cite{Biswas2021,Chen2024a,Chen2023,Guzman2023}, negative reflection \cite{BetancurOcampo2019}, perfect waveguides \cite{BetancurOcampo2020}, electronic cloaking \cite{MolinaValdovinos2022,LamasMartinez2024,LamasMartinez2023}, asymmetric Veselago lenses \cite{BetancurOcampo2018}, super-Klein tunneling modulated by linearly polarized light \cite{Chen2024}, and Goos-Hänchen effect \cite{Majari2023}.

\subsection{Super-Klein tunneling of pseudo-spin one systems}
\begin{figure*}
    \centering
    \begin{tabular}{cccc}
     (a) \qquad \qquad \qquad \qquad \qquad \qquad \qquad \qquad & (b) \qquad \qquad \qquad \qquad \qquad \qquad \qquad \qquad  & & (c) \qquad \qquad \qquad \qquad \qquad \qquad \qquad \qquad\\
     \includegraphics[width=0.3\linewidth]{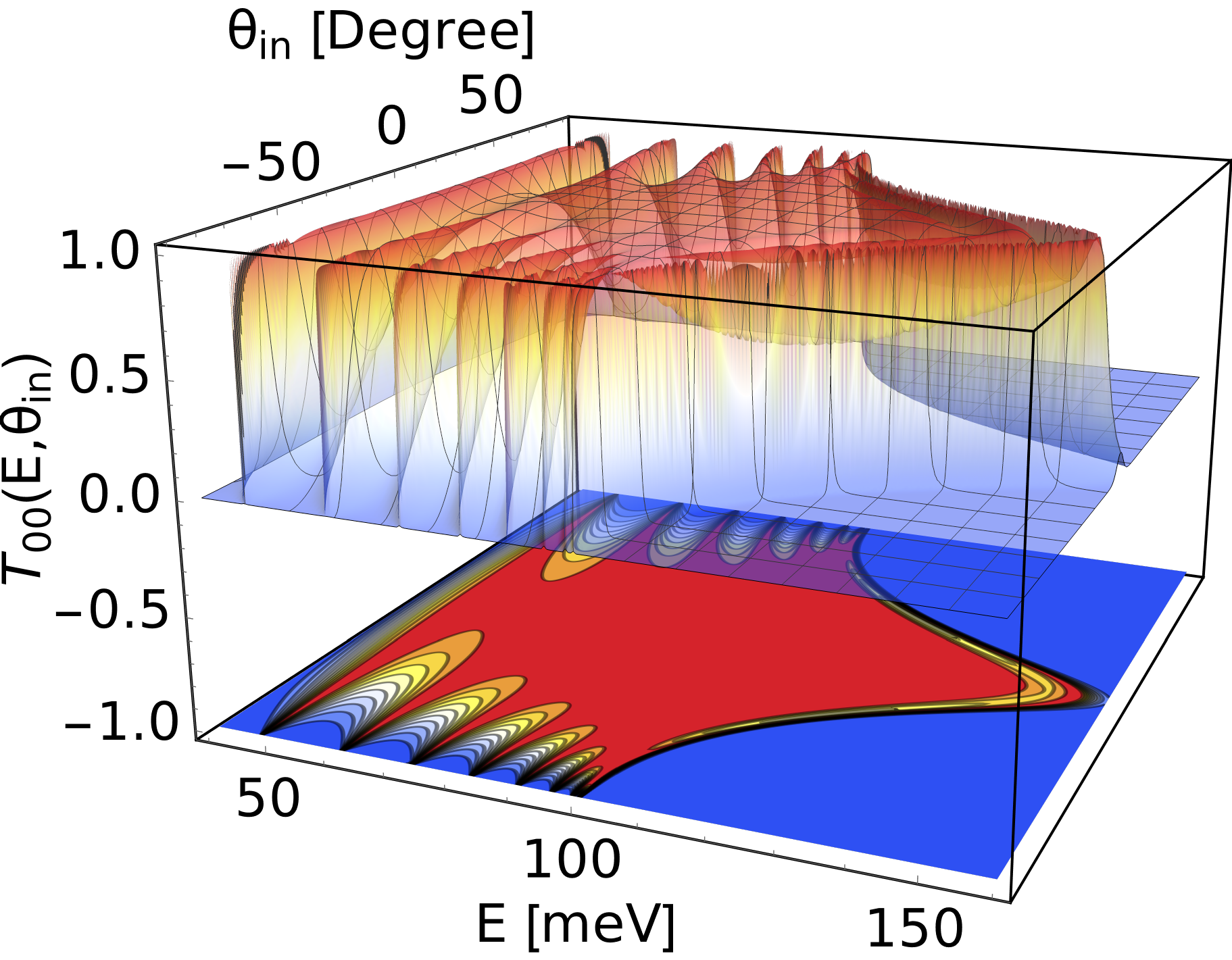}&
    \includegraphics[width=0.3\linewidth]{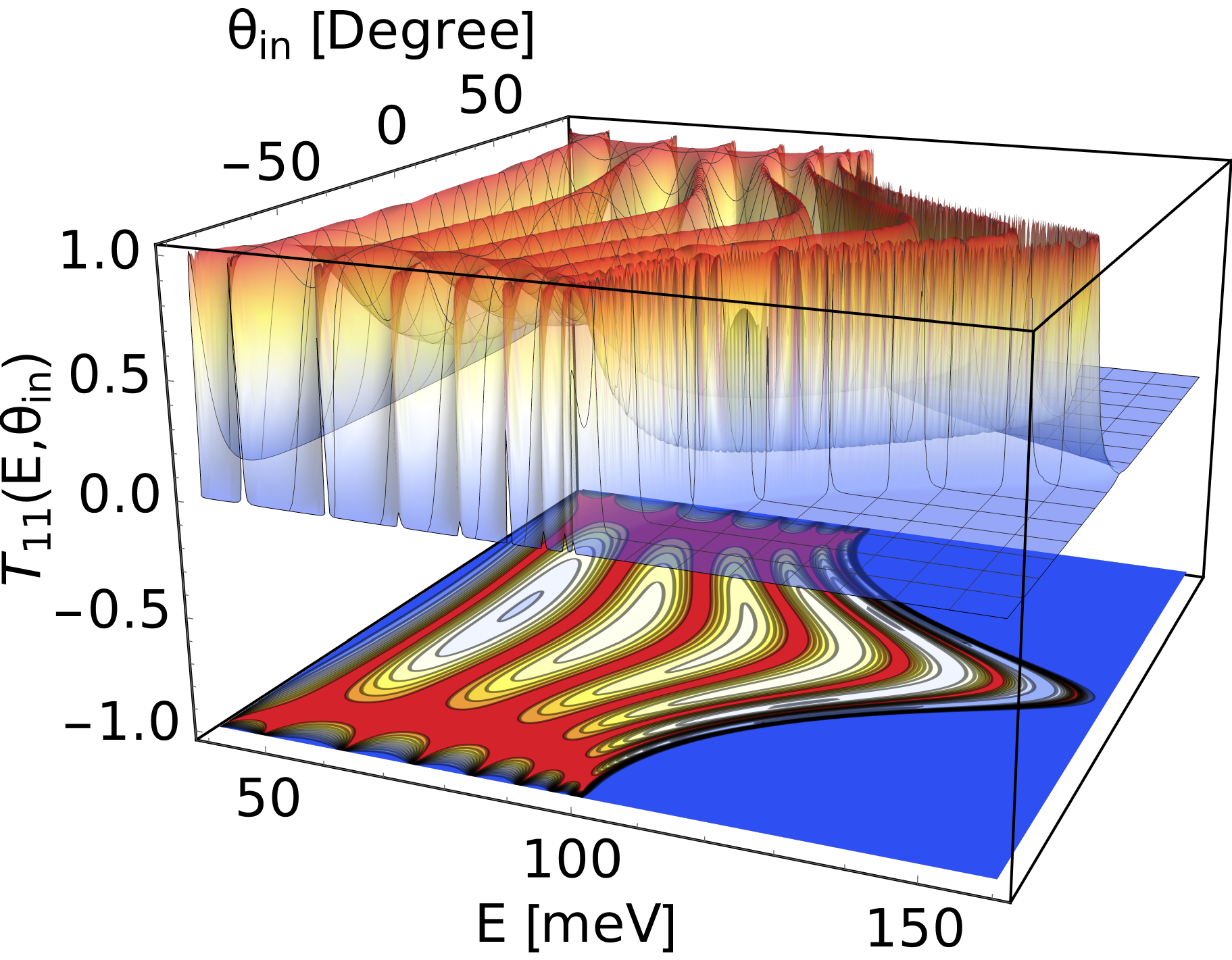}&
    \includegraphics[width=0.0275\linewidth]{Barra.pdf}&
    \includegraphics[width=0.35\linewidth]{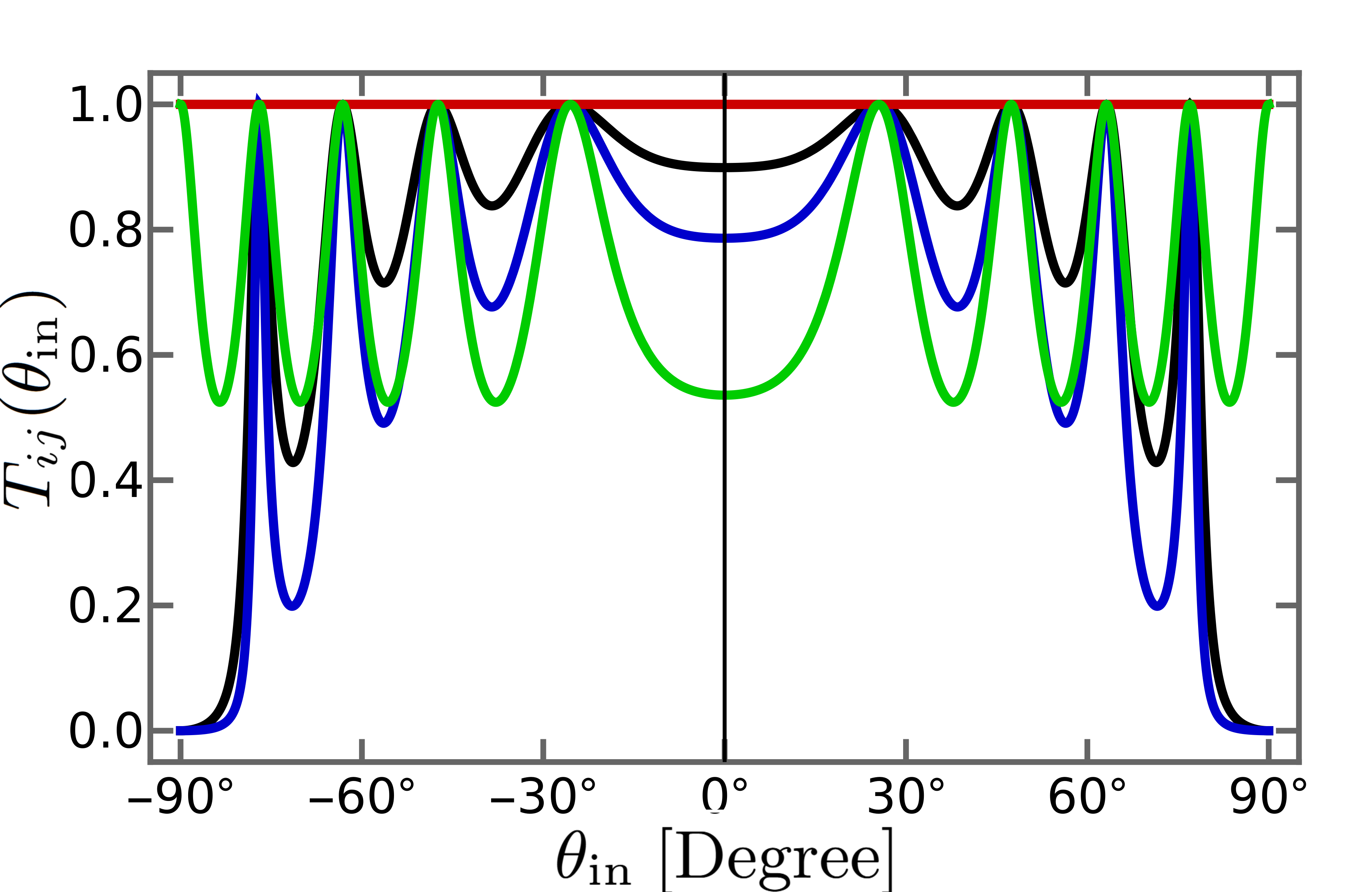}
    \end{tabular}
     \caption{(a) and (b) Interband transmission of massive pseudo-spin one particles as a function of incidence angle $\theta_\textrm{in}$ and energy $E$ for the configurations (0,0) and (1,1), where the flat band is located at the middle of the band gap and touching the conduction band, respectively. (c) Interband transmission for the focusing condition $V = 2E = 200$ meV as a function of incidence angle. Only the configurations $(-1,1), (1,-1)$, and $(0,0)$ (red line) present super-Klein tunneling of massive pseudo-spin one particles. Another curves correspond to the configuration $(1,0)$ in black, $(1,1)$ and $(-1,-1)$ in green, and $(-1,0)$ in blue.}
    \label{fig:Transm_pseudsp_one}
\end{figure*}
For crystals with three atoms in the unit cell, it is possible, in some particular cases, to relate the Bloch Hamiltonian in terms of spin-one matrices. Two outstanding examples are the dice and Lieb lattices \cite{BetancurOcampo2017a,Mandhour2026,Majari2021a,Iurov2021a,Wang2022,Zhu2023,Wu2024,Islam2017,Dey2019,Ye2020,Cunha2021,Korol2021,Korol2021,Weekes2021,Cheng2021,Wu2021,Filusch2021,Li2022c,Singh2022,Saleem2025,Parui2025,Lara2025,Fadil2025,Liu2025,Wang2025}. In such lattices, super-Klein tunneling can emerge for both massless and massive particles in the focusing condition $V = 2E$ of Veselago lenses.  In the continuum approximation, the Bloch Hamiltonian of the dice or Lieb lattice can be approximated to a spin-one Hamiltonian around a high symmetry point as

\begin{equation}
    H_1 = \hbar v_F\vec{S}\cdot\vec{k} + \Delta U,
\end{equation}

\noindent where $\vec{S} = (S_x,S_y)$ are the spin-one matrices and $U = S_z$ or $M$ as a matrix representing the mass term \cite{BetancurOcampo2017}, which is responsible for the gap opening. The dispersion relation of the electron under the Hamiltonian $H_1$ is $E_\pm = \pm\sqrt{\hbar^2v^2_Fk^2+\Delta^2}$ accompanied by a flat band $E_0 = 0$ or $\pm\Delta$ according to the sign of $U$. Pseudo-spin one systems have special matching conditions for the wave functions. In the low-energy regime, it is only necessary to consider the $n = 0$ term in the expansion \eqref{Qs}. Although the spinor in the Schrödinger equation for pseudo-spin one systems possesses three components $\Psi(\vec{r}) = (u_1(\vec{k}),u_2(\vec{k}),u_3(\vec{k}))\textrm{e}^{i\vec{k}\cdot\vec{r}}$, the matching conditions in Eq. \eqref{AryAt} involve an effective spinor of two components as

\begin{equation}\label{Pspinor}
    \vec{w}(\vec{k}_\textrm{in/r/t}) = (u_1(\vec{k}_\textrm{in/r/t})+u_3(\vec{k}_\textrm{in/r/t}),u_2(\vec{k}_\textrm{in/r/t})).
\end{equation}

For $npn$ bipolar junctions, the configuration $(i,j)$ with $i,j = -1,0,1$ denotes the position of the flat band in the regions I and II. Eq. \eqref{TCoeff} gives the transmission, where the wave vectors $\vec{k}_\textrm{in/r/t}$ in the interface are related to the scattering angles $\theta_\textrm{in/r/t}$

\begin{equation}
    \vec{k}_\textrm{in/r/t} = \frac{\sqrt{[E-V(x)]^2-\Delta^2}}{\hbar v_F}\left(\cos\theta_\textrm{in/r/t}, \sin\theta_\textrm{in/r/t}\right).
\end{equation}

\noindent The conservation of energy and the $y$-component of the wave vector allow us to establish the reflection and Snell's law

\begin{eqnarray}
    \theta_r & = & \theta_\textrm{in}, \nonumber\\
    \sin\theta_\textrm{t} & = & s s'\frac{\sqrt{E^2-\Delta^2}}{\sqrt{(E-V)^2-\Delta^2}}\sin\theta_\textrm{in},
\end{eqnarray}

\noindent where $s$ and $s'$ are the band indices in regions I and II, respectively. In this way, transmission is a function of incidence angle $\theta_\textrm{in}$, energy $E$, and the potential height $V$ of the bipolar junction.

In Fig. \ref{fig:Transm_pseudsp_one}, we showed this transmission for the systems $(i,j)$. Super-Klein tunneling appears when the focusing energy condition $E = V/2$ is reached for the configuration $(0,0)$, as shown in Fig. \ref{fig:Transm_pseudsp_one}(a). The unique configurations for the emergence of super-Klein tunneling are $(0,0)$, $(-1,1)$ and $(1,-1)$, corresponding to flat band in the middle band gap, and asymmetric location touching dispersive bands, as shown in Fig. \ref{fig:Transm_pseudsp_one}(c). These results have pointed out that not only massless particles present super-Klein tunneling, but also massive pseudo-spin one particles.

Further junction configurations with $(0,1)$, $(1,0)$, $(0,-1)$, and $(-1,0)$, the super-Klein tunneling is absent, as seen in Fig. \ref{fig:Transm_pseudsp_one}(b) and (c). Regardless of the flat band location inside the band gap, the limit of massless particles appears when $E >> \Delta$, and therefore, the transmission of massive particles is very similar to the massless case \cite{BetancurOcampo2017}. Klein tunneling occurs for a wide range of $\theta_\textrm{in}$, which is a distinctive feature of pseudo-spin-one particles. The effect of flat band location begins to be relevant for energies near the band gap, by considering propagation modes in the range $|E_c| < E < |V| - \Delta$ and $E > |V| + \Delta$, and incidence angle range $-\theta_c \leq \theta_\textrm{in} \leq \theta_c$, where $\theta_c = \arcsin(|E - V|/E)$ is the critical angle.

Most of the $pn$ junctions based on massive pseudo-spin 1 particles have Klein and super-Klein tunneling in contrast to pseudo-spin 1/2 particles \cite{Setare2010}. In Fig. \ref{fig:Transm_pseudsp_one}(c), we compare the transmission of massive pseudo-spin-one particles as a function of $E$ for the focusing condition $V  = 2E$. In the case of massless particles, Klein tunneling is independent of the energy by the conservation of pseudo-spin in $\theta_\textrm{in} = 0$. The transmission probability is the same for the configurations $(-1,-1)$ and $(1,1)$. Therefore, the flat band location does not affect the transmission under normal incidence in these configurations, but modifies the transmission when $\theta_\textrm{in} \neq 0$. Fabry-Pérot resonances emerge independent on the flat band location.

\subsection{Super-Klein tunneling of pseudo-spin $1/2$ systems}

The existence of super-Klein tunneling can also be observed in pseudo-spin-1/2 systems \cite{ContrerasAstorga2020}. In the model presented here, the electrostatic barrier does not possess translational invariance; instead, it is realized as a tunable chain of scatterers. The emergence of super-Klein tunneling in this framework can be attributed to its correspondence with a free-particle model, established through a supersymmetric transformation.

In quantum field theory, supersymmetric transformations, known as supercharges, convert bosons into fermions and vice versa. Supersymmetric quantum mechanics was introduced in \cite{Witten1981} as a simplified framework for realizing the supersymmetric algebra and exploring the spontaneous breaking of supersymmetry.

Within this model, the supersymmetric Hamiltonian is constructed as a diagonal operator  $${H_\textrm{SUSY}}=\begin{pmatrix}H_1&0\\0&H_0\end{pmatrix}$$ and supercharges are given as 
$$Q=\begin{pmatrix}0&L\\0&0\end{pmatrix},\quad Q^{\dagger}=\begin{pmatrix}0&0\\L^\dagger&0\end{pmatrix}.$$ They satisfy 	
\begin{equation}[H,Q]=[H,Q^{\dagger}],\quad \{Q,Q^{\dagger}\}=H_\textrm{SUSY}.\label{susyalg}\end{equation} In this framework, the supercharges do not relate bosons and fermions, they rather provide mapping between two orthogonal states of ${H_\textrm{SUSY}}$ corresponding to its degenerate eigenvalue.

Today, the framework is used mainly to generate new exact solvable models based on existing ones \cite{Cooper1995,Matveev1991}. 
It is based on the supersymmetric (SUSY) transformation $L$, which intertwines the operator $H_0$, whether it is a Schr\"odinger or Dirac operator of a known model, with a new operator $H_1$ that defines the transformed quantum system \cite{Correa2025}.

The commutator (\ref{susyalg}) implies in particular the following equality
\begin{equation}\label{genintertwining}
LH_0=H_1L,
\end{equation}  
that is called the intertwining relation for $H_0$  and $H_1$.
It reflects the fact that the eigenstates of $H_0$ can be mapped into the eigenstates of $H_1$ by $L$.   

The intertwining operator $L$ is known for a long time in the ana\-lysis of differential equations as Darboux transformation, see e.g. \cite{Matveev1991} and references therein. The intertwining relations for the stationary one-dimensional Dirac equation were discussed in \cite{Nieto2003}, for the non-stationary one in \cite{Pecheritsyn2005}. 

Let us briefly review the framework of non-stationary supersymmetric transformation for $1+1$D Dirac operators. The Hamiltonian of the initial system is given in the following form
\begin{equation}\label{td}
H_0\psi=(i\partial_t-i\sigma_2\partial_z-\mathbf{V}_0(z,t))\psi=0,
\end{equation}
where $\mathbf{V}_0(z,t)$ is a generic hermitian matrix whose entries can depend both on the spatial coordinate $z$ and on the time coordinate $t$. We assume that we know two solutions $\psi_1$ and $\psi_2$ of the Eq. (\ref{td}). We compose a $2\times 2$ matrix whose columns correspond to the solutions,   $\mathbf{u}=(\psi_1,\psi_2)$. By definition, the matrix satisfies 
\begin{equation}
H_0\mathbf{u}=0.
\end{equation}

Then we can define the operators $L$ and $H_1$ with the use of the matrix $\mathbf{u}$, 
\begin{eqnarray}\label{L}
L&=&\partial_z-\mathbf{u}_z\mathbf{u}^{-1},\quad L^\dagger=-\partial_z-(\mathbf{u}^\dagger)^{-1}\mathbf{u}^\dagger_z,\nonumber\\ \label{acheuno}
H_1&=&H_0-i[\sigma_2,\mathbf{u}_z\mathbf{u}^{-1}]\nonumber\\&=&i\partial_t-i\sigma_2\partial_z-\mathbf{V}_0(z,t)-\mathcal{V}^{(1)}_1(z,t)\sigma_1-\nonumber\\
&&\mathcal{V}^{(1)}_3(z,t)\sigma_3.
\end{eqnarray}
The operators satisfy the intertwining relation (\ref{genintertwining}). The operator $H_1$ has the new term $i[\sigma_2,\mathbf{u}_z\mathbf{u}^{-1}]$ that can be non-hermitian or singular. Either of these characteristics would compromise the possibility to consider it as the Hamiltonian of the new physical system. The properties of the new potential in $H_1$ depend on the choice of $\mathbf{u}$. Its proper selection leading to hermitian and singularity-free operator $H_1$ can be demanding. 

Let us present the case where judicious choice of the matrix $\mathbf{u}$ provides us with a physically acceptable Hamiltonian $H_1$.
We fix $H_0$ as the free-particle Hamiltonian,
\begin{equation}\label{free}
H_0\psi=(i\partial_t-i\sigma_2\partial_z-m\sigma_3)\psi=0.
\end{equation}
Its straightforward to find its explicit solutions,

\begin{align}\label{basis}
\psi_\pm(z,t)= e^{\pm k z}\bigg(\mp \frac{m \cosh ( \omega t)+i \omega  \sinh ( \omega t)}{k}, &\nonumber\\
\cosh ( \omega t)\bigg)^T \, ,&
\end{align}
where $k=\sqrt{\omega^2+m^2}$ and $\omega$ are real constants.
We fix the following solution $H\psi_1=0$,
\begin{eqnarray}\label{modelApsi1}
\psi_1(z,t)
&=&\frac{1}{2}\left(\psi_+(z,t)+\psi_-(z,t) \right),\nonumber\\
\end{eqnarray}
and compose the matrix $\mathbf{u}$ in this specific manner,
\begin{equation}
\mathbf{u}=(\psi_1,\sigma_1\psi_1).
\end{equation}
Then the intertwining operator $L$ has the following explit form
\begin{eqnarray}
L&=&\partial_z-\frac{1}{2D_1}\left(\frac{\omega^2\sinh(2kz)}{k}\sigma_0-2m\cosh^2 (\omega t)\, \sigma_1\right.\nonumber\\&&\left.+  \omega \sinh  (2\omega t) \,\sigma_2\right).
\end{eqnarray} 
Here, we abbreviated $D_1(z,t)=(m^2+k^2\cosh(2\omega t)+\omega^2\cosh (2kz))/2 k^2$. We can construct the new Dirac operator $H_1$ from the relation (\ref{acheuno}),
\begin{eqnarray}\label{h1}
H_1&=&i\partial_t-i\sigma_2\partial_z-m\sigma_3\nonumber\\
&&+\frac{4mk^2\cosh^2 (\omega t)}{m^2+k^2\cosh (2\omega t)+\omega^2\cosh (2k z)}\sigma_3.
\end{eqnarray}
It is intertwined with $H_0$ by $L$ via the Eq. (\ref{genintertwining}). The potential term represents a fluctuation of the constant mass,  exponentially localized both in time and space. In contrast with the original, free-particle system (\ref{free}), the new system possesses two solutions which are localized in space-time, see \cite{ContrerasAstorga2020}.
	
The derived system described by $H_1$  is $1+1$ dimensional. It can be converted into a stationary planar system by the means of Wick rotation. It is facilitated by the following change of the coordinates
\begin{equation}\label{x->kappa}
z=ix,\quad \partial_z=-i\partial_x,\quad t=y.
\end{equation}

Rewriting the equation (\ref{td}) in the new coordinates, multiplying it from the left by $\sigma_3$ and applying an additional gauge transformation $\mathbf{U} = e^{i\frac{\pi}{4}\sigma_1}$, we obtain a stationary two-dimensional equation at zero energy,
\begin{eqnarray}\label{2d2a}
\widetilde{H}_0(x,y)\widetilde{\psi}(x,y)&=&-\mathbf{U}\sigma_3H_0(ix,y)\mathbf{U}^{-1}\widetilde{\psi}(x,y)\nonumber\\&=&(-i\sigma_1\partial_x-i\sigma_2\partial_y+\nonumber\\
&&\qquad\widetilde{\mathbf{V}}_0(x,y))\mathbf{U}\psi(ix,y)\nonumber\\
&=&0.
\end{eqnarray}
The potential term is 
$
\widetilde{\mathbf{V}}_0(x,y)=\mathbf{U}\sigma_3\mathbf{V}_0(ix,y)\mathbf{U}^{-1}.
$
The solutions $\psi(z,t)$ of (\ref{td}) transform into the solutions $\widetilde{\psi}(x,y)$ of (\ref{2d2a}),
\begin{equation}\label{tildepsi}
\widetilde{\psi}(x,y)=\mathbf{U}\psi(ix,y).
\end{equation}

Now, suppose that two $1+1$-dimensional operators are related by the intertwining relation (\ref{genintertwining}). The Wick rotation, accompanied by the additional transformations, converts the intertwining relation into the following form

\begin{equation}\label{asymmetric}
\tilde{L}_1(x,y)\tilde{H}_0(x,y)=\tilde{H}_1(x,y)\tilde{L}_2,
\end{equation}
where the operators
\begin{eqnarray}
\tilde{L}_1(x,y)&=&\mathbf{U}\sigma_3L(ix,y)\sigma_3\mathbf{U}^{-1},\nonumber\\ \tilde{L}_2(x,y)&=&\mathbf{U}L(ix,y)\mathbf{U}^{-1}.\label{LL}
\end{eqnarray}
The equality (\ref{asymmetric}) differs from the standard intertwining relation by using two different intertwining operators (\ref{LL}). It allows us to map the solutions of $\tilde{H}_0(x,y)\psi_0(x,y)=0$ into the solutions of $\tilde{H}_1(x,y)\psi_1(x,y)=0$,
\begin{equation}
\tilde{H}_0(x,y)\psi_0(x,y)=0\Longrightarrow \tilde{H}_1(x,y)\tilde{L}_2(x,y)\psi_0(x,y)=0.
\end{equation}

The Wick rotation transforms the free-particle Dirac operator (\ref{free}) into the two-dimensional Hamiltonian of a massless particle,
\begin{equation}\label{WRH0}
\widetilde{H}_0=-i\sigma_1\partial_x-i\sigma_2\partial_y+m\sigma_0.
\end{equation}
Following (\ref{2d2a}), the mass term of (\ref{h1}) is converted into the electrostatic potential.  We get this stationary equation,
\begin{equation}\label{V1}
\widetilde{H}_1\widetilde\psi(x,y)=\left(-i\sigma_2\partial_y-i\sigma_1\partial_x+\widetilde{V}(x,y)\sigma_0\right)\widetilde{\psi}(x,y)=0,
\end{equation}
where the potential term is
\begin{equation}
\widetilde{V}(x,y)=m-\frac{4mk^2\cosh^2 (\omega y)}{m^2+k^2\cosh (2\omega y)+\omega^2\cos (2k x)}.
\end{equation}
For large values of $y$, the function rapidly approaches a constant,  $
\lim_{y \rightarrow \pm\infty} \widetilde{V}(x,y) = -m.
$
To account for this asymptotic behavior, we subtract the limiting value from the electrostatic potential and interpret it as a contribution to the fermion's energy. This leads us to define a modified potential $\mathcal{V}_A(x,y)=\widetilde{V}(x,y)+m,$ that reads explicitly
\begin{eqnarray}\label{Vm}
\mathcal{V}_A(x,y)\nonumber\\
&=&- \frac{4m\omega^2\sin^2(k x)}{m^2+k^2\cosh(2\omega y)+\omega^2\cos(2kx)}.\nonumber\\
\end{eqnarray}
It asymptotically approaches zero as $|y|$ becomes large. Its sign is determined by the sign of $m$, and it exhibits periodicity in $x$. Moreover, it is an even function under spatial reflection.

\begin{figure}[h!]
	\begin{center}
		\includegraphics[scale=.4]{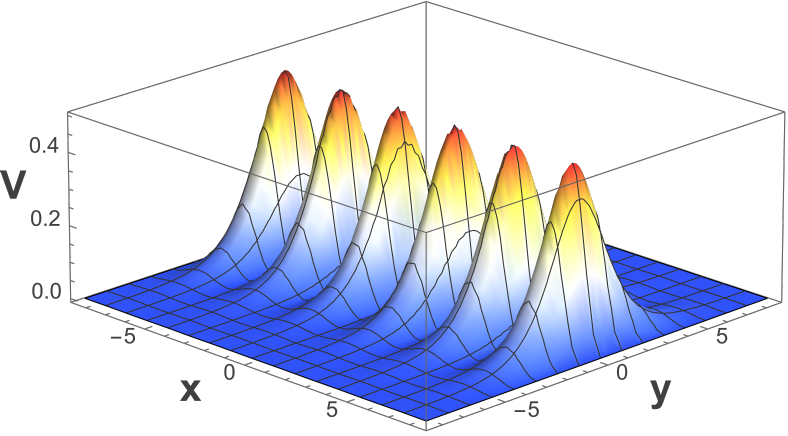}\\
        $a)$\\
		\includegraphics[scale=.4]{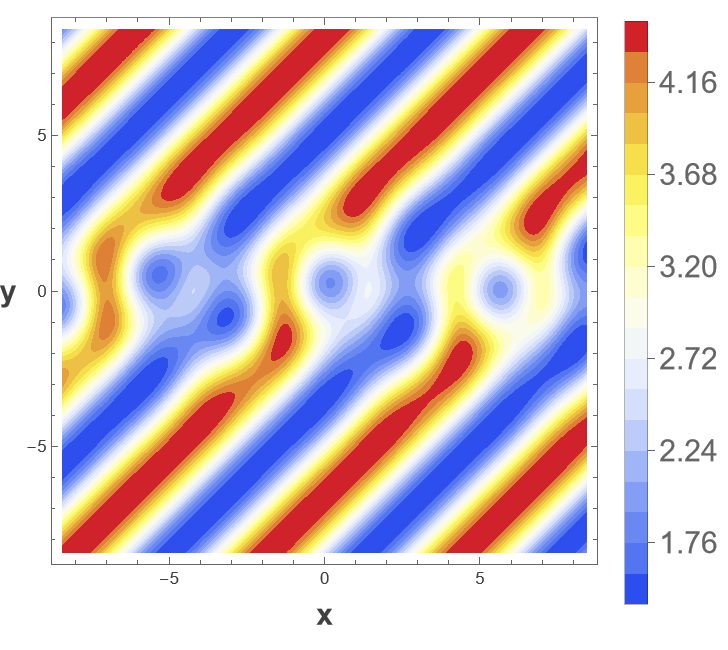}\\
        $b)$\\
		$ $ \hspace{38mm}  $ $\\
		\includegraphics[scale=.4]{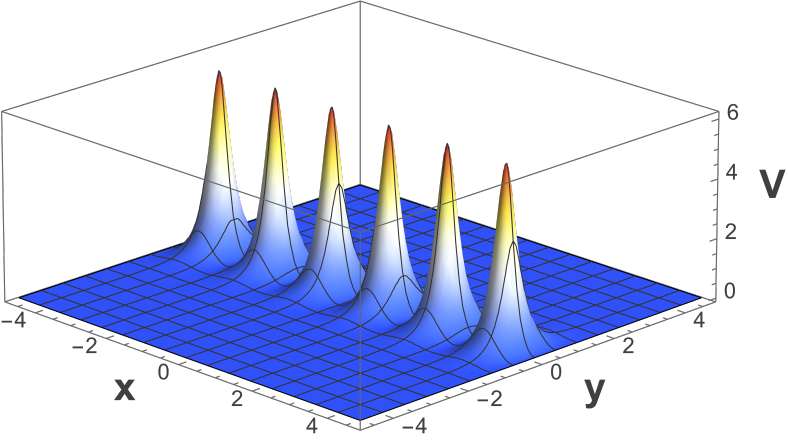}\\$c)$\\
		\includegraphics[scale=.4]{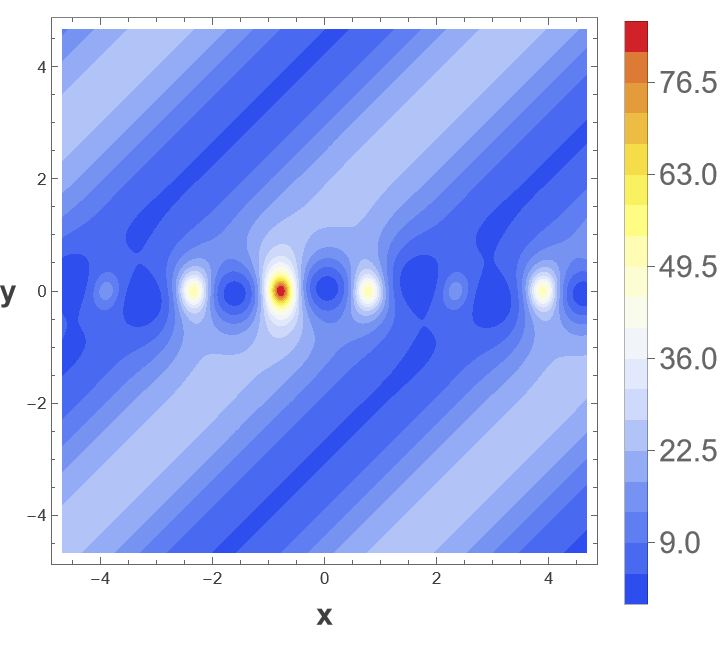}\\$d)$
	\end{center}
	\caption{(color online) The potential term $\mathcal{V}_A(x,y)$ (a) and c) and the density of probability b) and d) of a linear combination of asymptotically plane-wave solutions (\ref{Superposition}). There is $\omega=1.75$ in a), b) and $\omega=0.5$ in c) and d). We fixed $m=-1$ and $\phi_1=0$, $\phi_2=\frac{\pi}{2}$ in all plots. 
	}\label{modelAb}
\end{figure}

It creates a narrow electrostatic potential resembling a grating or a comb; see Fig.~\ref{modelAb}. Substituting (\ref{Vm}) into (\ref{V1}), we get the stationary equation 
for energy $E=m$, 
\begin{equation}\label{V1b}
\left(-i\sigma_1\partial_x-i\sigma_2\partial_y+\mathcal{V}_A(x,y)\sigma_0\right)\widetilde\psi(x,y)=m\widetilde{\psi}(x,y).
\end{equation}

We fix the wave vector $\vec{k}$ for an incoming plane wave as follows
\begin{equation}
k_x=m\sin\phi,\quad k_y=m\cos\phi,\quad\phi=(-\pi/2,\pi/2).
\end{equation}
Then the free-particle solutions of $\widetilde{H}_0\widetilde{\psi}_\phi(x,y)=0$ can be written as
\begin{eqnarray}\label{scatt0}
&&\widetilde{\psi}_\phi(x,y)=e^{i m(\sin\phi\, x+\cos\phi\,y)}\left(\begin{array}{c}1\\-i e^{-i\phi}\end{array} \right).
\end{eqnarray}
The parameter $\phi$ corresponds to the angle of incidence of the plane wave. 
The spinor $\widetilde{\psi}_\phi$ can be transformed into the scattering solution of Eq.(\ref{V1b}),
\begin{equation}
\left(-i\sigma_1\partial_x-i\sigma_2\partial_y+\mathcal{V}_A(x,y)\right)\widetilde{L}\widetilde\psi_\phi(x,y)=m\widetilde L\widetilde{\psi}_\phi(x,y).
\end{equation}

For large $|y|$, the operator $\widetilde{L}$ has the following asymptotic form
\begin{equation} \label{asymtotic}
\widetilde{L}\sim\begin{cases}-i\partial_{x}+m\sigma_1-\omega\sigma_3,\quad y\rightarrow-\infty,\\-i\partial_{x}+m\sigma_1+\omega\sigma_3,\quad y\rightarrow \infty.\end{cases}
\end{equation} 
Hence, the action of $\widetilde{L}$ on $\widetilde{\psi}_\phi$ is 
\begin{equation}\label{superklein}
\lim_{y\rightarrow\pm\infty}\widetilde{L}\widetilde{\psi}_\phi=e^{i m(\sin\phi\, x+\cos\phi\,y)}(\pm \omega-im\cos\phi)\left(\begin{array}{c}1\\i e^{-i\phi}\end{array}\right).
\end{equation}
The relation (\ref{superklein}) reveals that there is no reflection of the Dirac fermions on the potential, independent of the angle of incidence, provided that the particle has energy $E=m$. Therefore, the system has super-Klein tunneling. The scattering state $\widetilde{L}\widetilde{\psi}_\phi$ accumulates the phase shift when passing through the barrier,
\begin{equation}\label{phaseshift1}
\lim_{y\rightarrow+\infty}\widetilde{L}\widetilde{\psi}_\phi=\frac{ \omega-im\cos\phi}{ -\omega-im\cos\phi}\left(\lim_{y\rightarrow-\infty}\widetilde{L}\widetilde{\psi}_0\right).
\end{equation}
The phase shift depends both on the potential (determined by the values of $m$ and $\omega$) and on the incidence angle $\phi$. It is symmetric with respect to $\phi\rightarrow -\phi$ which is expected as the potential term is symmetric with respect to $x\rightarrow-x$.  The phase shift changes  the interference pattern of a linear combination of the plane waves. Taking the sum of two plane waves with incidence angles $\phi_1$ and $\phi_2$, 

\begin{equation}\label{Superposition}
F_A(x,y,\phi_1,\phi_2)=\widetilde{L}\widetilde{\psi}_0(x,y,\phi_1)+\widetilde{L}\widetilde{\psi}_0(x,y,\phi_2),
\end{equation} 
the interference pattern gets slightly shifted along the $x$ axis when passing through the potential barrier, see Fig.~\ref{modelAb} for an illustration. 

\begin{figure}[h!]
	\begin{center}
		\includegraphics[scale=.71,trim={6.3cm 10cm 1.5cm 8cm}, clip]{figure14a.pdf}\\
        $a)$\\
		\includegraphics[scale=.91,trim={2cm -.5cm 2.3cm -.5cm}]{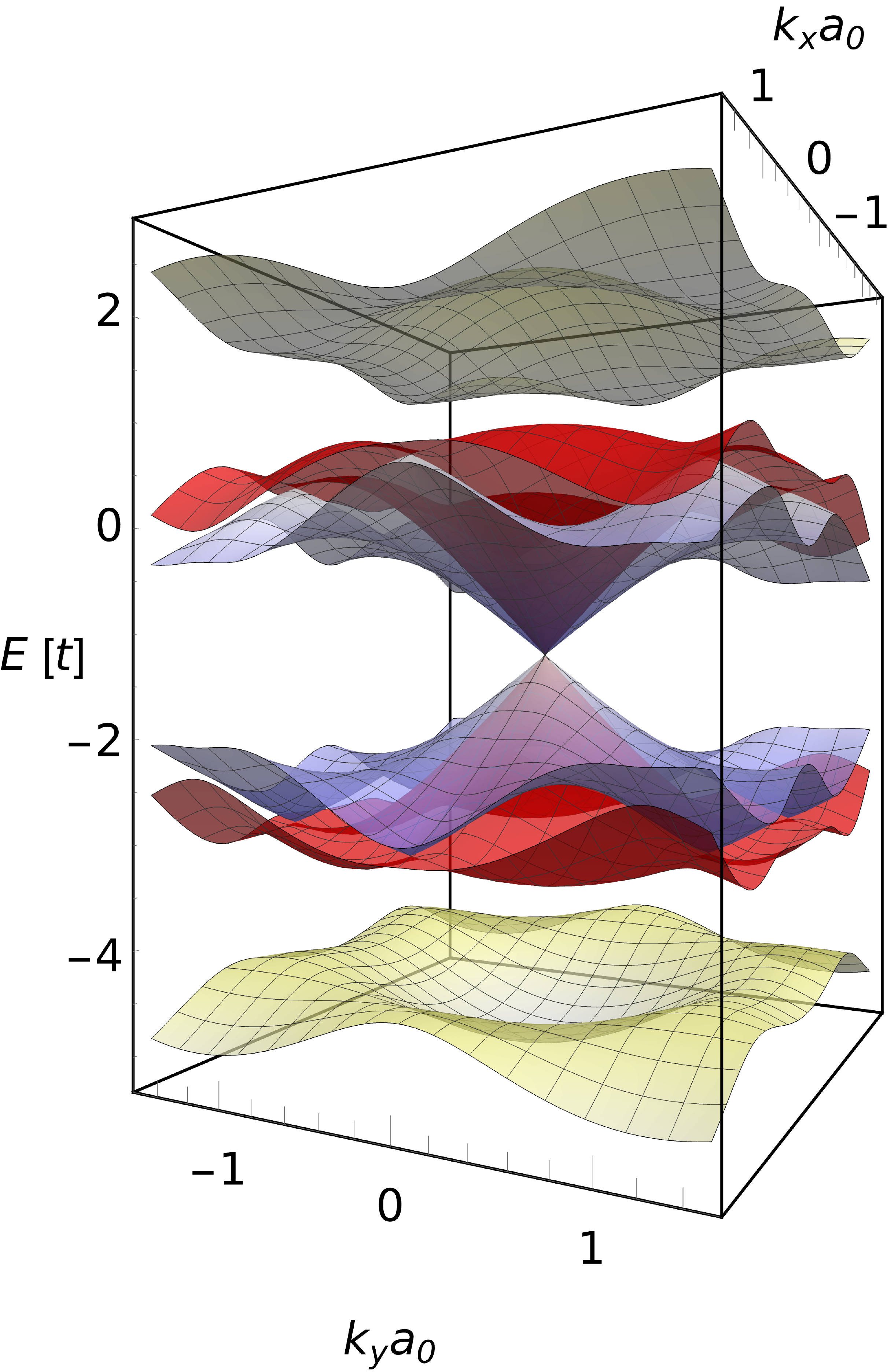}\\
        $b)$\\
		$ $ \hspace{38mm}  $ $\\
		\includegraphics[scale=.91,trim={2cm -.3cm 2.3cm 0cm}]{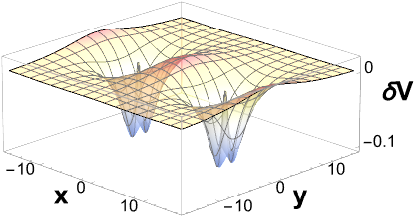}\\$c)$
	\end{center}
	\caption{(color online) a) Scheme of the STM comb. The electric field within the gray rectangle below the two central tips is illustrated in b) Potential $V_{STM}$ in (\ref{STMpotential}) created by four STM tips placed at $(x_n,0,z_1)$ with $n\in\{-2,-1,0,1\}$.  c)  The difference $\delta V=(\mathcal{V}_{A}-V_{STM})/\mbox{max}(\mathcal{V}_A)$. In the figures, we fixed $z_0=1$, $z_1=2$, $m=0.05$, $\omega=0.18$, $q=-1.939$.
	}\label{STM}
\end{figure}

\textcolor{black}{
The absence of backscattering at the potential barrier represented by $\mathcal{V}_A$ in (\ref{Vm}) is a result of a nontrivial interference between the incoming wave and those reflected from the structured potential $\mathcal{V}_A$. This remarkable property was endowed to the system at its very construction; through a supersymmetric transformation it is linked to the free‑particle system, where any backscattering is absent. The exceptional characteristics of the system are less surprising taking into account that supersymmetric transformation (known also as Darboux transformation) plays central role in the theory of integrable systems and  solitons  \cite{Matveev1991}. It serves there to generate new integrable models from the known ones while preserving their sol\-va\-bility. In this context, the potential $\mathcal{V}_A$ has solitonic nature and the backscattering is suppressed. }

\textcolor{black}{A natural question arises as to whether such a potential can represent a real physical scenario. In Fig.\ref{STM}a) a schematic of a possible experiment is shown, where scanning tunneling microscopy (STM) tips are arranged in a line so that they form a comb above the graphene sheet. Beneath the sheet, there is a grounded plate. The electric potential generated by the STM tips on the graphene surface can be calculated using the mirror‑image method in the following form:
\begin{align}
V_{STM}=&q\sum_{n}\left(\frac{1}{\sqrt{(x-x_n)^2+y^2+(z_0-z_1)^2}}\right.\nonumber\\&\left.-\frac{1}{\sqrt{(x-x_n)^2+y^2+(z_0+z_1)^2}}\right),\label{STMpotential}
\end{align}
where $(x,y,z_0)$ are the points on the graphene sheet and $(x_n,0,z_1)$ are coordinates of the STM tips. The points $x_n=\frac{(2n+1)\pi}{2\sqrt{m^2+\omega^2}}$ correspond to extremal points of  $\mathcal{V}_{A}$. In the Fig.\ref{STM}b) and c), there is shown the potential $V_{STM}$ and the scaled difference $\delta V=\frac{\mathcal{V}_{A}-V_{STM}}{\mbox{max}{\mathcal{V}_A}}$, respectively. It illustrates that there can be found a reasonable match between the $\mathcal{V}_A$ and $V_{STM}$ with a difference $\sim 10\%$. Although the matching is not exact, it is reasonable to expect that the realistic potential would show enhanced transmission across a wide interval of incidence angles for the energy $E=m$.}

\subsection{Valley-cooperative Klein tunneling in Kekulé graphene superlattices}

\begin{figure*}
    \centering
    \begin{tabular}{ccc}
     (a) \qquad \qquad \qquad \qquad \qquad \qquad \qquad \qquad & (b) \qquad \qquad \qquad \qquad \qquad \qquad  & (c) \qquad \qquad \qquad \qquad \qquad \qquad \qquad \qquad\\
         \includegraphics[width=0.35\linewidth]{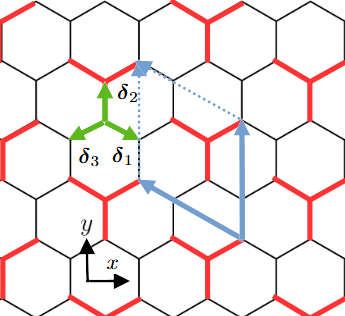}&
         \includegraphics[width=0.25\linewidth]{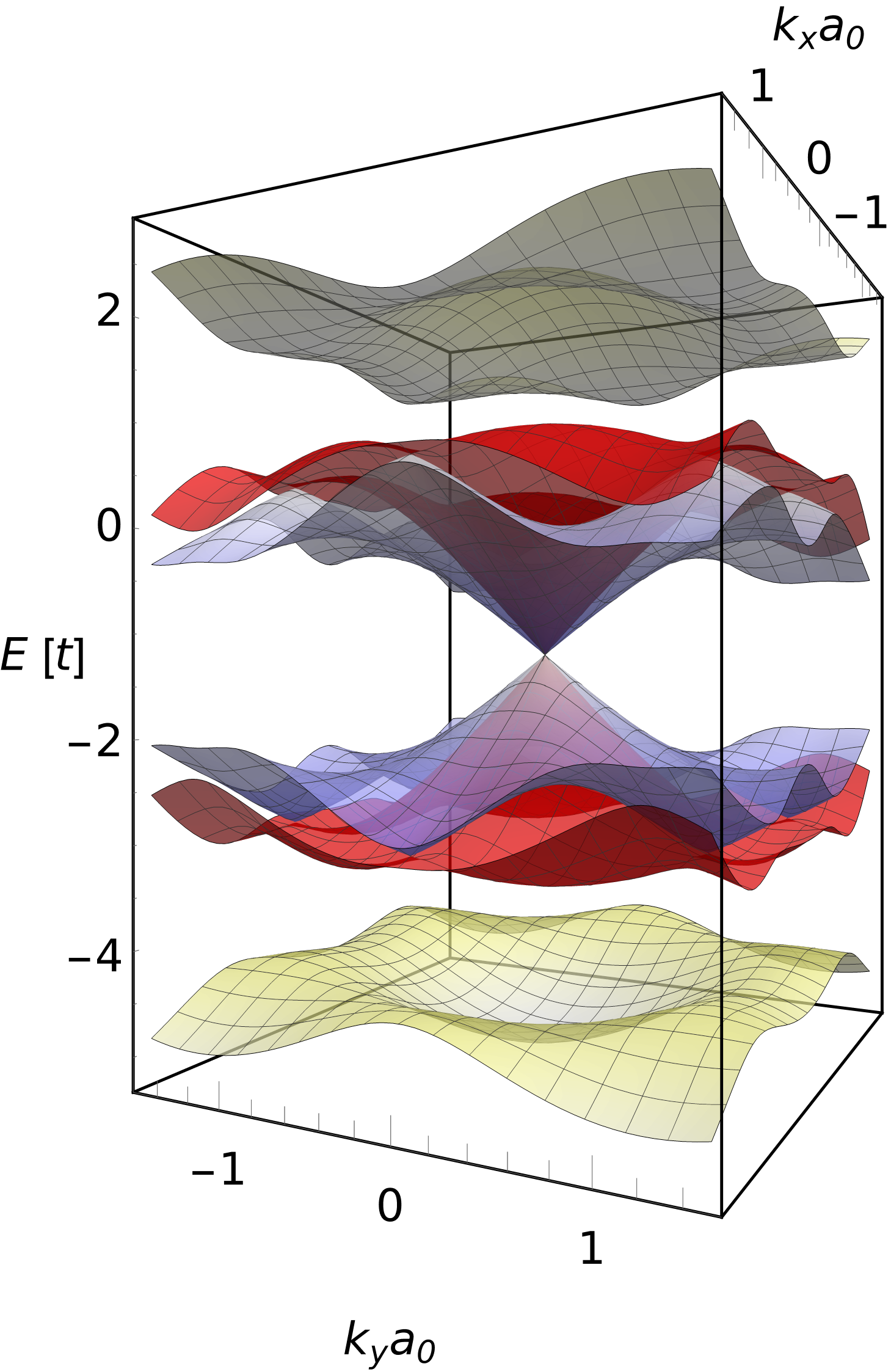}&
         \includegraphics[width=0.39\linewidth]{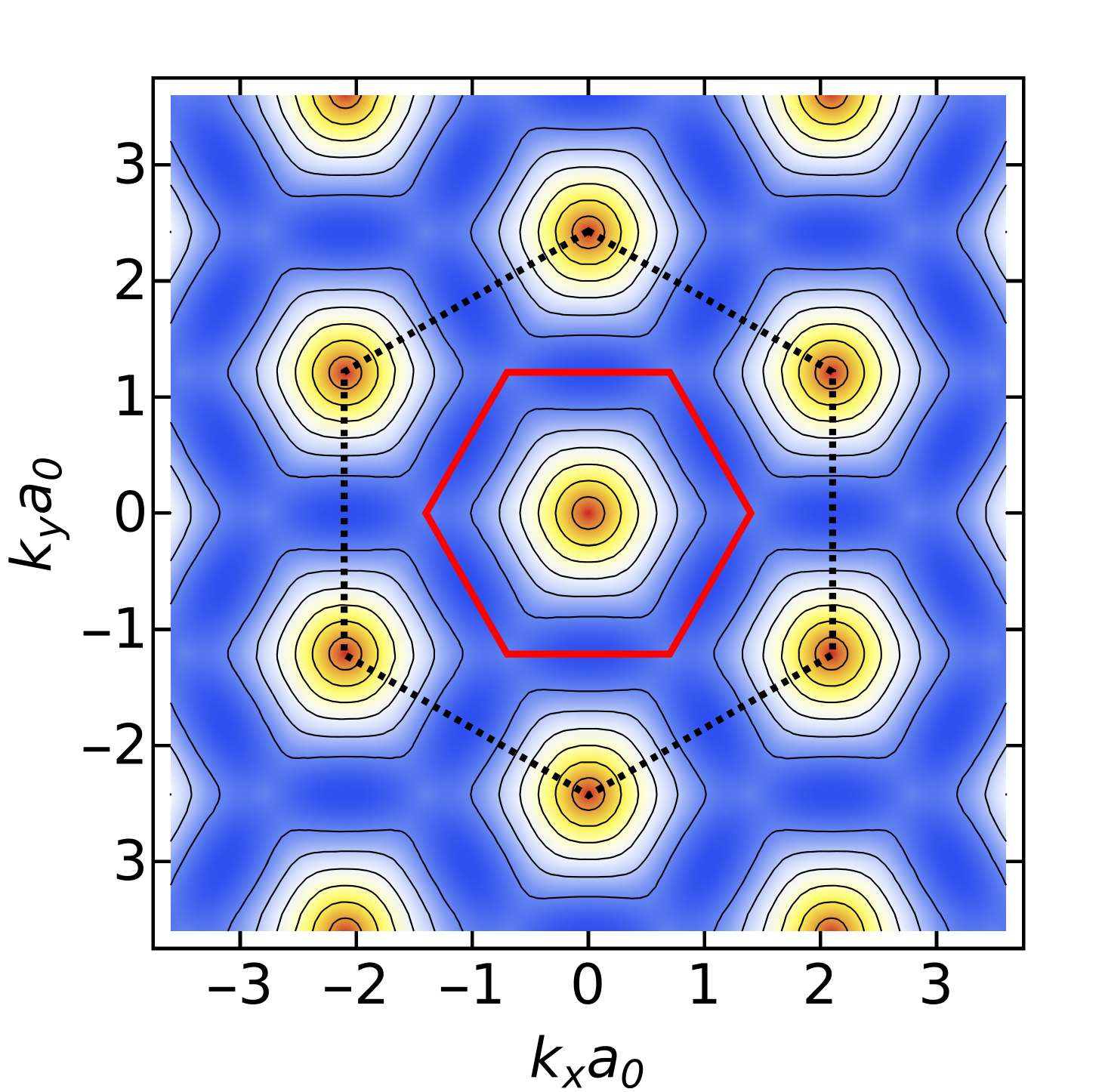}
    \end{tabular}
    \caption{(a) Lattice structure of Kekulé-Y graphene. The periodic distortion in Y is indicated by the hopping parameter $t'$ and the simple one by $t$. The unit cell contents six Carbon atoms. (b) The electronic band structure from the eigenenergies of the Bloch Hamiltonian in Eq. \eqref{HBKGS} as a function of wave vector $\vec{k}$. Two Dirac cones appear in the center of the first Brillouin zone with Fermi velocities $v_\tau \pm v_\sigma$. (c) Energy contours of the red conduction band in (b).}
    \label{fig:KekY}
\end{figure*}

Valley-cooperative Klein tunneling is a variant of the conventional Klein tunneling in graphene that arises when the two Dirac cones at the $K^+$ and $K^-$ valleys possess different Fermi velocities \cite{Garcia2022}. \comA{This effect can be understood if we consider the electron scattering of a double conduction band involving two wave vectors $\vec{k}^\pm_\textrm{in}$. Due to the existence of two Dirac cones with different Fermi velocities, the incident wave can have a superposition of two states of group velocities $v_\pm$ and amplitudes $A^\pm_\textrm{in}$. When this wave impinges normally at the interface, the conservation of pseudo-spin leads to the absence of backscattering, namely, $A^\pm_\textrm{r}=0$ for the reflected waves. However, the transmitted wave has a superposition of two states in valence or conduction bands, but the amplitudes $A^\pm_\textrm{t} \neq A^\pm_\textrm{in}$.} \comA{Unlike conventional valley-selective transport effects in graphene, including trigonal warping \cite{GarciaPomar2008} and strain-induced pseudo-magnetic fields \cite{Zhai2011, Stegmann2017}, which primarily lift valley degeneracy, where the valley degree of freedom is manipulated through valley-filtering to be used as information carrier. Remarkably, the valley-cooperative Klein tunneling involves a valley-flip process that nevertheless conserves the pseudo-spin for the perfect tunneling across the interface.}  The Kekulé-Y graphene superlattice provides an ideal platform for observing this effect \cite{Wang2018,Zhang2023,Wang2024g,Andrade2022,Ding2023,Santacruz2022,Andrade2025}.

Following the procedure of tight-binding approach to nearest neighbors outlined in Eqs. \eqref{BWF} to \eqref{TBH}, the Bloch Hamiltonian of Kekulé-Y graphene, with six Carbon atoms in the unit cell as illustrated in Fig. \ref{fig:KekY}(a), can be written as 

\begin{eqnarray}
    H^\textrm{B}_\textrm{Kek-Y}(\vec{k}) = \qquad \qquad \qquad \qquad \qquad \qquad \qquad \qquad \qquad&  & \nonumber\\
    \small{\left(\begin{matrix}
    0 & t'\textrm{e}^{i\vec{k}\cdot\vec{\delta}_2} & 0 & t\textrm{e}^{i\vec{k}\cdot\vec{\delta}_3} & 0 & t\textrm{e}^{i\vec{k}\cdot\vec{\delta}_1}\\
     t'\textrm{e}^{-i\vec{k}\cdot\vec{\delta}_2} & 0 & t'\textrm{e}^{-i\vec{k}\cdot\vec{\delta}_3} & 0 & t'\textrm{e}^{-i\vec{k}\cdot\vec{\delta}_1} & 0\\
     0 & t'\textrm{e}^{i\vec{k}\cdot\vec{\delta}_3} & 0 & t\textrm{e}^{i\vec{k}\cdot\vec{\delta}_1} & 0 & t\textrm{e}^{i\vec{k}\cdot\vec{\delta}_2}\\
     t\textrm{e}^{-i\vec{k}\cdot\vec{\delta}_3} & 0 & t\textrm{e}^{-i\vec{k}\cdot\vec{\delta}_1} & 0 & t\textrm{e}^{-i\vec{k}\cdot\vec{\delta}_2} & 0\\
     0 & t'\textrm{e}^{i\vec{k}\cdot\vec{\delta}_1} & 0 & t\textrm{e}^{-i\vec{k}\cdot\vec{\delta}_2} & 0 & t\textrm{e}^{i\vec{k}\cdot\vec{\delta}_3}\\
     t\textrm{e}^{-i\vec{k}\cdot\vec{\delta}_1} & 0 & t\textrm{e}^{-i\vec{k}\cdot\vec{\delta}_2} & 0 & t\textrm{e}^{-i\vec{k}\cdot\vec{\delta}_3} & 0
    \end{matrix}\right)}. &&\nonumber\\
    & &
    \label{HBKGS}
\end{eqnarray}
\noindent Upon diagonalization of this Hamiltonian, the band structure exhibits two Dirac cones with different Fermi velocities at the center of the first Brillouin zone, as seen in Fig. \ref{fig:KekY}(b). The original first Brillouin zone of graphene is folded, yielding a reduced hexagonal Brillouin zone, where energy bands are mapped onto the $\Gamma$ point. As illustrated in Fig. \ref{fig:KekY}(c), the energy contours for the red conduction band displayed in Fig. \ref{fig:KekY}(b) indicate that this Dirac cone is isotropic in the vicinity of the $\Gamma$ point. Expanding the Bloch Hamiltonian of Eq. \eqref{HBKGS} around this point and neglecting the upper and lower bands leads to an effective $4 \times 4$ continuum Hamiltonian, valid to linear order in momentum $\vec{p} = \hbar \vec{k}$

\begin{eqnarray}
    H_\textrm{Kek-Y} & = & \left(\begin{matrix}
    0 & v_\sigma p_- & v_\tau p_- & 0\\
     v_\sigma p_+ & 0 & 0 & v_\tau p_-\\
     v_\tau p_+ & 0 & 0 & v_\sigma p_-\\
     0 & v_\tau p_+ & v_\sigma p_+ & 0
    \end{matrix}\right),
    \label{HKGS}
\end{eqnarray}

\noindent where $v_\sigma$ is the Fermi velocity and $v_\tau$ quantifies the periodic distortion of the Kekulé graphene. We define the linear momentum components as $p_\pm = p_x \pm ip_y$. The dispersion relations have a simple linear form $E_{\nu,s} = s(v_\sigma + \nu v_\tau)p$, where $s = \pm 1$ and $\nu = \pm 1$ are the band and valley index, respectively, and $p = \sqrt{p^2_x + p^2_y}$ is the magnitude of linear momentum. The velocities $v_\tau$ and $v_\sigma$ are related with the hopping parameters $t$ and $t'$ \cite{Gamayun2018,Andrade2025}.

\begin{figure}
    \centering
    \includegraphics[width=1\linewidth]{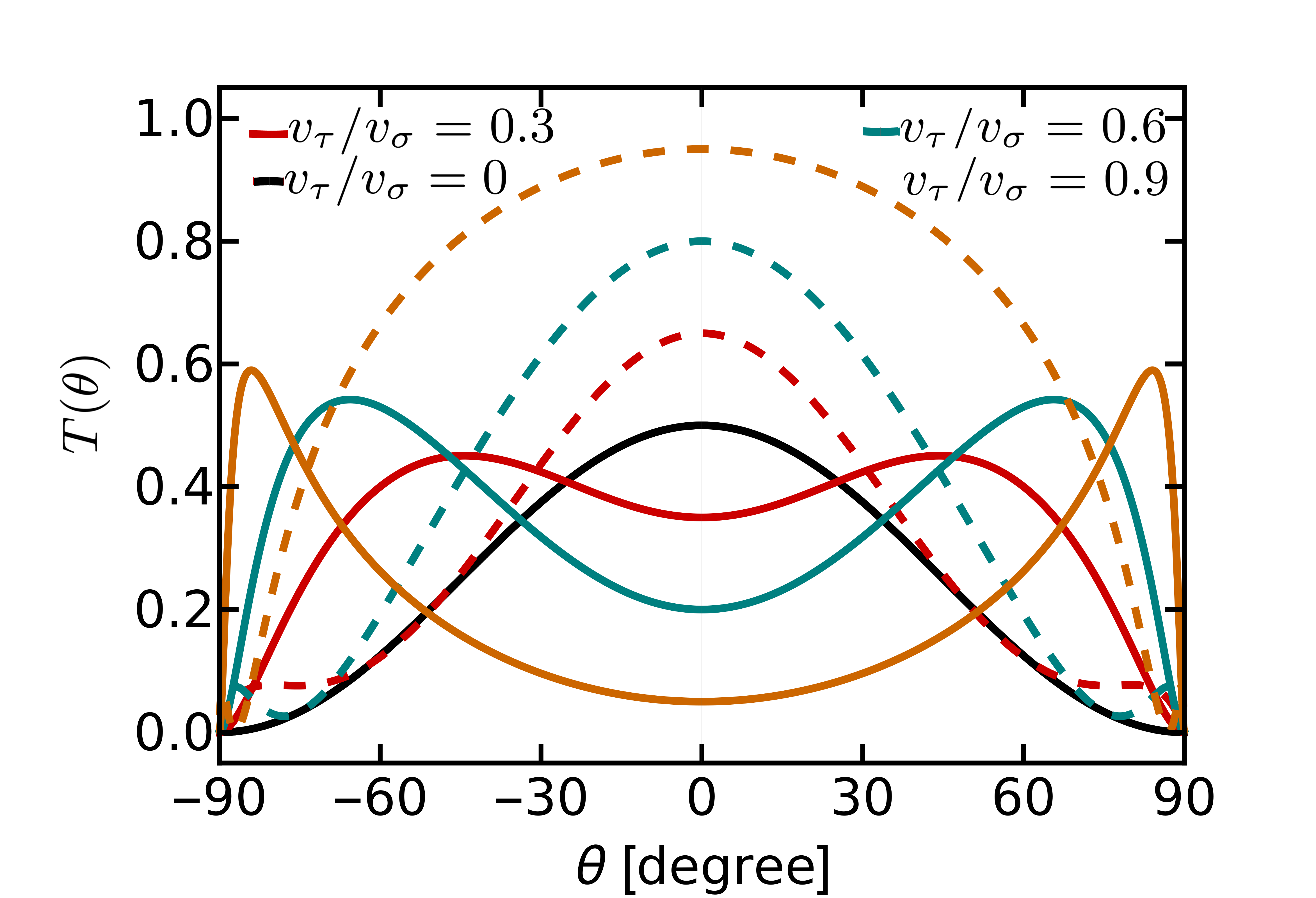}
    \caption{Electron transmission in $pn$ junctions of Kekulé-Y graphene as a function of incidence angle $\theta$. The different curves, solid and dashes with the same color, correspond to the valleys $K^+$ and $K^-$, respectively. By increasing the Kekulé distortion, which is quantified through the ratio $v_\tau/v_\sigma$, the valley-flip is more evident for normal incidence. The conservation of pseudo-spin guarantees that $T^+(0) + T^-(0) = 1$.}
    \label{fig:VCKT}
\end{figure}

Using the wave functions \eqref{PsiI} and \eqref{PsiII} for a $pn$ junction, we compute the transmission coefficient via Eq. \eqref{Transm}, and presents the results in Fig. \ref{fig:VCKT}. The incident state is prepared with equal weights $50 \%$ in the $K^+$ and $K^-$ valleys. At normal incidence, perfect tunneling appears as a consequence of pseudo-spin conservation. For oblique incidence, $\theta \neq 0\deg$, electrons have non-zero probabilities for the reflection and transmission in each valleys.  Remarkably, an unconventional transmission behavior occurs at normal incidence. Although back-scattering is completely suppress ($R^\nu=0$), the transmitted current is not equally distributed between the two valleys. Instead, for the ratio $v_\tau/v_\sigma = 0.2$, the transmitted fractions are $0.6$ and $0.4$ in the $K^-$ and $K^+$ valleys, respectively. This means that $10 \%$ of the electrons initially in the $K^-$ valley undergo a valley-flip, enabling cooperative transmission across the interface without backscattering. We refer to this counterintuitive transport mechanism as valley-cooperative Klein tunneling \cite{Garcia2022}.

The valley flip process is controlled only by the Kekul\'e distortion. Increasing the velocity $v_\tau$, the difference of $T_+ - T_-$ under normal incidence also increases, as shown in Fig.~\ref{fig:VCKT}. This effect is due to the Kekul\'e distortion and the conservation of the pseudo-spin, which is independent of the Fermi level and electrostatic potential. To analyze the valley-cooperative Klein tunneling, we calculate the transmission probabilities for normal incidence and weights $a^\pm$
\begin{equation}
T^\pm(0) = \frac{(v_\sigma \pm v_\tau)|a^\pm|^2}{\sum_\nu(v_\sigma + \nu v_\tau)|a^\nu|^2},
\label{VCKT}
\end{equation}

\noindent where the pseudo-spin angles are $\phi^-_i = \phi^+_i = 0$, $\phi^-_r = \phi^+_r = \pi$, and $\phi^-_t = \phi^+_t = (s-s')\pi/2$. For the reflection coefficients, we have $R^{\pm} = 0$. For $|a^\nu|^2 = 0.5$ the Eq.~\eqref{VCKT} can be simplified to $T^\pm(0) = \tfrac{1}{2}(1 \pm \tfrac{v_\tau}{v_\sigma})$. Clearly, we observe that $T^-(0) + T^+(0) = 1$. We recover the conventional Klein tunneling in graphene making $v_\tau = 0$, and as expected, unpolarized current under normal incidence crosses perfectly the electrostatic potential \cite{Katsnelson2006, Allain2011}. The extreme case $v_\sigma = v_\tau$ indicates that all the electrons in the $K^-$ valley perform a valley-flip to cross perfectly the junction, resulting in a fully valley polarized current. This behavior is evidenced in Fig. \ref{fig:VCKT} for different Kekulé-Y distortions $v_\tau/v_\sigma$. Moreover, $v_\sigma = v_\tau$ merges the conduction and valence bands in the $K^-$ valley to a flat band, giving rise to super-Klein tunneling, which is a typical phenomenon in the scattering of pseudo-spin-one particles \cite{Shen2010, Urban2011, Xu2016, BetancurOcampo2017, Nandy2019, Kim2020, ContrerasAstorga2020, Xu2021, Wang2021, Mandhour2020,Mandhour2026}. 

When chiral symmetry is broken, massive Dirac fermions exhibit a finite reflection probability. However, the sum of probabilities $R^++T^+$ remains greater than 0.5. If the band gap is valley-dependent, the junction works as a perfect valley filter. In the pseudo-ultra-relativistic regime (i.e. at high energies), the transmission displays the same behavior as in the gapless case \cite{Garcia2022}. For normal incidence, both reflection and transmission are constant and independent of the energy due to pseudo-spin conservation. In particular, the valley-flip mechanism underlying valley-cooperative Klein tunneling persists even when the chiral symmetry is broken in both valleys, since the valley-flip process depends solely on the Fermi velocities $v_\sigma + \nu v_\tau$.

To quantify the degree of polarization $P$ in KekGr $pn$ junctions, we define
\begin{equation}
    P = \frac{\langle T^+ \rangle - \langle T^- \rangle }{\langle T^+ \rangle + \langle T^- \rangle },
\end{equation}

\noindent where $\langle T^\nu \rangle$ is the angular averaged transmission. The polarization in junctions for massless Dirac fermions remains almost constant for intraband transmission ($E > V$) and has the approximated value of $P \approx v_\tau/v_\sigma=0.1$. This asymptotic value of $P$ remains approximately constant with the smoothness of the junction, and it is a signature of the valley-flip process that may be detected experimentally. Fully valley polarized current flow is achieved by breaking the chiral symmetry. The polarization can be changed by tuning the Fermi energy within the different gaps of the valleys, making these systems a perfect valley switch. 

\section{Conclusions and final remarks}

Klein tunneling is a robust transport phenomenon that persists across a wide variety of physical platforms and, \comA{as emphasized throughout this review,} can be understood beyond the Dirac equation framework. We have presented a unifying perspective based on a tight-binding formalism, where periodic systems are naturally described by Bloch Hamiltonians. \comA{With this approach, all the versions of Klein and anti-Klein tunneling are governed by the conservation of an effective pseudo-spin-1/2, providing a unified explanation applicable to low-dimensional crystals and metamaterials alike.} 

In this \comA{general} framework, Klein tunneling is addressed using both the tight-binding approach and supersymmetric quantum mechanics. \comA{Importantly, the full Bloch Hamiltonian contains higher-order terms in the wave vector, and therefore,} the matching conditions at interfaces depend on the underlying crystal structure. Consequently, they may differ from the conventional continuity of the wave function and its derivative prescribed by the Schrödinger equation. \comA{Hence, the emergence of Klein tunneling goes beyond the paradigm that linear dispersion relation of massless fermions, normal incidence, and conservation of pseudo-spin are the essential elements. A remarkable example is the realization of Klein tunneling in one-dimensional chains, where its description required the full Bloch Hamiltonian rather than a low-energy Dirac approximation. The introduction of a reduced pseudo-spin 1/2 allowed us to explain anomalous Klein tunneling, super Klein tunneling, anti-Klein tunneling, anti-super-Klein tunneling, and valley-cooperative Klein tunneling. We have highlighted artificial crystals and engineered materials as promising platforms for the experimental realization of yet unobserved Klein tunneling regimes, as summarized in Table \ref{tab:KT_comparison}. These systems offer unprecedented control over lattice geometry, coupling strengths, and disorder, opening new avenues for exploring generalized relativistic transport phenomena in condensed-matter and wave-based systems.}

Most importantly, we identified that the emergence of Klein and anti-Klein tunneling is governed by the conservation of an effective pseudo-spin one-half imposed by the matching conditions. This identification broadens the range of systems in which these phenomena can arise, from synthesized materials such as graphene, borophene, and phosphorene, as well as artificial lattices, among them, phononic and photonic crystals \cite{Jiang2020,Gao2020,FillionGourdeau2025}. Moreover, Klein tunneling is not restricted to electronic systems, but also it can also emerge for acoustic and electromagnetic waves, extending well beyond the Dirac paradigm \cite{Gao2020,Zhang2022c}.

Historically, Klein tunneling was originally predicted by Oskar Klein within the context of the Dirac equation \cite{Klein1929}. Remarkably, perfect tunneling is also possible for spinless particles governed by the Klein–Gordon equation \cite{Kim2019}. The introduction of the angle $\gamma(\vec{k}_\textrm{in} , \vec{k}_\textrm{t})$ in Eq. \eqref{gamma} enabled us to identify Klein tunneling independently of the specific nature of the pseudo-spin, whether originating from sublattice, intrinsic spin, or other internal degrees of freedom \cite{BetancurOcampo2024}. For instance, massive pseudo-spin-one particles obey the conservation of an effective and reduced pseudo-spin one-half given by the state $\vec{w}(\vec{k}_\textrm{in/r/t})$, as shown in Eq. \eqref{Pspinor}. Similarly, spinless particles acquire an effective pseudo-spin one-half through scattering at an interface. In general, when an incoming wave impinges on an interface, it partially reflects and partially transmits. Under the matching conditions, an effective spinor $\vec{w}(\vec{k})$ can be constructed, as defined in Eq. \eqref{AryAt}. The appearance of Klein or anti-Klein tunneling is therefore strongly dictated by the intrinsic properties of the system and is independent of the applied electrostatic potential. We discussed the diversity of Klein tunneling phenomena across systems of different dimensionalities. In Su–Schrieffer–Heeger chains, the topological phase transition leads to an interchange between Klein tunneling in the trivial phase and topological phase for interband and intraband regimes.

Anisotropic materials, such as borophene and strained graphene, gives rise to anomalous Klein tunneling \cite{BetancurOcampo2018,Chen2023,Zhou2019a}, where perfect transmission occurs at oblique incidence rather than normal incidence. The angular deviation of this anomalous tunneling depends on the Fermi velocity tensor components, as shown in Eq. \eqref{thKT_eq}. Despite its fundamental importance, Klein tunneling has not yet been experimentally realized in high-energy physics. Its first experimental observation occurred in condensed-matter systems followed by the discovery of graphene \cite{Young2009,Stander2009}. More recently, experimental realizations in acoustic and phononic crystals have revealed Klein tunneling to be an omnipresent consequence of the wave nature of matter and classical fields \cite{Jiang2020,Zhang2022,Zhu2023}. Several related phenomena, including super-Klein tunneling in massive pseudo-spin-one and one-half systems \cite{BetancurOcampo2017,ContrerasAstorga2020}, anti-super-Klein tunneling \cite{BetancurOcampo2019,ParedesRocha2021,LizarragaBrito2025}, one-dimensional chains \cite{BetancurOcampo2024}, and valley-cooperative Klein tunneling \cite{Garcia2022}, remain experimentally unexplored. Nevertheless, the rapid development of artificial lattices, such as elastic metamaterials \cite{Yu2023,BetancurOcampo2024,ManjarrezMontanez2025}, topolectrical circuits \cite{Dong2021a,Lee2018,Albooyeh2023}, ultracold atoms in optical lattices \cite{Schaefer2020,Boudjada2020}, and photonic \cite{Nakatsugawa2024,Segev2020,Tang2022,FillionGourdeau2025,Poblete2021,Palmer2020} and phononic \cite{Wu2024,Wang2020b,Sirota2021,Yang2020a,Zhu2023,Jiang2020,Laforge2020,Ma2021} crystals provides promising platforms for realizing these unconventional effects. Finally, Klein tunneling holds significant potential for technological applications, including ultrafast transistor operations \cite{Wilmart2014}, high-efficiency electron-optics devices, and advanced wave-manipulation components based on collimators \cite{Wang2022,Xu2023}, Veselago and super-diverging lenses \cite{Heinisch2013}.

\section*{Acknowledgments}
Y.B.-O. gratefully acknowledges financial support from UNAM-PAPIIT project IA 102125. \comA{VJ acknowledges the support of the project "Topological Defects Dynamics: New analytical and Numerical Developments with Applications" no. PID2023-148409NB-I00.}

\bibliography{RevKT}

\end{document}